 \documentclass[11pt,preprint]{emulateapj}

\slugcomment{Submitted to The Astrophysical Journal}

\shorttitle{UNSTABLE BAR AND SPIRAL MODES OF DISK GALAXIES}
\shortauthors{M. A. JALALI and C. HUNTER}

\begin{document}


\title{UNSTABLE BAR AND SPIRAL MODES OF DISK GALAXIES}


\author{Mir Abbas Jalali\altaffilmark{1} and 
        C. Hunter\altaffilmark{2}}
\affil{Department of Mathematics, Florida State University,
Tallahassee, FL 32306-4510}


\altaffiltext{1}{E-mail: mjalali@math.fsu.edu
}

\altaffiltext{2}{E-mail: hunter@math.fsu.edu}


\begin{abstract}
We study bisymmetric modes of angular wavenumber $2$ for flat stellar
disks in potentials with smooth cores. Stars either all circulate in
the same direction or a small fraction may counter-rotate.  The
bisymmetric modes are unstable unless there is a sufficiently large
external halo or bulge. We find two modes for each disk: a more
central fundamental mode and a more extensive and more spiral
(trailing) secondary mode. The fundamental mode is particularly
sensitive to the orbital population in the central part of the
disk. Depending on that population, it varies from a small compact bar
to a trailing spiral that is almost as wound as the secondary
mode. All modes rotate too rapidly for there to be an inner Lindblad
resonance. They transfer angular momentum from the central to the
outer regions of the disk. Most of them release gravitational energy
and convert it to kinetic energy, which also flows outwards through
the disk.
\end{abstract} 




\section{INTRODUCTION}
The publication by Lin \& Shu (1964) of a density wave theory of
spiral structure generated wide interest, and has led to a broad
literature.  One strand of that literature consists of stellar dynamic
studies of the instabilities of flat stellar systems. Important
ingredients of the analysis of Lin \& Shu are the approximations of
tight--winding, or near--axisymmetry, of the waves, and of the
near--circularity of the stellar orbits of the unperturbed disk. A key
result obtained using these approximations is that of Toomre (1964)
who showed that the stability of stellar disks to axisymmetric waves
requires that the parameter $Q=\kappa \sigma_R/3.36G\Sigma_{\rm D}
>1$. Here $\kappa$ is the epicyclic frequency, $\sigma_R$ is the
radial velocity dispersion, $\Sigma_{\rm D}$ is the disk density,
and $G$ is the gravitational constant. This result has proved to be
remarkably robust and widely applicable, despite the simplifying
approximations used in its derivation. Yet it was soon recognized that
Toomre's criterion does not provide a complete stability criterion for
stellar disks when $N$--body experiments (Miller, Prendergast \& Quirk
1970, Hohl 1971, Ostriker \& Peebles 1973) revealed that disks with
$Q$ safely greater than $1$ are prone to large--scale barlike
instabilities.

Kalnajs (1971,1977) developed the matrix method for using linear
perturbation theory to study the
stability of stellar systems. His method is quite general and does not
impose any restrictions on the forms of either the instability, or the
unperturbed stellar orbits. The matrix method is sufficiently
complex that it has not been widely used, though that is gradually
changing. The first comprehensive study using it was that of Zang
(1976) who analyzed the instabilities of scale--free singular
isothermal disks. Some of the results of his PhD thesis are given in
papers by his supervisor (Toomre 1977, 1981). Zang's methods were
re-used and extended by Evans \& Read (1998a,b) who analyzed singular
scale--free disks with other power--law rotation laws. Scale--free
disks have the property that the orbits at any energy are simply
scaled versions of those at any other energy. Though this is indeed a
simplification, even applications of the matrix method to scale--free
disks are by no means simple.

We study stellar disks with smooth and non--singular cores. 
One of the main purposes of our work is
to understand the influence that orbital populations have on the
responses of the disks to density perturbations, by studying
the responses of a variety of disks.
That variety allow us to study the effects of the mass and extent of
the disk, the orbital population, and a central bulge.
In \S\ref{sec::linear-pert-theory} we use the matrix method to derive the 
eigenvalue problem whose outcome is the pattern speed,
growth rate and mode shape of an unstable galactic disk.
We show in \S\ref{sec::boundary-integrals} 
that a boundary integral, which has been unjustifiably 
ignored in some earlier work, may occur with unidirectional disks. 
In \S\ref{sec::angularmomentumenergy} we extend the linear perturbation
theory to second order. Those results are needed for studying how modes 
transfer angular momentum and potential and kinetic energy.
In \S\ref{sec::P04theory} we discuss Polyachenko's (2004)
simplified theory of spiral and bar--like modes.
In later sections we compare our findings with its predictions.

We introduce the three axisymmetric potentials which we use for our models
in \S\ref{sec::our-models}. They are those of Kuzmin's disk,
the isochrone disk, and the cored logarithmic potential. We use known
distribution functions (DFs) for stellar disks with the first two potentials.
However, none are currently available for cored exponential disks embedded in
the cored logarithmic potential, and so we construct some DFs for them in
\S\ref{subsec::DF-exponential}. Lastly in \S\ref{subsec::bulges} we discuss a
way to model the effects of hot central bulges by modifying DFs by
cutting out a part of their orbital population.

We describe our computational procedures in
\S\ref{sec::computational-procedures}.  In \S\ref{sec::numresults} we
give the results of our computations of modes of angular wavenumber
$m=2$ obtained using the matrix method.  Some of our models are
new, while others are the same as those studied in prior work. The
prior work, some of which was done using carefully applied $N$--body
simulations, includes that of Kalnajs (1978), Earn \& Sellwood (1995)
and Pichon \& Cannon (1997) for the isochrone potential, and that of
Athanassoula \& Sellwood (1986), Sellwood \& Athanassoula (1986),
Hunter (1992), and Pichon \& Cannon (1997) for Kuzmin's potential.
Vauterin \& Dejonghe (1996) studied modes of a cored exponential disk, like
that we use in \S\ref{subsec::expdisk_in_coredlog}, in a potential
with a nearly flat rotation curve, which they obtained by combining
two Kuzmin potentials. Sellwood (1989) studied modes of the uncored and mildly
singular exponential disk in the same cored logarithmic potential as
we use for our models in \S\ref{sec::example-exponential}.

In \S\ref{sec::discussion} we discuss and interpret our computational
results. We summarize our results in \S\ref{sec::summary}.

\section{PERTURBATION THEORY}
\label{sec::dynamical-theory}
This section presents the dynamical theory on which our work is based.
\S\ref{sec::linear-pert-theory} carries out an Eulerian linear
perturbation analysis of the collisionless Boltzmann equation in
action--angle variables. We derive a matrix eigenvalue problem for
finding modes, following Kalnajs (1971,1977). We apply this analysis
to unidirectional disks in which all stars circulate in the same
direction in \S\ref{sec::boundary-integrals}, and show that the matrix
formulation must then include additional boundary integral terms. We
extend the perturbation theory to second order in
\S\ref{sec::angularmomentumenergy} to the extent necessary for
discussing the transfer of angular momentum, and kinetic and potential
energy. In \S\ref{sec::P04theory} we discuss orbits and their
responses, and relate the analysis of \S\ref{sec::linear-pert-theory}
to Polyachenko's (2004) theory.

\subsection{Linear Perturbation Theory}
\label{sec::linear-pert-theory}
We study the stability of a collisionless stellar disk
composed of a distribution of stars moving in
orbits in a circularly symmetric potential $V_0(R)$. The unperturbed
system is described by a DF $f_0(E,L)$ (Binney \& Tremaine 1987) where
\begin{equation}
\label{eq::integrals}
E=\frac 12 \left (v_R^2+v_{\phi}^2 \right ) +V_0(R), ~~ L=Rv_{\phi},
\end{equation}
are the energy and angular momentum which remain
constant along an orbit. The density corresponding to the unperturbed DF is
\begin{equation}
\label{eq::sigmadisk}
\Sigma _{\rm D}=\int f_0 {\rm d}{\bf v},
\end{equation}
where ${\rm d}{\bf v}$ denote an element of the two-dimensional 
velocity space. This density is the one that produces the potential 
$V_0(R)$ only in the fully self-consistent case; in our work it 
often provides only a part of that potential. 
The development of the DF for the perturbed
state is described by the collisionless Boltzmann equation
\begin{equation}
\label{eq::CBE}
\partial f/\partial t+[f , {\cal H}]=0,
\end{equation}
where $[,]$ denotes a Poisson bracket and ${\cal H}$ is the Hamiltonian
\begin{equation}
\label{eq::hamiltonian}
{\cal H}=\frac 12 \left (v_R^2+v_{\phi}^2 \right ) +V(R,\phi,t).
\end{equation}
We expand the DF as $f=f_0+f_1+f_2+ \cdots$ and the Hamiltonian as
${\cal H}={\cal H}_0+V_1+V_2 + \cdots$, where $V=V_0+V_1+V_2 + \cdots$
is the corresponding expansion of the potential. Collecting terms of
first and second orders then gives the equations
\begin{eqnarray}
{\partial f_1\over \partial t}+[f_1,{\cal H}_0] &=& 
                 -[f_0,V_1], \label{eq::Boltzmann-equationone} \\
{\partial f_2\over \partial t}+[f_2,{\cal H}_0] &=&
       -[f_0,V_2]-[f_1,V_1]. \label{eq::Boltzmann-equationtwo}
\end{eqnarray}
It is necessary that the perturbed densities due to the changes from the
unperturbed DF are precisely those needed to produce the corresponding
components of the perturbed density, so that
\begin{mathletters}
\begin{eqnarray}
V_j({\bf x},t) &=&  -G\int  {\Sigma _j({\bf x}^{\prime},t) 
          {\rm d}{\bf x}^{\prime} 
   \over |{\bf x}-{\bf x}^{\prime}|}, ~~ j \geq 0, \label{eq::Poisson-eq} \\
   \Sigma _j({\bf x},t) &=& \int f_j({\bf x},{\bf v},t) 
          {\rm d}{\bf v},~~ j>0. 
                             \label{eq::first-moment-eq}
\end{eqnarray}
\end{mathletters}
Here ${\rm d}{\bf x}$ and ${\rm d}{\bf v}$ denote elements of the 
two-dimensional position and velocity spaces. Equation (\ref{eq::Poisson-eq})
is true when $j=0$ because we shall use $\Sigma_0$ to denote the
self-consistent density for the potential $V_0$. 
Equation (\ref{eq::sigmadisk}) replaces
the $j=0$ case of equation (\ref{eq::first-moment-eq}).

It is convenient to work with the action variables of the unperturbed motion,
defined by
\begin{equation}
\label{eq::action-variables}
J_R=\frac 1{2\pi} \oint v_R {\rm d}R,~~
J_{\phi}=L.
\end{equation}
The first equation here defines $J_R$ as a function of $E$ and $L$.
The actions provide an alternative pair of integrals of motion, in terms
of which we can express the unperturbed DF $f_0(J_R,J_{\phi})$ 
and Hamiltonian ${\cal H}_0(J_R,J_{\phi})$. The advantage of using
action variables is that their conjugate angle variables 
$(\theta _R, \theta _{\phi})$ increase uniformly with time
\begin{equation}
\label{eq::equations-of-motion}
{{\rm d} \theta _R \over {\rm d}t} = 
{\partial {\cal H}_0\over \partial J_R}
=\Omega _R (J_R, J _{\phi}),~
{{\rm d} \theta _{\phi} \over {\rm d}t} = 
{\partial {\cal H}_0\over \partial J_{\phi}}
=\Omega _{\phi}(J_R, J_{\phi}).
\end{equation}
Perturbations must be periodic in the angles, and this allows us to use
Fourier series in these angles for them (Kalnajs 1971). 
It is convenient to write these Fourier series in the complex form
\begin{eqnarray}
f_1= e^{i(m\theta _{\phi}-\omega t)} \sum_{l=-\infty}^{\infty}
      \tilde f_{l}(J_R,J_{\phi})e^{il\theta _R}, 
      \label{eq::expansion-f} \\
V_1=e^{i(m\theta _{\phi}-\omega t)} \sum_{l=-\infty}^{\infty}
      \tilde V_{l}(J_R,J_{\phi})e^{il\theta _R},
      \label{eq::expansion-V}
\end{eqnarray}
with the understanding that it is their real parts which give 
the physical solution. Following Landau (1946), we suppose that
the frequency $\omega$ is complex with real and imaginary parts 
\begin{equation}
\label{eq::omegasplit}
\omega=m\Omega _p+is,
\end{equation}
with $\Omega _p$ representing an angular pattern speed 
and $s$ a growth rate. We suppose initially that $s>0$ so that we
have a growing disturbance which was infinitesimally small infinitely
long ago. The possibility of stationary oscillations and real values
of $\omega$ has to be considered via analytical continuation to $s=0$
from $s>0$. 

The solutions (\ref{eq::expansion-f})
and (\ref{eq::expansion-V}) are those for a perturbation of 
a single angular wavenumber $m$. Perturbations of all angular
wavenumbers are possible and could be considered 
(Kalnajs 1971, Lynden-Bell \& Kalnajs 1972) but we forgo that
generality. Different wavenumbers do not interact at the first
order when as here, the unperturbed state is axisymmetric.
In fact we shall be concerned almost entirely with the case of $m=2$.

Substituting expansions (\ref{eq::expansion-f}) and (\ref{eq::expansion-V})
into equation (\ref{eq::Boltzmann-equationone}) and matching Fourier
coefficients yields the relation
\begin{equation}
\label{eq::f-coeff-intermsof-V-coeff}
\tilde f_{l}(J_R,J_{\phi}) = { \tilde V_{l}(J_R,J_{\phi}) \over 
       l\Omega _R+m\Omega _{\phi} -\omega} \left ( 
       l{\partial f_0 \over \partial J_R}+
       m{\partial f_0 \over \partial J_{\phi}} \right ).
\end{equation}

The potential $V_1$, and the density $\Sigma _1$ which causes it and
is obtained from integrating $f_1$ as in equation (\ref{eq::first-moment-eq}),
can also be expanded in position space as
\begin{eqnarray}
V_1 &=& e^{i(m\phi-\omega t)}\sum_{j=0}^{\infty} c_j \psi^m_j(R),
                        \label{eq::pot-exp} \\
\Sigma _1 &=& e^{i(m\phi-\omega t)}\sum_{j=0}^{\infty} c_j \sigma^m_j(R),
                        \label{eq::density-exp}  
\end{eqnarray}
where $\psi^m_j(R)$ and $\sigma^m_j(R)$ are a complete set of real
basis functions, and $c_j$ are constant coefficients. We multiply
equation (\ref {eq::density-exp}) by $e^{i(\omega t - m\phi)}\psi^m_j(R)$
and integrate over position space to get 
$\Sigma_{k=0}^{\infty}D_{jk}c_k$, where
\begin{equation}
\label{eq::Djk-definition}
D_{jk}(m)= 2\pi \int\limits_{0}^{\infty} \psi^m_j(R)\sigma^m_k(R)R{\rm d}R,
\end{equation}
are the components of a constant matrix ${\bf D}(m)$.
It is diagonal if $\psi^m_j$ and $\sigma^m_k$ form a biorthogonal 
set. Alternatively, we can rewrite $\Sigma _1$ using its integral
(\ref{eq::first-moment-eq}), and then carry out the integration over
phase space using action and angle variables. This requires that we 
calculate Fourier coefficients of the basis potential functions
\begin{eqnarray}
\Psi^m_{l,j}(J_R,J_{\phi}) &=& {1\over \pi}\int\limits_{0}^{\pi}
\psi^m_j(R)     \nonumber \\
&\times& \cos [l\theta _R+m(\theta _{\phi}-\phi)]{\rm d}\theta _R,
 \label{eq::fourier-coeffs} \\
\tilde V_{l} &=& \sum_{j=0}^{\infty}c_j \Psi^m_{l,j}, 
 \label{eq::expansion-coeffs-of-hamiltonian} 
\end{eqnarray}
(Kalnajs 1977; Tremaine \& Weinberg 1984). Using equation 
(\ref{eq::f-coeff-intermsof-V-coeff}) to relate the Fourier
coefficients of $f_1$ to those of $V_1$ yields
\begin{equation}
\label{eq::matrix-equation}
[{\bf M}(m,\omega)-{\bf D}(m)]{\bf c}={\bf 0},
\end{equation}
where the components of the matrix ${\bf M}(m,\omega)$ are defined as
\begin{equation}
\label{eq::matrix-elements}
{M_{jk}\over 4\pi ^2}  =  \sum_{l=-\infty}^{\infty} 
           \int\limits_{0}^{\infty}  {\rm d}J_R 
           \int\limits_{-\infty}^{\infty} 
        { \left( l{\partial f_0\over \partial J_R}+
          m{\partial f_0\over \partial J_{\phi}} \right)
        \over l\Omega _R+m\Omega _{\phi} -\omega} 
             \Psi^m_{l,j} \Psi^m_{l,k}  {\rm d}J_{\phi}.
\end{equation}
The integration is over the whole of action space
[cf equation (\ref{eq::action-variables})], and we suppose that the
integrand decays sufficiently rapidly as $J_R \to \infty$ and
$J_{\phi} \to \pm \infty$ for the integrals to converge.
This system of linear equations (\ref{eq::matrix-equation}) has a
non-trivial solution for the coefficient vector ${\bf c}$ only if
\begin{equation}
\label{eq::dispersion-relation}
{\cal M}(m,\omega)\equiv \vert {\bf M}(m,\omega)-{\bf D}(m) \vert =0.
\end{equation}
This determinantal equation defines a nonlinear eigenvalue problem for
$\omega$. Details of how to solve it are discussed in 
\S\ref{sec::computational-procedures}. Once $\omega$ is found, 
its eigenvector ${\bf c}$ gives the physical shape of the 
perturbation.

\subsection{Boundary Integrals}
\label{sec::boundary-integrals}
The DF of a unidirectional disk for which all the stars rotate in the 
prograde direction has the form
\begin{equation}
\label{eq::sharp-cut-DF}
f_0(J_R,J_{\phi})=f_0^P(J_R,J_{\phi})H(J_{\phi}),
\end{equation}
where $H$ is the Heaviside function. The derivative of this DF with respect
to $J_{\phi}$ is
\begin{equation}
\label{eq::sharp-cut-DFderiv}
{\partial f_0 \over \partial J_{\phi}}=
{\partial f_0^P \over \partial J_{\phi}}H(J_{\phi})+
f_0^P(J_R,0)\delta(J_{\phi}).
\end{equation}
The matrix ${\bf M}(m,\omega)$ then has two components
\begin{equation}
\label{eq::matrix-elements-withBI}
{\bf M}(m,\omega)={\bf M}^{\rm A}(m,\omega)
+{\bf M}^{\rm B}(m,\omega),
\end{equation}
whose elements are
\begin{eqnarray}
\!\!{M^{\rm A}_{jk} \over 4\pi ^2} \! &=& \! \sum_{l=-\infty}^{\infty} 
           \int\limits_{0}^{\infty}  {\rm d}J_R 
           \int\limits_{0}^{\infty} 
        { \left( l{\partial f_0^P \over \partial J_R}+
          m{\partial f_0^P \over \partial J_{\phi}} \right)
        \over l\Omega _R+m\Omega _{\phi} -\omega} 
             \Psi^m_{l,j} \Psi^m_{l,k}  {\rm d}J_{\phi}, 
        \label{eq::MsupA-elements} \\
\!\!{M^{\rm B}_{jk} \over 4\pi ^2} \! &=& \! 
        \sum_{l=-\infty}^{\infty} \int\limits_{0}^{\infty}
       {\rm d}J_R  \left [ { m f_0^P(J_R,0) \Psi^m_{l,j} \Psi^m_{l,k}
          \over l\Omega _R+m\Omega _{\phi} -\omega } \right ]_{J_{\phi}=0}.
          \label{eq::MsupB-elements}
\end{eqnarray}
Because $\Omega _{\phi}=\Omega _R/2$ for radial orbits, the denominator
of the boundary integral (\ref{eq::MsupB-elements}) reduces to $-\omega$
for $l=-m/2$ when $m$ is even. Modes with $\omega=0$ are therefore precluded
when $f_0^P(J_R,0) \not\equiv 0$. 
More generally, this special component of the boundary integral can be
incorporated into an iterative scheme for computing eigenvalues, 
which we describe in \S\ref{sec::eigenvaluesearch}.

The DF (\ref{eq::sharp-cut-DF}) also drops abruptly to zero at the
circular orbit limit $J_R=0$ so that one should also include an extra
$H(J_R)$ factor in the DF. Differentiation of $f_0$ with respect to
$J_R=0$ then gives a $\delta(J_R)$.  However it gives no boundary
integral, regardless of the value of $f_0$ at $J_R=0$. The reason is
that its integrand, for which the $m f_0^P(J_R,0)$ of equation
(\ref{eq::MsupB-elements}) is replaced by $l f_0^P(0,J_{\phi})$,
vanishes at $J_R=0$.  That is because the Fourier coefficients
$\Psi^m_{l,j}(0,J_{\phi})$ vanish for all $l\neq 0$ because orbits with
$J_R=0$ are circular, and the remaining non-zero $l=0$ Fourier
coefficient is annulled by the factor $l$.  Hence the simpler form
(\ref{eq::sharp-cut-DF}) of the DF suffices.

The boundary integral (\ref{eq::MsupB-elements}) does not arise if the
unperturbed DF contains no radial orbits so that
$f_0^P(J_R,0)=0$. This is the case for the unidirectional DFs used by
Zang (1976) and Evans \& Read (1998a,b) because their DFs contain
positive powers of $L=J_{\phi}$ as factors, and hence contain no
radial orbits. They work with scalefree potentials which are singular
as $R \to 0$, whereas ours are not.  Only orbits with low angular
momenta penetrate near the center of a cored potential. As Gerhard
(1991) discusses for the analogous problem of spherical systems, most
DFs for smooth potentials which produce finite densities in their
cores tend to isotropy there and so have radial orbits. His reasoning
applies to thin disks too.  Only orbits with low angular momenta
penetrate near the center, and the only alternative to radial orbits
is a singular distribution of non-radial orbits, as when all orbits
are circular.  The need for radial orbits to provide a non--zero
central density disappears if the density of the stellar disk drops to
zero in the center, as it does with the cutouts which we discuss in
\S\ref{subsec::bulges}.

The omission of the boundary integral terms (\ref{eq::MsupB-elements})
in the cases computed by Hunter (1992) invalidates his results. As we
shall see in \S\ref{subsec::example-Kuz-disk}, the effect is
substantial. Pichon \& Cannon (1997) confirmed those results, but only
because they repeated Hunter's error of neglecting the boundary
integral. As we show in Appendix \ref{app::lagrangian}, its omission
means neglecting the contributions to potential energy and
angular momentum which arise from the perturbation of radial orbits.

Boundary integrals are avoided if one uses a Lagrangian instead of 
an Eulerian perturbation theory (Kalnajs 1977). The two forms of the
theory are complementary and we show in Appendix \ref{app::lagrangian}
that the Lagrangian results relevant to this work can be derived simply
from the Eulerian theory.

\subsection{Angular Momentum and Energy}
\label{sec::angularmomentumenergy}

The perturbations to angular momentum and energy are of second order, and
so their calculation requires considering the second
order terms from (\ref{eq::Boltzmann-equationtwo}). 
Summing the contributions from all elements in phase space gives
\begin{equation}
{\cal L}=\int\!\!\!\int J _{\phi}f {\rm d}{\bf J}{\rm d}{\Theta},
\end{equation}
for the total angular momentum, and
\begin{equation}
{\cal K}=\frac{1}{2}\int\!\!\!\int (v^2_R+v^2_{\phi})f 
{\rm d}{\bf J}{\rm d}{\Theta},
\end{equation}
for the total kinetic energy.
To compute the gravitational energy, we must distinguish between the
part $V_0^{\rm D}$ of the unperturbed gravitational potential $V_0$ 
which is due to the stars of the disk, and the remainder $V_0^{\rm ext}$ 
which is provided by some other and external source. 
The perturbational terms $V_j$, $j>0$, of the potential all
arise from the perturbed DF, and so all belong to the self-gravitational
potential. The double contribution of the external potential to the
gravitational energy is taken care of by writing it as
\begin{equation}
{\cal W}=\frac{1}{2}\int\!\!\!\int(V+V_0^{\rm ext})f 
{\rm d}{\bf J}{\rm d}{\Theta}. 
\end{equation}
The leading corrections to these quantities due to the
perturbations are of second order because all the first order terms
vanish when integrated over $\theta_{\phi}$. They are
\begin{eqnarray}
{\cal L}_2 &=& \int\!\!\!\int J _{\phi}f_2 
{\rm d}{\bf J}{\rm d}{\Theta}, \\
{\cal K}_2 &=& \int\!\!\!\int ({\cal H}_0-V_0)f_2 
{\rm d}{\bf J}{\rm d}{\Theta}
             ={\cal K}_{2,1}+{\cal K}_{2,2},
\end{eqnarray}
after using the zeroth order part of equation (\ref{eq::hamiltonian}) 
to substitute 
for the velocities, and, because $V_0^{\rm ext}=V_0-V_0^{\rm D}$,
\begin{eqnarray}
\label{eq::Wtwo}
{\cal W}_2 &=& \frac{1}{2}\int\!\!\!\int(V_2f_0+V_1f_1
                                 -V_0^{\rm D}f_2+2V_0f_2)
               {\rm d}{\bf J}{\rm d}{\Theta}, \nonumber \\
           &=& \frac{1}{2}\int\!\!\!\int f_1V_1
               {\rm d}{\bf J}{\rm d}{\Theta}
               +\int\!\!\!\int V_0f_2 {\rm d}{\bf J}{\rm d}{\Theta}, 
                                                  \nonumber \\
           &=& {\cal W}_{2,1}+{\cal W}_{2,2}.
\end{eqnarray}
The component ${\cal W}_{2,2}=-{\cal K}_{2,2}$ because their defining
integrals match. The step to the second line of equation
(\ref{eq::Wtwo}) uses the fact that the first and third terms on the
first line cancel. This is seen by transforming the integrations to
$({\bf x,v})$ space, and then using equations (\ref{eq::sigmadisk}),
(\ref{eq::first-moment-eq}) for $j=2$, and Poisson integrals like
(\ref{eq::Poisson-eq}) to relate potentials to densities, to express
them as identical integrals of the product of the densities
$\Sigma_{\rm D}$ and $\Sigma_2$.

The simplest integral to calculate is that for the first component of
${\cal W}_{2,1}$ in equation (\ref{eq::Wtwo}). As we noted earlier,
the physical parts of our solutions are given by the real parts of our
complex solutions.  The real parts of $f_1$ and $V_1$ are given by the
sums $\frac{1}{2}(f_1+\bar f_1)$ and $\frac{1}{2}(V_1+\bar V_1)$,
where a bar denotes a complex conjugate. Therefore
\begin{eqnarray}
\label{eq::Wtwoone}
{\cal W}_{2,1} &=& \frac{1}{8}\int\!\!\!\int 
          (f_1+\bar f_1)(V_1+\bar V_1)
          {\rm d}{\bf J}{\rm d}{\Theta}, \nonumber \\
  &=& \frac{1}{8}\int\!\!\!\int (f_1\bar V_1 +V_1 \bar f_1)
      {\rm d}{\bf J}{\rm d}{\Theta},     \nonumber \\
  &=& e^{2st} \sum_{l=-\infty}^{\infty} W_{2,1}^l,
\end{eqnarray}
where the components $W_{2,1}^l$ are defined by
\begin{eqnarray}
\label{eq::W_2_1_ell}	
W_{2,1}^l &=& \pi^2 \int {\rm d}{\bf J}
 \left(l{\partial f_0\over \partial J_R}
       +m{\partial f_0\over \partial J_{\phi}} \right) \nonumber \\
 &\times&      {[l\Omega _R+m(\Omega _{\phi}-\Omega_p)]|\tilde V_{l}|^2
       \over |l\Omega _R+m\Omega _{\phi} -\omega|^2}.
\end{eqnarray}
We have used here, and will again, the fact that the only products which 
do not vanish on integration over $\theta_{\phi}$ are those which pair 
a conjugate with a non-conjugate quantity.

We show in appendix \ref{app::angmomenergy} that
\begin{equation}
\label{eq::Ltwo}
 {\cal L}_2(t)= e^{2st} \sum_{l=-\infty}^{\infty} L_2^l,~~
 {\cal K}_{2,1}(t)= e^{2st} \sum_{l=-\infty}^{\infty} K_{2,1}^l,
\end{equation}
where the components of these sums are defined by the integrals
\begin{eqnarray}
 L_2^l &=& -m\pi^2 \int {\rm d}{\bf J}
       \left(l{\partial f_0\over \partial J_R}
       +m{\partial f_0\over \partial J_{\phi}} \right) \nonumber \\
    &\times& {|\tilde V_{l}|^2
        \over |l\Omega _R+m\Omega _{\phi} -\omega|^2}, \label{eq::L_2_ell} \\
 K_{2,1}^l &=& -\pi^2 \int {\rm d}{\bf J}
 \left(l{\partial f_0\over \partial J_R}
       +m{\partial f_0\over \partial J_{\phi}} \right) \nonumber \\
    &\times& {(l\Omega _R+m\Omega _{\phi})|\tilde V_{l}|^2
        \over |l\Omega _R+m\Omega _{\phi} -\omega|^2}. \label{eq::K_2_1_ell}
\end{eqnarray}
To each of the area integrals (\ref{eq::W_2_1_ell}), 
(\ref{eq::L_2_ell}) and (\ref{eq::K_2_1_ell})
must be added the boundary integrals given by the delta function term
of equation (\ref{eq::sharp-cut-DFderiv})
for the prograde DF (\ref{eq::sharp-cut-DF}).
The integrals can be combined to give the simple relation
\begin{equation}
\label{eq::KWLcomponentbalance}
K_{2,1}^l+ W_{2,1}^l=\Omega_p L_2^l,
\end{equation}
between the separate components. Although each of these components can be
found directly from the first order solution, computing 
${\cal W}_{2,2}=-{\cal K}_{2,2}$ requires more, as we show in
Appendix \ref{app::angmomenergy}.

The second order corrections ${\cal L}_2$ and ${\cal W}_{2,1}$ to the 
angular momentum and gravitational energy have simple representations in terms
of the real and imaginary parts of the matrix ${\bf M}={\bf M}_{\rm R}
+i{\bf M}_{\rm I}$. Combining equations (\ref{eq::L_2_ell})
and (\ref{eq::W_2_1_ell})  with expansion
(\ref{eq::expansion-coeffs-of-hamiltonian}), we get the quadratic forms
\begin{eqnarray}
{\cal L}_2(t)  &=&-\frac{m}{4s}e^{2st}\bar{\bf c}^T{\bf M}_{\rm I}{\bf c},
  \label{eq::quadformL} \\
{\cal W}_{2,1}(t) &=& \frac{1}{4}e^{2st}\bar{\bf c}^T{\bf M}_{\rm R}{\bf c},
  \label{eq::quadformWtwoone}
\end{eqnarray}
where the superscript $T$ denotes transposition of the column vector
${\bf c}$ to generate a row vector. These expressions are real because
the matrix ${\bf M}$ is symmetric. Moreover multiplying equation 
(\ref{eq::matrix-equation}) by $\bar{\bf c}^T$ and separating real and
imaginary parts (${\bf D}$ is also symmetric) gives
\begin{equation}
\label{eq::twoquadforms}
\bar{\bf c}^T{\bf M}_{\rm R}{\bf c}=\bar{\bf c}^T{\bf D}{\bf c},~~
\bar{\bf c}^T{\bf M}_{\rm R}{\bf c}=0.
\end{equation}
The second relation shows that ${\cal L}_2(t)=0$, which it must be because 
the disk is not subject to any external torques,
and hence its angular momentum is conserved.
There is no such restriction on sizes of the separate Fourier components
represented by the terms for different $l$ in the sum
(\ref{eq::Ltwo}), other than that they must sum to zero. Similarly the
sizes of the different components of the potential and kinetic energy
vary, because only the total energy ${\cal E}$ is constrained to be
zero, with ${\cal E}_2(t)={\cal K}_2(t)+{\cal W}_2(t)=\Omega_p{\cal
L}_2(t)=0$. Hence the two components of the kinetic and potential energy
are related in the same way: 
\begin{equation}
\label{eq::KWbalance}
{\cal K}_{2,1}(t)=-{\cal W}_{2,1}(t), ~~ {\cal K}_{2,2}(t)=-{\cal W}_{2,2}(t).
\end{equation}

The fact that the second order components grow twice as fast as those
of first order is not paradoxical. It reflects the fact that our analysis
can describe only the early stages of the growth of an instability.
If we introduce a small ordering parameter $\varepsilon$ into our
expansion $f=f_0+\varepsilon f_1+\varepsilon^2 f_2+ \cdots$ of the
DF to measure the size of the perturbation relative to that of the
unperturbed state, then we see that the linearization breaks down,
and our analysis is unreliable, after a time $t$ such that
$\varepsilon e^{st}=O(1)$. The second order 
components are then of magnitude $\varepsilon^2 e^{2st}$, which is
also $O(1)$. Our perturbation theory has then ceased to be valid.
It is useful only when our expansion is well-ordered,
that is for times for which $\varepsilon e^{st}$ is small.

\begin{figure}
\plotone{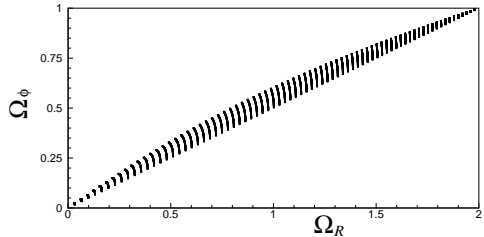}
\caption{The orbital frequency space of the cored logarithmic potential
(\ref{eq::pot-coredlog}) with frequencies in units of $v_0/R_{\rm C}$. The lines 
are curves of constant $E$ in equal steps in $e^{-E}$. 
They show that $\Omega_R$ depends only weakly on 
the angular momentum $L$, which varies from zero at their lower limit on the
boundary $\Omega_{\phi}=0.5\Omega_R$ which corresponds to radial orbits,
to $L_c(E)$ on the curved upper limit which corresponds to circular orbits. 
The largest value achieved by $\Omega_i=\Omega_{\phi}-0.5\Omega_R$ is $0.106$.
Similarly narrow lens-shaped plots are obtained for other cored potentials. 
When normalized to the same range of frequencies as in Fig.\ref{fig1}, the largest value of $\Omega_i$ is $0.130$ for Kuzmin's disk and
$0.119$ for the isochrone. $\Omega_R$ depends only on $E$ for
the isochrone, so that curves of constant $E$ are then exactly straight.
\label{fig1}}
\end{figure}

\subsection{Orbits and Polyachenko's Unified Theory}
\label{sec::P04theory}

Every orbit in a disk with an axisymmetric potential is a rosette
(Binney \& Tremaine 1987). One extreme case is that of circular orbits
with the maximum angular momentum $L_c(E)$ for the energy $E$. They have 
no radial motion and their radial action $J_R=0$. The other extreme is
that of radial orbits with zero angular momentum $J_{\phi}$. For them
the polar angle $\phi$ remains constant, except for discontinuous
changes by $\pi$ whenever the orbit passes through the
center. Intermediate orbits oscillate in $R$ between maximum and
minimum orbital radii $R_{\rm max}$ and $R_{\rm min}$. Figure
\ref{fig1} shows
an orbital frequency space for prograde orbits. Its shape is 
characteristic of those of other cored potentials. The upper curved
boundary is formed by circular orbits for which
$\Omega_{\phi}=V_c(R)/R$, where $V_c(R)$ is the circular velocity at
radius $R$, and $\Omega_{R} =\kappa(R)$ is the epicyclic
frequency. The straight lower boundary is formed by radial orbits for
which $\Omega_{\phi}=\Omega_{R}/2$.  The slimness of the region in
between these boundaries, where all the intermediate orbits lie, shows
how limited is the range of $\Omega_{\phi}-\Omega_{R}/2$ for all
orbits.

The frequency $\Omega_i=\Omega_{\phi}-\Omega_{R}/2$ and its narrow range
has long been recognized as dynamically important (Lindblad 1959).
Lynden-Bell (1979) pointed out that the response of an orbit to a
weak bar-like potential, which rotates with a pattern speed $\Omega_p$
for which $|\Omega_p-\Omega_i|$ is small, is to align the orbit with the 
bar if $\Omega_i$ decreases as $J_{\phi}$ decreases when $J_f=J_R+J_{\phi}/2$,
the action corresponding to the frequency $\Omega_i$,  
is fixed. This is the case when 
\begin{equation}
\label{eq::abnormalcondition}
\left( {\partial \over \partial J_{\phi}}-\frac{1}{2}
       {\partial \over \partial J_R} \right) \Omega_i=
       {\partial \Omega_i \over \partial L}+
       \Omega_i {\partial \Omega_i \over \partial E} > 0.
\end{equation}
Lynden-Bell labeled such orbits as abnormal, as opposed to normal or
donkey orbits which respond contrarily to a bar-like force. It is
straightforward to classify orbits in any potential according to the
criterion (\ref{eq::abnormalcondition}); abnormal orbits are those
which lie below the dashed lines in Figure \ref{fig4}.  All
sufficiently radial orbits are abnormal, but so too are all orbits
which remain in the central regions of cored potentials. The dividing
curve for the cored logarithmic potential in Figure \ref{fig4} tends
to the asymptote $L/L_c=0.723$ as $E \to \infty$ 
(See \S \ref{subsec::P04results}). That asymptote,
which is barely evident due to the compressed scale
at high energies, is the dividing curve between normal and abnormal
orbits for the singular logarithmic potential at all energies.

Polyachenko (2004) has outlined a unified theory of spiral and
bar-like modes of disk galaxies based on Lynden-Bell's (1979)
analysis. The analytical basis of his theory is a simplified version
of the analysis given in \S\ref{sec::dynamical-theory}. He uses the
smallness of $|\Omega_p-\Omega_i|$ to deduce that, to leading order,
the perturbed DF is given by a single $l=-m/2$ component of the
Fourier series (\ref{eq::expansion-f}) so that, in our notation,
\begin{equation}
\label{eq::shortexpansion-f}
f_1 \simeq e^{i[m(\theta_{\phi}-\theta_R/2)-\omega t]}
              \tilde f_{-m/2}(J_R,J_{\phi}).
\end{equation}
Polyachenko's equation (11) relates the perturbed DF to its potential
when averaged over the `fast' angle $\theta_R$.  That averaged
potential is given in our notation by the single $l=-m/2$ component of
the Fourier series (\ref{eq::expansion-V}). Polyachenko's equation
(11) is therefore equivalent to the relation
\begin{equation}
\label{eq::simpleFouriermatch}
(m\Omega_i-\omega)\tilde f_{-m/2}=m\left(
{\partial f_0 \over \partial J_{\phi}}-\frac{1}{2}
   {\partial f_0 \over \partial J_R}\right) \tilde V_{-m/2}.
\end{equation}
He has a different method for solving his equation (11) which has the
advantage that he is able to derive a linear eigenvalue problem for
$\omega$, albeit one which must be solved for an unknown function of
the two action variables. Equation (\ref{eq::simpleFouriermatch})
shows that his method is equivalent to the Kalnajs matrix formulation
when that formulation is simplified by replacing the sum over $l$ in
the coefficient equation (\ref{eq::matrix-elements}) by the single
$l=-m/2$ term, or just the $l=-1$ term in the important $m=2$
case. Hence the matrix method can be applied to check the accuracy and
validity of Polyachenko's approximations by comparing its results with those
obtained when only the $l=-1$ terms of the matrix ${\bf M}(2,\omega)$
are used.  Such comparisons are made throughout \S\ref{sec::numresults},
and discussed in \S \ref{subsec::P04results}.

\section{MODELS FOR GALACTIC DISKS}
\label{sec::our-models}
In \S\ref{subsec::kuzminsdisk}, \S\ref{subsec::isochronedisk}, and
\S\ref{subsec::coredlogpot} we give the three potentials which we use
for stability calculations; Kuzmin's disk, the isochrone, and the
cored logarithmic potential (Kuzmin 1956, Binney \& Tremaine 1987).
The rotation curves for both Kuzmin's disk and the isochrone rises at
small radii, peaks at a characteristic radius, and then falls like a
Keplerian potential at large distances. That for the cored logarithmic
potential gives a flat rotation curve at large distances.  All
potentials have smooth cores.  In \S\ref{subsec::expdisk_in_coredlog}
we discuss constraints on the sizes of exponential stellar disks 
which lie in cored logarithmic
potentials.  In \S\ref{subsec::DF-exponential} we construct DFs for
such stellar disks.  In \S\ref{subsec::bulges} we discuss cutout
functions which can be applied to DFs to model hot central bulges.

\subsection{Kuzmin's Disk}
\label{subsec::kuzminsdisk}
Kuzmin's disk is known also as the Kuzmin-Toomre disk because it is
Model $1$ of a family given by Toomre (1963). Its potential
and self-consistent density are
\begin{equation}
\label{eq::pot-den-KT}
V_0(R)={-GM\over \sqrt{R^2+R^2_{\rm C}}},~
\Sigma _0(R) = {MR_{\rm C}\over 2\pi (R^2+R^2_{\rm C})^{3/2}},
\end{equation}
where $R_{\rm C}$ is the core radius of the potential. 
Kuzmin's disk has been widely used for stability
studies because of its simplicity. Sellwood \& Athanassoula (1986)
and Athanassoula \& Sellwood (1986) used it for $N$--body 
simulations, while Hunter (1992), Pichon \& Cannon (1997) and 
Polyachenko (2004) have used it for analyses based on the 
collisionless Boltzmann equation. To have results which can be 
compared with those of earlier work, we use the DFs given by 
Miyamoto (1971) and those used in Athanassoula \& Sellwood's simulations.
 
\subsection{The Isochrone Disk}
\label{subsec::isochronedisk}

The potential and self-consistent density of the isochrone disk are
\begin{eqnarray}
\label{eq::pot-den-isochrone}
\!\!\! V_0(R) &=& {-GM\over R_{\rm C}+\sqrt{R^2+R^2_{\rm C}}}, \\
\!\! \Sigma _0(R) &=& {MR_{\rm C} \over 2\pi R^3} 
\!\!\! \left [ \ln \!\! \left ( \!\!{R+\sqrt{R^2+R_{\rm C}^2} \over R_{\rm C}} \right ) 
\!\! - \! {R\over \sqrt{R^2+R_{\rm C}^2}} \! \right ].
\end{eqnarray}
Again $R_{\rm C}$ is the core radius of the potential. 
We use the DFs given in Kalnajs (1976b),
modified so as to allow for a population of retrograde stars in the
manner specified in Earn \& Sellwood (1995). The stability of these
models was investigated by Kalnajs (1978), Earn \& Sellwood (1995) and
Pichon \& Cannon (1997).

\subsection{The Cored Logarithmic Potential}
\label{subsec::coredlogpot}

The cored logarithmic potential, which has been widely used in 
galactic studies because of its flat rotation curve, is 
\begin{equation}
\label{eq::pot-coredlog}
V_0(R)= v_0^2 \ln \sqrt{1+R^2/R^2_{\rm C}},
\end{equation}
where $v_0$ is the flat rotation velocity, and $R_{\rm C}$ is again the
core radius. The disk density needed to produce the potential
(\ref{eq::pot-coredlog}) is
\begin{equation}
\label{eq::den-coredlog}
\Sigma _0(R)=\frac{v_0^2}{2\pi G \sqrt{R^2+R^2_{\rm C}}}
             E\left(\frac{R^2}{R^2+R^2_{\rm C}}\right), 
\end{equation}
where $E$ is the complete elliptic integral of the second kind. This
self-consistent density may be derived by Toomre's (1963) Bessel
function method, using in turn formulas (6.566.2), (6.576.3),
(9.131.1), and (8.114.1) of Gradshteyn \& Ryzhik (1980, hereafter GR).
It can also be obtained by taking the $\beta \to 0$ limit in equations
(4.7) of Qian (1992) for $m=0$ and $\gamma_1=0.5$. The limit is
straightforward for the density, but it is necessary first to subtract
the constant value of the potential at $R=0$ before taking the $\beta
\to 0$ limit for the potential.

The potential--density pair (\ref{eq::pot-coredlog}) and
(\ref{eq::den-coredlog}) is similar to that of the Rybicki disk 
(Zang 1976; Evans \& Collett 1993) for which
\begin{eqnarray}
V_0(R)&=& v_0^2 \ln \left ( 1+\sqrt{1+R^2/R^2_{\rm C}} \right ),
\label{eq::pot-rybicki} \\
\Sigma _0(R)&=&\frac{v_0^2}{2\pi G \sqrt{R^2+R^2_{\rm C}}}.
\label{eq::den-rybicki}
\end{eqnarray}
Rybicki's disk is obtained by subtracting Toomre's (1963) Model $0$
from a singular Mestel (1963) disk. The density
(\ref{eq::den-coredlog}) for the cored logarithmic potential is larger
than that of the Rybicki disk in the center, which is why it has
somewhat larger circular velocities.  Both disks resemble the singular
Mestel disk, whose stability has been studied by Zang (1976) and Evans
\& Read (1998a,b), at large distances.  Both tend to that singular
limit as the core radius $R_{\rm C} \to 0$.

\subsection{Exponential Disks in the Cored Logarithmic Potential}
\label{subsec::expdisk_in_coredlog}

It is widely accepted that the densities of the stellar disks of
spiral galaxies decay exponentially with distance (Freeman 1970). 
We study cored disks with densities of the form
\begin{equation}
\label{eq::den-expdisk}
\Sigma_{\rm D}(R)=\Sigma_s \exp
           \left[-{\sqrt{R^2 + R_{\rm C}^2} \over R_{\rm D}}\right],
\end{equation}
where $R_{\rm D}$ is the length--scale of the exponential decay, and
$\Sigma_s$ is a density scale. We include the core radius $R_{\rm C}$ of
the potential in the density (\ref{eq::den-expdisk}) to avoid the
logarithmic singularity of the uncored exponential disk 
$\Sigma_{\rm D} = \Sigma_s \exp(-R/R_{\rm D})$ (Freeman 1970).
Vauterin \& Dejonghe (1996) did likewise.

The self--gravitational potential due to the cored disk density 
(\ref{eq::den-expdisk})
can be derived by the method of Evans \& de Zeeuw (1992) as
\begin{eqnarray}
V_0^{\rm D}(R) &=& -2\pi G \Sigma_s R_{\rm D} \nonumber \\
&\times& \! \int\limits_{0}^{\infty} \!\!\!
{x^2 J_1(x) {\rm d}x \over \sqrt {x^2+\lambda ^2}
\sqrt {x^2+\lambda ^2+R^2/R_{\rm D}^2} }.
\label{eq::potential-self-exp-disk}
\end{eqnarray}
Here $\lambda =R_{\rm C}/R_{\rm D}$ and $J_1$ is the Bessel function of the first
kind. Beware that this integral must be evaluated with great care
because it is oscillatory and its amplitude decays only as
$x^{-1/2}$ as $x \to \infty$.  $V_0^{\rm D}(R)$ reduces to the
potential of the exponential disk in the limit of
$R_{\rm C} \to 0$ when $\lambda \to 0$.  

\begin{figure}
\plotone{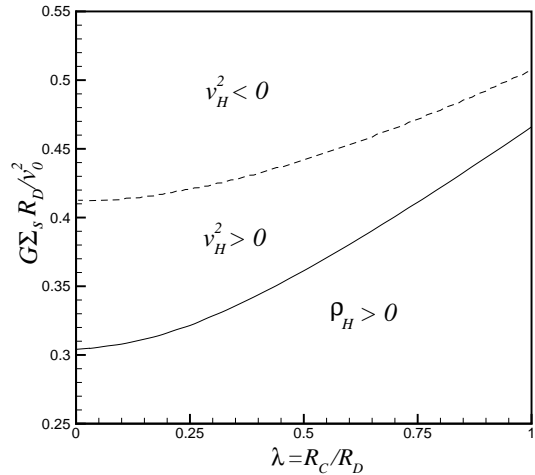}
\caption{The boundary of physical exponential disks (solid line) in 
the parameter space. Models below the solid line have spherical halos 
with $\rho _{\rm H}>0$ at all radii. 
\label{fig2}}
\end{figure}

There are limits on the scales of a cored exponential disk
(\ref{eq::den-expdisk}) lying in the cored logarithmic potential
(\ref{eq::pot-coredlog}).  The difference $[V_0(R)-V_0^{\rm D}(R)]$
between the total potential and that of the disk must be provided by
some halo or bulge component of the galaxy, whose density must
be everywhere positive. If that halo/bulge is spherical, then its 
density is given by Poisson's equation as
\begin{eqnarray}
\rho_{\rm H}(r) & = & {1\over 4\pi G} {1\over r^2}
{{\rm d}\over {\rm d}r} \left [ rv_{\rm H}^2(r) \right], 
\label{eq::halodensity}  \\
v_{\rm H}^2(R) & = & R \left [ {{\rm d}V_0(R) \over {\rm d}R}-
            {{\rm d}V_0^{\rm D}(R) \over {\rm d}R} \right] \ge 0,
\label{eq::halovelsquared}
\end{eqnarray}
(Zang 1976; Vauterin \& Dejonghe 1996) where $r=\sqrt{R^2+z^2}$.  Here
$v^2_{\rm H}(R)$ gives the amount by which the square of the circular
velocity for the galaxy exceeds that of the disk alone. The physically
necessary condition that $\rho _{\rm H}(r)>0$ requires that $v^2_{\rm
H}(R)>0$, but is more restrictive. The region below the solid curve in
Figure \ref{fig2} gives the range of the dimensionless combinations of
parameters $G\Sigma_sR_{\rm D}/v_0^2$ and $R_{\rm C}/R_{\rm D}$ which it allows. The
boundary value $G\Sigma_sR_{\rm D}/v_0^2=0.304$ of the solid curve at
$\lambda=0$ applies to the limit of the uncored exponential disk in a
singular logarithmic potential. An alternative statement of the
condition $\rho _{\rm H}(r)>0$ in this limit is
$v_0(R_{\rm D}/GM_{\rm D})^{1/2}>.723$, where $M_{\rm D}$ 
is the mass of the uncored exponential disk.

Sellwood (1989), and unpublished work by Toomre described there,
studied modes of the uncored exponential disk in the cored logarithmic
potential (\ref{eq::pot-coredlog}). The critical value of $v_0(R_{\rm
D}/GM_{\rm D})^{1/2}$ needed to ensure that $\rho_{\rm H}>0$ for this
case is not greatly changed when the logarithmic potential is cored.
It is reduced only to $0.691$ when $\lambda=R_{\rm C}/R_{\rm D}=0.5$.  For
$\lambda=0.2$ and $v_0(R_{\rm D}/GM_{\rm D})^{1/2} =0.6$ as in the
model displayed in Sellwood's Figures 1 and 2, not only does
$\rho_{\rm H}$ become negative, but $v^2_{\rm H}(R)<0$ for
$1.4<R/R_{\rm D}<3.0$. The fixed halo therefore exerts an outward
force in this range. (The singular nature of the potential of the
uncored exponential disk also causes $\rho_{\rm H}<0$, but only for
$R/R_{\rm D} \ll 1$ which is well below the softening length used in
computations.)

\begin{figure}
\plotone{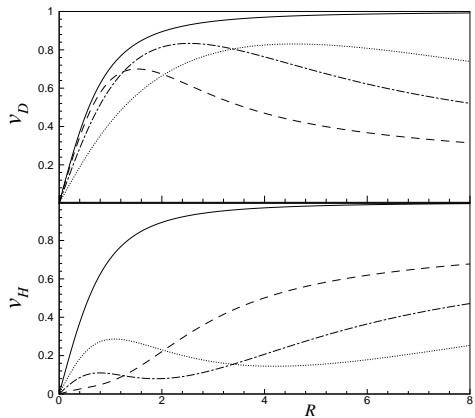}
\caption{The rotational velocities due to the disk (upper panel)
and halo/bulge (lower panel) components of the exponential disk, in
units in which $R_{\rm C}=v_0=1$. In each case
the dashed, dot-dashed, and dotted lines correspond to
$(R_{\rm D},G\Sigma _s R_{\rm D})=(0.5,0.6)$, $(1,0.42)$ and $(2,0.32)$,
respectively, while the the solid line shows the 
circular velocity of the cored logarithmic potential.
\label{fig3}} 
\end{figure}

Giovanelli \& Haynes (2002) analyzed a large number of rotation curves
and have found that the ratio of the scale length of the steep inner
rise of the rotation curve to the scale length of the exponential
varies from 0.63 for the most luminous galaxies, to 1.28 for their
least luminous. They fit a parametric model in which the rotation
curve decays exponentially towards its form in the outer regions, and
hence their ratios provide only the approximate guidance that it is to
reasonable to use values of order unity for our ratio $R_{\rm
C}/R_{\rm D}$.  Figure \ref{fig3} plots total circular velocity, and
the parts provided by the disk and the halo/bulge for three different
relative sizes of the exponential disk which more than cover their
range. The value of $G\Sigma_sR_{\rm D}/v_0^2$ is close to 90\% of the
maximum allowable for that $R_{\rm C}/R_{\rm D}$ in each case. The
least extensive disk makes the largest contribution to the rotation
curve at the center, but its relative contribution then declines
rapidly. The disk's contribution tracks the rotation curve
considerably further in the intermediate case with $R_{\rm D}=R_{\rm
C}$, and requires only a relatively small contribution from a central
halo/bulge to make up the deficit. As $R_{\rm D}$ increases, the disk
becomes less maximal and an increasingly large spherical central
halo/bulge is needed.

\begin{figure*}
\plottwo{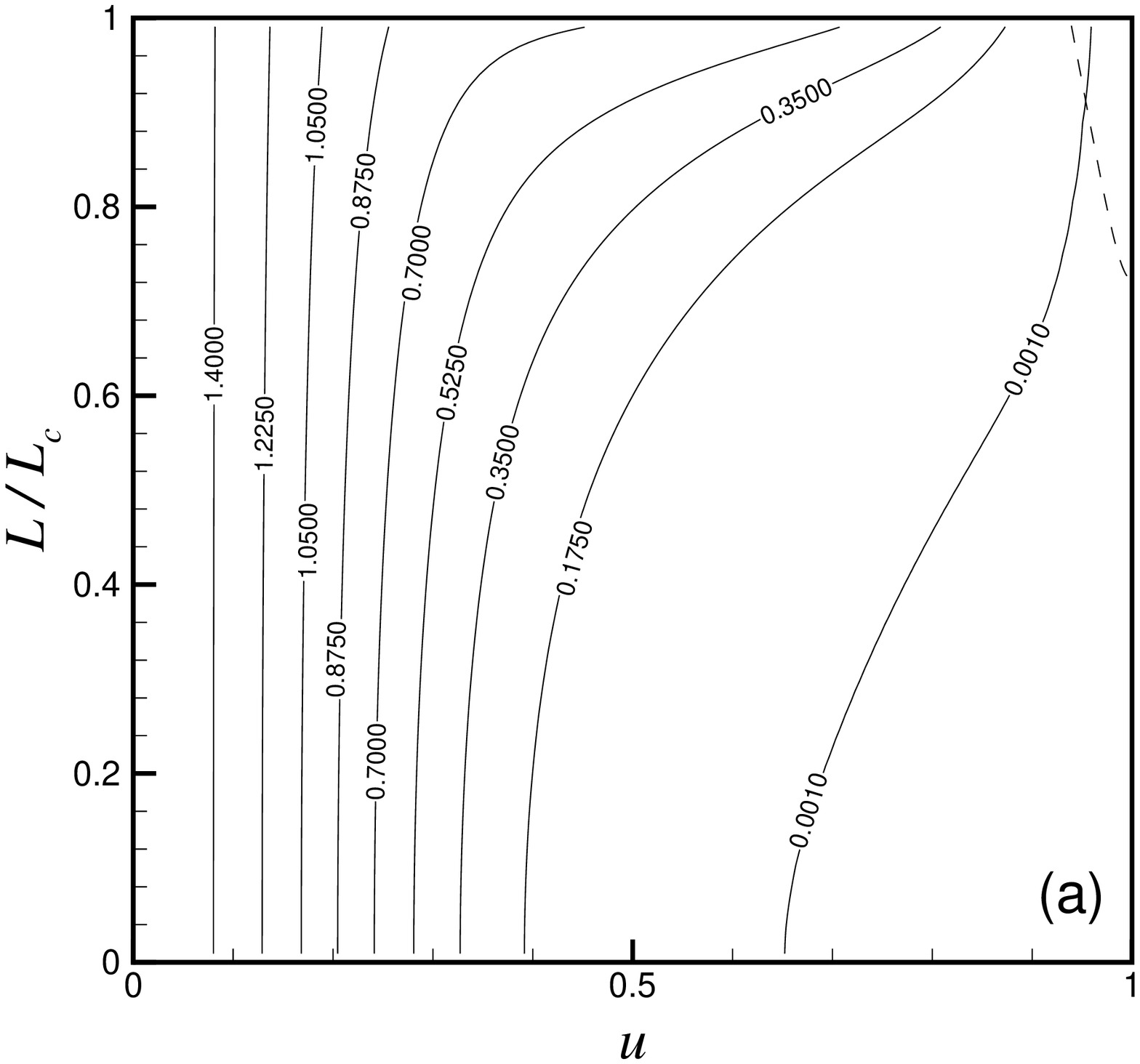}{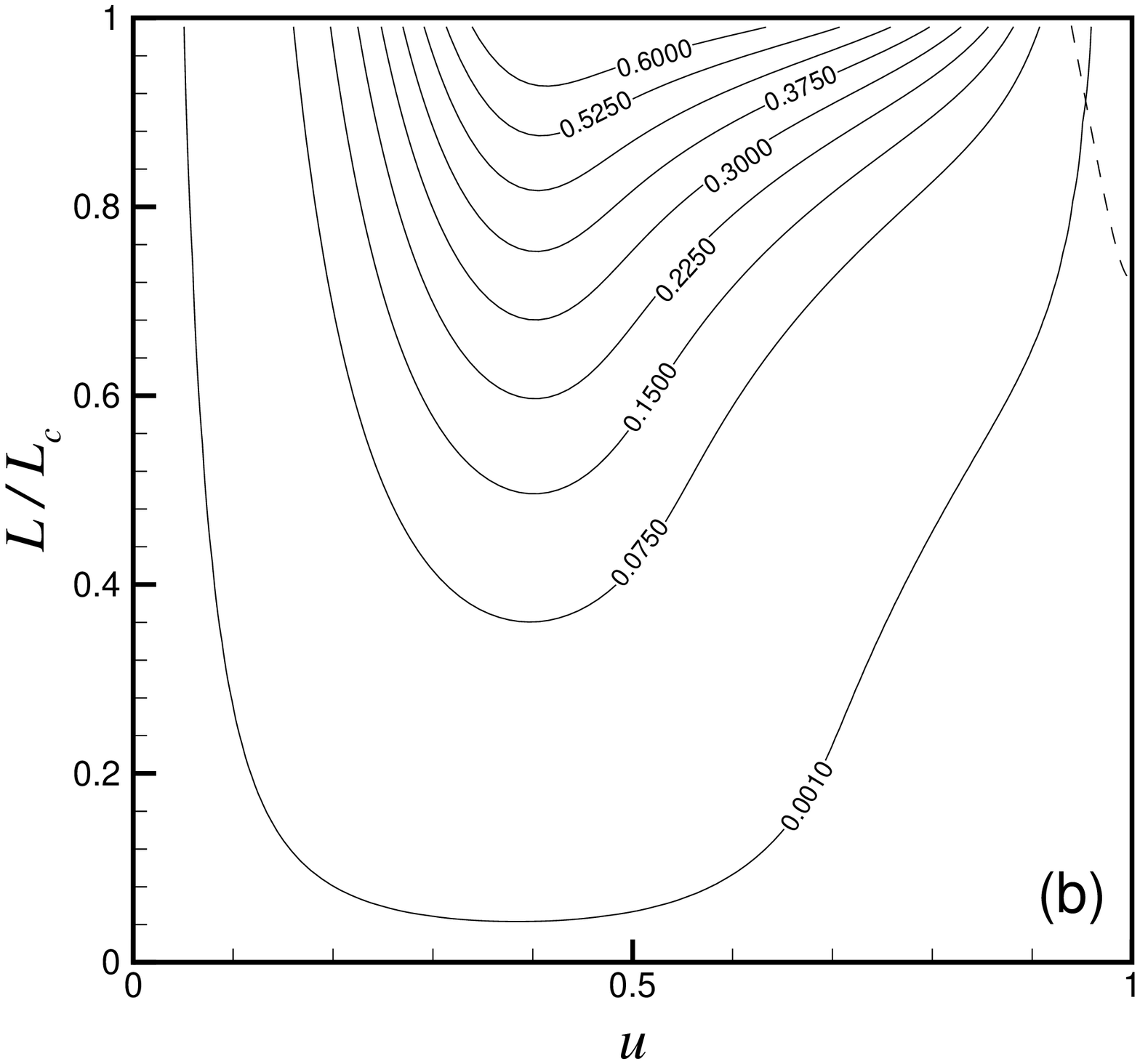}
\figcaption{(a) The DF $f_0^P$ of the exponential disk for 
$N=6$ and $R_{\rm C}=R_{\rm D}=1$, and (b) that DF after applying
the cutout (\ref{eq::ourcutout}) with $L_0=0.1$.
The abscissa $u=\sqrt{1-e^{-E}}$ is that used in numerical work,
as defined in equation (\ref{eq::map-rising}). It ranges from $u=0$
at the center to $u=1$ at infinite distances.
Contours are labeled in units of $\Sigma_s$, and show slopes which 
increase with increasing $N$. The dashed line shows the curve
$\partial \Omega_i/\partial L+\Omega_i \partial \Omega_i/\partial E=0$.
Only orbits which lie above this curve are normal in
Lynden-Bell's (1979) classification.
\label{fig4}}
\end{figure*}

\begin{figure*}
\plottwo{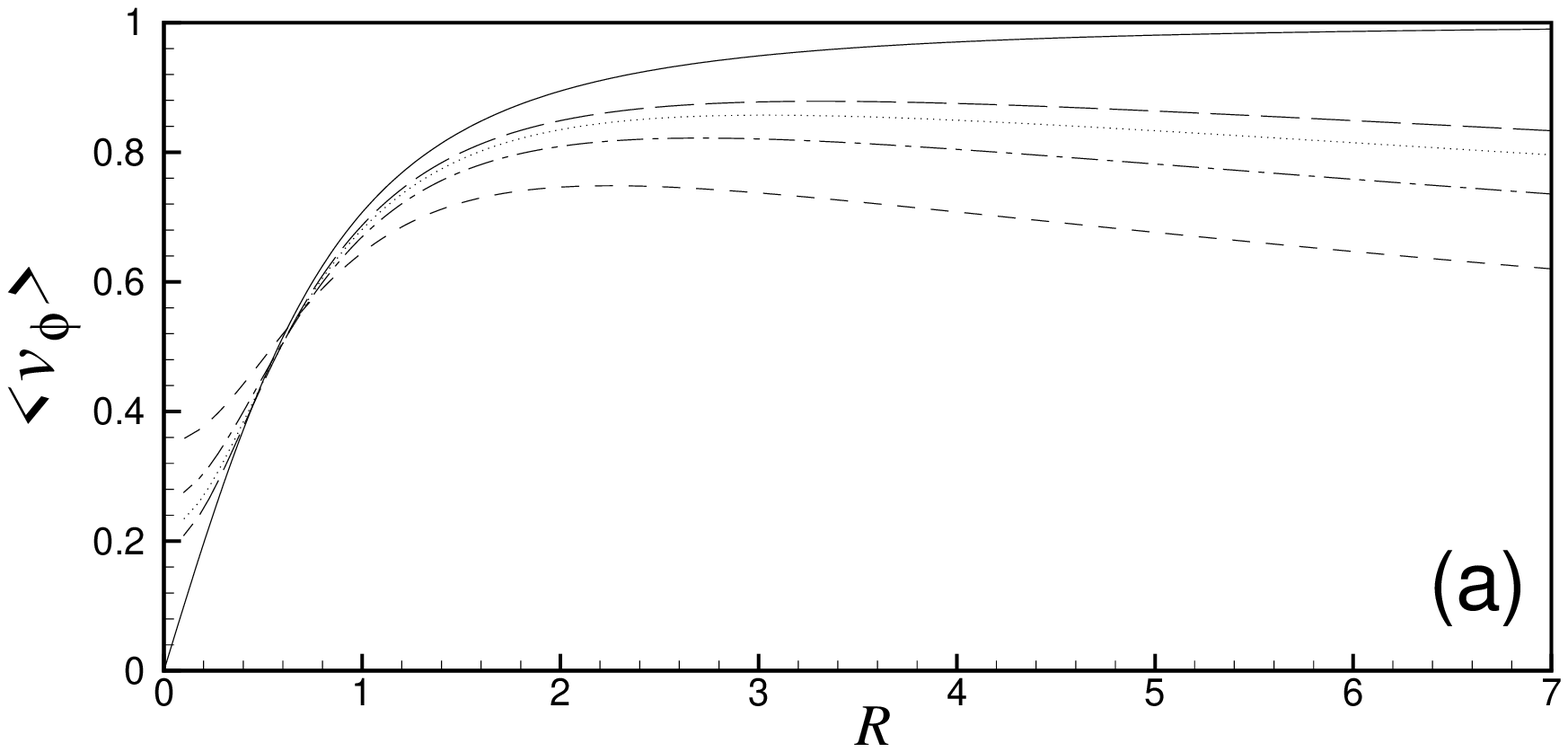}{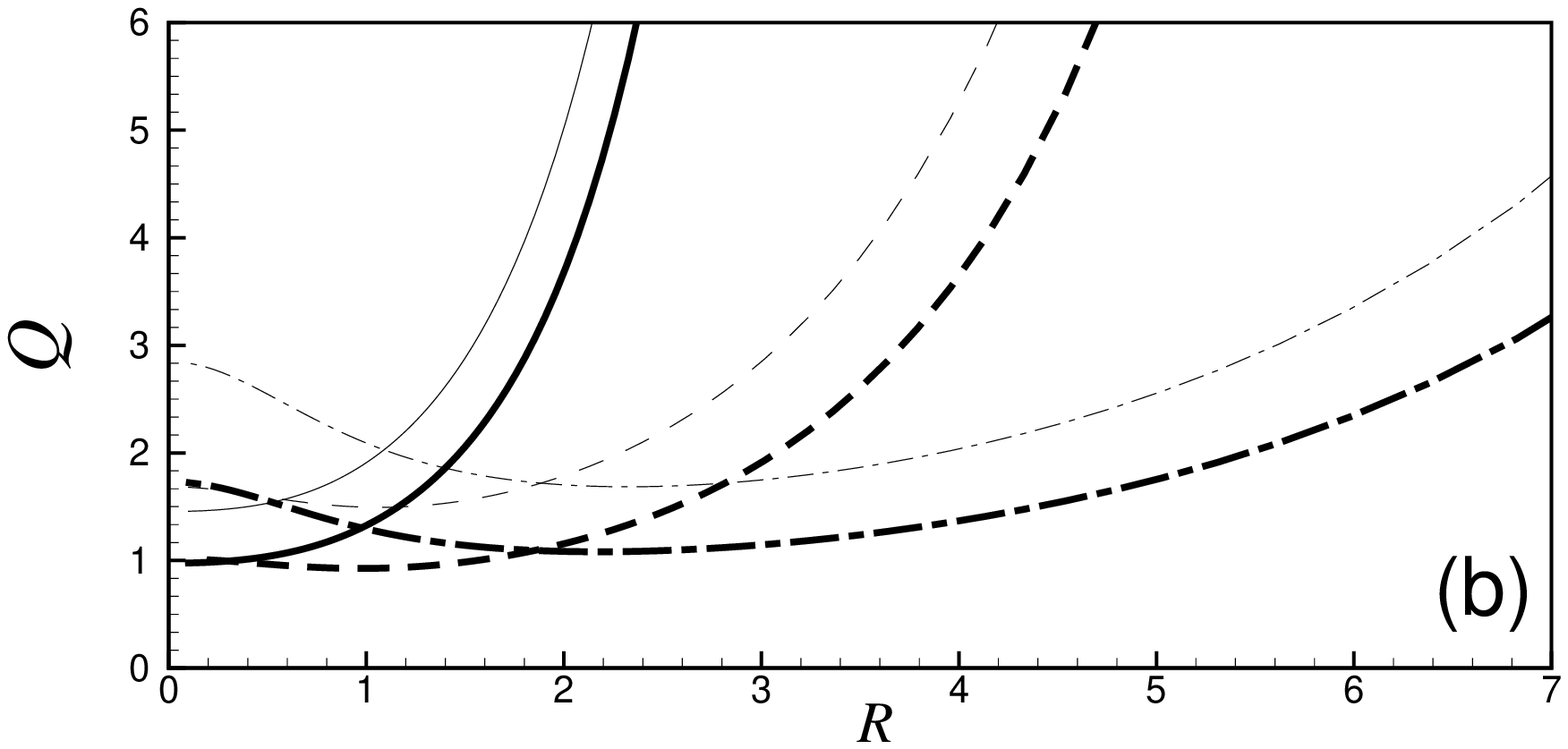}
\caption{(a) Circular velocity (solid line) and
mean rotation velocities of the exponential disk. Dashed, dot-dashed,
dotted and long-dashed lines respectively show the mean rotation velocity
$\langle v_{\phi} \rangle$ of exponential disks with $N=2$, 4, 6 and 8.
All curves are for $R_{\rm C}=R_{\rm D}=v_0=1$.
(b) Toomre's $Q$ for the exponential disk for $R_{\rm C}=1$ with 
$N=2$ (thin lines) and $N=6$ (thick lines). Solid, dashed and dot-dashed 
lines here correspond to $(R_{\rm D},G\Sigma _sR_{\rm D})=(0.5,0.6)$, 
$(1,0.42)$ and $(2,0.32)$, respectively as in Figure \ref{fig3}. 
\label{fig5}}
\end{figure*}

\subsection{Distribution Functions for the Exponential Disk}
\label{subsec::DF-exponential}

We construct DFs by using the identity
\begin{equation}
\label{eq::logpot-identity}
e^{-\Phi}\sqrt{1+{R^2 \over R_{\rm C}^2}}=1, ~~ \Phi=\frac{V_0}{v_0^2},
\end{equation}
to partition the density (\ref{eq::den-expdisk}) in the form
\begin{equation}
\Sigma _{\rm D}(R)=\Sigma_s e^{-\lambda e^{\Phi}}e^{-2N\Phi}
          \left(1+\frac{R^2}{R_{\rm C}^2}\right)^N, ~~\lambda=\frac{R_{\rm C}}{R_{\rm D}}.
\end{equation}
Here $N$ is an integer parameter which allows us to generate a family
of models. Binomial expansion of $(1+R^2/R_{\rm C}^2)^N$
gives $\Sigma _{\rm D}$ as a sum of powers of $R^2$ multiplied by
functions of the potential $\Phi$. We then use Sawamura's (1988) method,
following Evans \& Collett (1993) \S 3.1, to find the DF as the series
\begin{equation}
\label{eq::new-DF--EXP}
f^P_0(E,L) = \Sigma_s \sum_{n=0}^{N}
{N\choose n} \left({L \over R_{\rm C}}\right)^{2n}g_n(E),
\end{equation}
where the functions $g_n(E)$ are given by
\begin{eqnarray}
g_n(E)  &=&  { (-1)^{n+1} \over 
         2^n \sqrt{\pi} \Gamma (n+1/2)}    \nonumber \\
       &{}&  \qquad  \times {{\rm d}^{n+1} \over {\rm d}E^{n+1}}
         \left ( e^{-2NE/v_0^2} e^{-\lambda e^{E/v_0^2}} \right ).  
                            \label{eq::EXP-gn}
\end{eqnarray}
Figure \ref{fig4}{\em a} plots an $N=6$ case of 
equation (\ref{eq::new-DF--EXP}).

Figure \ref{fig5}{\em a} plots
the mean rotation velocity $\langle v_{\phi} \rangle$ 
for four different values of $N$, and shows that the disks become
increasingly cool and centrifugally supported with increasing $N$.
The mean rotation velocity exceeds the circular velocity near the center,
due to the $n=0$ isotropic term in the DF (\ref{eq::new-DF--EXP}).
Its density, which is
\begin{equation}
\label{eq::bulge-dens-EXP}
\Sigma _{\rm iso} = \Sigma_s \left(1+\frac{R^2}{R_{\rm C}^2}\right)^{-N}
                    \exp\left[-{\sqrt{R^2 + R_{\rm C}^2} \over R_{\rm D}}\right],
\end{equation}
is confined to the central region. It is compressed as $N$ increases, 
and disappears as $N \to \infty$ along with all radial orbits.
The other interesting feature of Figure \ref{fig5}{\em a} 
is that $\langle v_{\phi} \rangle$ 
declines away from $v_0$ at large distances. Evans \& Collett's (1993)
models for a simple uncored exponential disk in a singular logarithmic
potential have the same property - see their Figure 2a - and for the
same reason. Although $\langle v_{\phi}\rangle \to v_0$ 
as $N \to \infty$, their equation (3.17) also shows that 
$\langle v_{\phi} \rangle$ decreases for increasing $R$ for 
fixed $N$, as in their Figure 2a and our Figure \ref{fig5}{\em b}.

Figure \ref{fig5}{\em b} plots the Toomre stability
parameter $Q$ for two different $N$ values and for the same three
sizes of the exponential disk as in Figure \ref{fig3}.
The cooler $N=6$ models have regions of varying extent in which $Q$ is
close to its marginal value of $1$. The growth of $Q$ with increasing $R$  
is due primarily to decreasing $\Sigma_{\rm D}$ because the radial 
velocity dispersion $\sigma_R=\sqrt{\langle v^2_R \rangle}$ falls off only 
mildly.

\subsection{Simulating a Hot Bulge}
\label{subsec::bulges}
Most spiral galaxies of S type have a three--dimensional central bulge
which stands out from the disk. Kormendy (1977) proposed modeling the
disk with an inner--truncated exponential so that the disk has a hole
in its center, and others have followed.  We construct disks with
central holes by removing stars which penetrate to the center. We use
a cutout function $H_{\rm cut}(L)$ to change an unperturbed DF $f_0(E,L)$
to $H_{\rm cut}(L)f_0(E,L)$. In so doing, we make the tacit assumption
that the part of the gravitational potential which the cutout stars
had previously provided, is instead provided by a bulge which is so hot
that it does not respond at all to disturbances in the disk.

Cutouts were introduced first in Zang's (1976) pioneering study of
DFs in a singular scalefree logarithmic potential. His purpose was to
eliminate stars with short dynamical time scales and high dynamical
frequencies. Dynamical frequencies in
cored potentials are bounded (cf Figure \ref{fig1}) and 
so avoid that problem. 
Zang's cutout also removes stars on nearly radial orbits,
regardless of their dynamical frequencies. Evans \& Read (1998a)
went further and also used an outer cutout to remove stars that spend
so much time far out in the disk that they do not respond to its
changing potential. Their disks are infinitely massive whereas ours
are not; ours have sufficiently few such stars that they do not pose
a problem. 

We apply a cutout function
\begin{equation}
\label{eq::ourcutout}
H_{\rm cut}(L) = 1-e^{-(L/L_0)^2},
\end{equation}
where $L_0$ is some angular momentum scale. 
Like Zang's, it removes stars for which $L$
is significantly less than $L_0$, but has no effect on stars with $L \gg L_0$.
It also removes stars on radial and near--radial orbits, as well as
stars of low energy because they too have low angular momenta.
The result of the cutout (\ref{eq::ourcutout}) is
to give an active surface density
\begin{equation}
\label{eq::sigma_act}
\Sigma_{\rm act}=\int H_{\rm cut}(L)f_0(E,L)d{\bf v},
\end{equation}
which tends to zero at the center, and hence models a central hole. 
Figure \ref{fig4}{\em b} shows the effect of an $L_0=0.1$ cutout
on the DF of Figure \ref{fig4}{\em a}.
No boundary integrals (\ref{eq::MsupB-elements}) arise for such cutout 
unidirectional disks because their unperturbed DF vanishes at $L=0$. 
The active surface density corresponding to the cutout DF of 
Figure \ref{fig4}{\em b} is shown in Figure \ref{fig14}{\em c}.
It shows that our truncation is far less sharp
than that given by Kormendy's (1977) inner--truncated exponential formula.

\section{COMPUTATIONAL PROCEDURES}
\label{sec::computational-procedures}

This section gives details on the numerical methods and algorithms 
we used. Those uninterested in these topics should skip to 
\S \ref{sec::numresults} where we present the results of our
computations.

\subsection{Choice of Basis Functions}
\label{sec::basisfunctions}

We use the set of basis functions introduced by Clutton-Brock (1972),
and simplified by Aoki \& Iye (1978), to relate densities and potentials.
They are
\begin{eqnarray}
\psi^m_j &=& -{1\over (1+R^2/b^2)^{1/2}}P^m_{m+j}(\xi), 
               \label{eq::CBfunctions-pot}  \\
\sigma ^m_j &=& {(2m+2j+1)\over 2\pi Gb(1+R^2/b^2)^{3/2} }P^m_{m+j}(\xi),
               \label{eq::CBfunctions-den} 
\end{eqnarray}
where $\xi=(R^2-b^2)/(R^2+b^2)$ and the $P^m_{m+j}(\xi)$ are
associated Legendre functions. The variable $\xi$ ranges from $-1$ at
$R=0$ to $1$ as $R \to \infty$. The functions
(\ref{eq::CBfunctions-pot}) and (\ref{eq::CBfunctions-den}) form a
complete biorthogonal set over the range $0 \leq R < \infty$, and so
are useful for disks of infinite extent. The parameter $b$ is a length
scale which can be chosen so as to match the oscillatory behavior of
the Legendre functions with the regions in which modes vary most. We
have found that repeating calculations with different $b$ values provides 
a helpful check on their accuracy.

The Clutton-Brock functions with the choice $b=R_{\rm C}$ are a simple
choice to use with Kuzmin's disk. The self-consistent
potential--density pair (\ref{eq::pot-den-KT}) is then given by
$(V_0,\Sigma_0)=(GM/R_{\rm C})(\psi^0_0, \sigma^0_0)$. However we have also
found Clutton-Brock functions to work well for determining the modes
of the exponential disk (\ref{eq::den-expdisk}), even though its
density decays more rapidly than the $R^{-3}$ decay of the
$\sigma^0_j$ functions. This seems to be because modes decay rapidly
in the outer regions and are little influenced by matter there. For
instance, the frequencies which Pichon \& Cannon (1997) computed for a
Kuzmin's disk truncated at $R=5R_{\rm C}$ differ little from those computed
by Hunter (1992) who did not truncate them.  Unfortunately, the
Clutton-Brock functions are not suitable for computing reliable values
of the gravitational energy component ${\cal W}_{2,2}$ using the sum
(\ref{eq::W22-equation}). The reason is that the combination of the
growth of the logarithmic potential $V_0$ at large $R$
with the $R^{-3}$ decay of the $\sigma^0_j$ causes the integrals in
that sum to grow slowly with increasing $j$.  The convergence of the
sum is poor, and we do not have accurate values for ${\cal
W}_{2,2}$. No such problem arises with the more rapidly decaying
potentials of the Kuzmin and isochrone disks.

Kalnajs (1976a) gave a set of biorthogonal Abel-Jacobi functions 
which are suitable for disks of finite radius,
and were used by Earn \& Sellwood (1995) and Pichon \& Cannon (1997).
No recursive relations have been given for them, and so they must be computed 
using explicit formulae. That requires high--precision arithmetic 
because of cancellations between large coefficients of opposite sign.
Clutton-Brock functions are computed accurately using
simple recursive formulae, and do not need high--precision arithmetic.

\subsection{Action Space Integrations}
\label{sec::actionspaceintegrations}

Since DFs are usually given in terms of the energy $E$ and the angular
momentum $L$, it is more convenient to use $E$ and $L$ as the
integration variables for integrations over action space. We do this
and replace the Jacobian ${\rm d}J_R{\rm d}J_{\phi}$ with ${\rm
d}E{\rm d}L/\Omega _R$.  To evaluate the surface integral in the
$(E,L)$--space, we adopt the trapezoidal rule in the $E$-direction and
an extended open scheme [equation (4.1.18) of Press et al. (1992)] in
the $L$-direction. With this latter choice, we avoid the circular and
radial orbit boundaries where the calculation of orbital frequencies
and their derivatives with respect to the actions has some
computational difficulties.  The Fourier coefficients $\Psi^m_{l,j}$,
and hence the integrands, all vanish at $E_{\rm min}$ and $E_{\rm
max}$ because the potential functions $\psi^m_j(R)$ vanish at $R=0$
(for $m>0$) and as $R \to \infty$.  For the exponential disk for which
$0\le E < \infty$, we transform from $E$ to $u$ where
\begin{equation}
\label{eq::map-rising}
E=-\ln (1-u^2),
\end{equation}
and map $0\le E < \infty$ onto the finite range $0\le u < 1$. 
We generate a uniform $n_G\times m_G$ mesh in either $(E,L/L_c)$--space
or $(u,L/L_c)$--space, and evaluate integrands at grid points.   

Accurate evaluation of the boundary integrals ${\bf M}^{\rm B}$ requires
an especially careful treatment of the central region, especially
for centrally concentrated modes. More basis functions are needed there,
up to 15 in some cases, and a fine grid to allow for the oscillations
of Fourier coefficients near the center where $f_0^P$ is largest.
We use a uniform grid in $u=\sqrt{1+E}$ for boundary integrals for
Kuzmin's disk; the transformation (\ref{eq::map-rising}) still works well for
the exponential disk.

Formulas for the orbital frequencies $\Omega _R$ and $\Omega _{\phi}$
defined in (\ref{eq::equations-of-motion}) are
obtained by differentiating equation (\ref{eq::action-variables}).
Computationally convenient forms of the integrals which occur in them
are given by Evans \& Read (1998a). An alternative and compact form is
obtained by expressing $E$ and $L$ in terms of the
maximum and minimum orbital radii, $R_{\rm max}$ and $R_{\rm min}$.
Then if we define $x=R^2$ and $U(x)=2R^2V_0(R)$, we get
\begin{equation}
\label{eq::divdiff}
2R^2(E-V_0)-L^2=(x_{\rm max}-x)(x-x_{\rm min})
                          U[x_{\rm min},x,x_{\rm max}].
\end{equation}
Here $U[x_{\rm min},x,x_{\rm max}]$ denotes a second order divided difference,
as in de Zeeuw \& Hunter (1990). Using this result, and then the substitution
$x=x_{\rm min}\cos^2 \varphi + x_{\rm max}\sin^2 \varphi$, the orbital 
frequencies are given by
\begin{equation}
\Omega _R(J_R,J_{\phi})={\pi \over I_0},
  ~~\Omega _{\phi}(J_R,J_{\phi}) = {J_{\phi} I_1\over I_0},
\end{equation}
where
\begin{equation}
I_k=\int\limits_{0}^{\pi/2}
    {{\rm d}\varphi \over x^k\sqrt{U[x_{\rm min},x,x_{\rm max}]}}.
\end{equation}
These integrals are free of singularities, and they can be differentiated
analytically with respect to $E$ and $L$ to obtain the derivatives
of the frequencies needed in \S\ref{sec::P04theory}. 
 
\subsection{Fourier Coefficients}
\label{sec::fouriercoeffcalc}

We use the trapezoidal rule to evaluate the Fourier coefficients
(\ref{eq::fourier-coeffs}) because it is a highly accurate method for
integrating periodic functions if a fine enough gridding is used
(e.g., Davis \& Rabinowitz 1984). Closed form expressions for $R(t)$
and $\phi (t)$ are known for orbits in the planar isochrone
(Boccaletti \& Pucacco 1996). Otherwise orbits must be obtained by
numerical integration.  We integrate the equations of motion for a
half period, starting from the initial conditions
$t=\theta_R=\theta_{\phi}=v_R=\phi=0$, $R=R_{\rm min}$, and record the
values at equal steps in $\theta_R$ for use with the trapezoidal rule.
    
\subsection{Evaluating Matrix Elements}
\label{subsec::matrixelements}

The denominator term $(l\Omega _R+m\Omega _{\phi}-\omega)$ in the integrals
(\ref{eq::matrix-elements}) for the matrix elements $M_{jk}$ may vanish
if $\omega$ becomes real during the course of the iterative search.
When it does, the integral becomes singular. 
We use Hunter's (2002) method to handle such resonances. It is designed to
ensure that matrix elements are calculated according to Landau's (1946)
rule, and also provides an efficients way for the repeated evaluations
of the matrix elements needed in an iterative search for eigenvalues. 
It requires the evaluation of the integrals
\begin{eqnarray}
\alpha_n(j,k,l) &=& {4(2n+1)\pi^2 \over
(\eta_{{\rm max}}-\eta_{{\rm min}})}
\int\!\!\!\int \left[l{\partial f_0 \over \partial J_R}
+m{\partial f_0 \over \partial J_{\phi}}\right] \nonumber \\
                & \times & \Psi^m_{j,l}\Psi^m_{k,l} P_n[w(l)]
                {\rm d}J_R {\rm d}J_{\phi},
\end{eqnarray}
where $\eta=l\Omega_R+m\Omega_{\phi}$ and
$w(l)=2(\eta-\eta_{{\rm min}})/(\eta_{{\rm max}}-\eta_{{\rm min}})-1$.
The extreme values of $\eta$ depend on $l$ as is seen from 
Figure \ref{fig1}. Matrix elements are then formed as sums
\begin{equation}
\label{eq:matrixelementsum}
M_{jk}(m,\omega)=\sum\limits^{\infty}_{l=-\infty}
\left\{-2\sum\limits^{\infty}_{n=0}\alpha_n
\left(j,k,l\right)Q_n\left[\lambda(l)\right]
\right\}.
\end{equation}
where 
$\lambda(l)=2(\omega-\eta_{{\rm min}})/(\eta_{{\rm max}}-\eta_{{\rm min}})-1$.
Beware that the sign of the matrix ${\bf M}$ used here is the
opposite of that in Hunter (2002). Both $P_n$ and $Q_n$ denote the
usual Legendre functions. The multivalued $Q_n$ functions must be
evaluated for real $\omega$ by taking the limit $s={\rm Im}(\omega)\to 0$
through positive values. The coefficients $\alpha$ are computed and 
stored for the whole ranges of $n$, $j$, $k$ and $l$. Weinberg (1994)
gives an alternative method of handling resonances.

\subsection{Searching for Eigenvalues}
\label{sec::eigenvaluesearch}

We search for eigenvalues $\omega$ using the modified Newton method of
Stoer \& Bulirsch [1993, eq. (5.4.1.7)].  We calculate ${\cal
M}(m,\omega)$ using an LU decomposition as in Press et al. (1992). We
compute the derivative ${\rm d}{\cal M}/{\rm d}\omega$, needed by
Newton's method, numerically using central differences.  Whether
Newton's method converges or not, and the rapidity with which it
converges if it does, depends on the choice $\omega _0$ of a starting
approximation. We are most interested in finding the modes with the
largest growth rates.  The set of all initial guesses of $\omega_0$
which lead to an eigenvalue is the basin of attraction of its
mode. Our computations show that the $\omega_0$--space is dominated by
the basin of attraction of the fastest growing mode, but that there
are also regions which lead to other modes. We have found two modes
for most of our models.

We have found the following search procedure to be useful when there
are large boundary integral terms.
As we noted in \S\ref{sec::boundary-integrals}, the $l=-1$ component of
the boundary integral ${\bf M}^{\rm B}$ of (\ref{eq::MsupB-elements}) is
\begin{equation}
\label{eq::component-l=-1}
-{4\pi^2\over \omega}
   \int\limits_{0}^{\infty} {\rm d}J_R
   \left [2 f^P_0 \Psi^2_{-1,j} \Psi^2_{-1,k}
  \right ]_{J_{\phi}=0}= -{1\over \omega} A_{jk},
\end{equation}
where $A_{jk}$ are the components of a positive definite matrix ${\bf A}$
which is independent of $\omega$. We define the matrices
${\bf B}$ and ${\bf E}$ as
\begin{equation}
\label{eq::BE-matrix-def}
{\bf B}={\bf M}+{1\over \omega}{\bf A}, ~~ 
{\bf E}={\bf A}^{-1} \cdot ({\bf B}-{\bf D}).
\end{equation}
The eigenvalue problem (\ref{eq::matrix-equation}) can be recast as
\begin{equation}
\label{eq::newmatrix-equation}
\left({\bf E}-{1\over \omega}{\bf I}\right){\bf c}={\bf 0},
\end{equation}
so that $1/\omega$ is an eigenvalue of $\omega$--dependent matrix
${\bf E}$. We find that it is simple to find roots of the reduced
equation $\vert {\bf E}\vert =0$ by Newton's method 
because they are insensitive to the initial choice of $\omega$.
We use a continuation scheme to find eigenvalues of the recast
eigenvalue problem (\ref{eq::newmatrix-equation}) by introducing
a parameter $\mu$ and defining a sequence of problems
\begin{equation}
\label{eq::auxiliary-equation-mu}
\left \vert {\bf E}(m,\omega) -
              {\mu \over \omega}{\bf I} \right \vert =0.
\end{equation}
We proceed by continuation in $\mu$ from the simple $\mu=0$ case
to the $\mu=1$ case of equation (\ref{eq::newmatrix-equation}),
using our solutions at one stage as the initial estimate for the solution
at the next stage. We also apply continuation methods when we use
eigenvalue estimates obtained from matrix truncations of one size
as initial estimates for larger matrix truncations.

Getting a converged solution needs some controls on our series
truncations and integration grids. Using a fine grid at the energies
corresponding to distant stars, requires too many Fourier coefficients
to expand the basis functions. In turn, a large $l_{\rm max}$ ($\vert
l_{\rm min}\vert$) increases the cost of computations while such an
attempt does not capture further physics of the problem.  In fact,
unstable waves are confined to the central regions of stellar disks
and they lose much of their power as they reach the corotation
and outer Lindblad resonances. So, large-amplitude orbits,
which spend much time in outer regions, do not respond to density
perturbations. This is the reason why outer--truncated disks have
been used successfully by others (Athanassoula \& Sellwood 1986,
Pichon \& Cannon 1997). We allow disks to be infinite.  
We start the computation of $\omega$ with $40 \times 40$ grids 
and $j_{\rm max}=2$. We simultaneously increase $j_{\rm max}$
and grid size until the relative accuracy in computing $\omega$
becomes better than $5\times 10^{-3}$. We find in most cases that this
accuracy is achieved when $n_G\approx 75$ and $j_{\rm max}=10$.
An independent check of accuracy is to see whether the condition 
$e^{-2st}\Omega _p {\cal L}_2=0$ is satisfied after finding $\omega$ 
and ${\bf c}$. In all of our unstable models, the normalized value 
of this quantity remains below $5\times 10^{-3}$. 
 
After an eigenvalue $\omega$ has been found, we need its eigenvector
${\bf c}$. A theorem of linear algebra states that, when
$\omega$ is an eigenvalue, then any column of the adjoint matrix ${\rm
adj}[{\bf M}(m,\omega)-{\bf D}(m)]$ is an eigenvector associated with
$\omega$ (Brogan 1990).  The computation of the adjoint matrix
requires the computation of the minor determinants of ${\bf
M}(m,\omega)-{\bf D}(m)$.  It can be done rapidly and accurately
because we work with matrices of relatively small size.

\begin{deluxetable*}{rrrrrrrrrrr}
\tablecolumns{11}
\tablewidth{0pc}
\tablecaption{Eigenvalues for $m=2$ modes of unidirectional Miyamoto models
for Kuzmin's disk.\label{table1}}
\tablehead{
\colhead{} & 
\colhead{} & 
\colhead{} & 
\colhead{} & 
\multicolumn{5}{c}{Full Model} &   \colhead{}   &
\multicolumn{1}{c}{$l=-1$ only} \\
\cline{5-9} \cline{11-11} \\
\colhead{$n_{\rm M}$} & 
\colhead{$L_0$} &
\colhead{$M_{\rm act}$} &
\colhead{mode} &       
\colhead{$\Omega _p$} &
\colhead{$s$} & 
\colhead{${\cal K}_{2,2}/{\cal K}_{2,1}$} & 
\colhead{$R_{\rm CR}$} & 
\colhead{$R_{\rm OLR}$} & 
\colhead{} &
\colhead{$\Omega _p$} }
\startdata

3 & 0 & 1.000 & 1 & 0.825 & 0.939 & 1.58 & 0.541 & 1.296 &  & 
                    0.649 \\ 

3 & 0 & 1.000 & 2 & 0.418 & 0.265 & 1.88 & 1.483 & 2.246 &  & 
                    0.270 \\ \\

5 & 0 & 1.000 & 1 & 0.913 & 1.216 & 1.61 & 0.359 & 1.176 &  & 
                    0.738 \\ 

5 & 0 & 1.000 & 2 & 0.530 & 0.409 & 1.86 & 1.154 & 1.878 &  & 
                    0.323 \\ \\

7 & 0 & 1.000 & 1 & 0.963 & 1.465 & 2.58 & 0.227 & 1.115 &  & 
                     0.805 \\

7 & 0 & 1.000 & 2 & 0.643 & 0.588 & 2.64 & 0.895 & 1.609 &  & 
                     0.372 \\ \\

3 & 0.2 & 0.868 & 1 & 1.023 & 0.114 & -3.26 & \nodata & 1.045 &  & 
                    \nodata \\ 

3 & 0.2 & 0.868 & 2 & 0.336 & 0.203 & -1.67 & 1.811 & 2.632 &  & 
                    0.259 \\ \\

5 & 0.2 & 0.882 & 1 & 1.049 & 0.222 & 0.53 & \nodata & 1.017 &  &
                    \nodata \\ 

5 & 0.2 & 0.882 & 2 & 0.384 & 0.259 & -0.64 & 1.607 & 2.390 &  & 
                    0.294 \\ \\

7 & 0.2 & 0.892 & 1 & 1.067 & 0.311 & 1.27 & \nodata & 0.997 &  & 
                    \nodata \\ 

7 & 0.2 & 0.892 & 2 & 0.430 & 0.302 & -0.17 & 1.443 & 2.200 &  & 
                    0.321 \\ 

\enddata
\end{deluxetable*}

\section{m=2 MODES OF DISKs}
\label{sec::numresults}

The following three subsections present our numerical results for the
different DFs we have studied. Properties of $m=2$ modes are listed in the
tables, and selected ones are displayed.  The final column in each
table gives the pattern speed $\Omega_p$ computed by Polyachenko's
simplified theory described in \S\ref{sec::P04theory}. No growth rate
$s$ is listed for this case because it is always $0$. 

We have found two modes for most of the models we have investigated.
Classifying them is less straightforward than it is for simpler
physical systems, for which the fundamental mode has the lowest
frequency and simplest structure, and modes of successively higher
order have higher frequencies and more complex structure. The most
important instability is that with the highest growth rate. Often it
also has the simplest structure. However, we find that small changes
in the orbital population have a much greater effect on relative
growth rates than they do on radial structure, with the result that
the mode with the simplest structure is not always the fastest
growing. For that reason we have chosen structure, rather than growth
rate, as our criterion for determining which modes are
fundamental. Fundamental modes are labeled $1$ and secondary modes $2$
in our tables.

The displays include a contour plot of the perturbed density
\begin{equation}
\label{eq::modedens}
\Sigma _1 = P(R)\cos \left[2\phi-\omega t+\vartheta(R)\right].
\end{equation}
It is obtained from the real part of equation (\ref{eq::density-exp})
after writing its $R$--dependent part in the form
$P(R)e^{i\vartheta(R)}$ for some real functions $P(R)$ and
$\vartheta(R)$.  Eigenvectors are arbitrary to within a complex
constant multiple, with the result that the phase $\vartheta(R)$ is
arbitrary to within an additive constant, so that modes are oriented
arbitrarily.  As usual, we draw only the contours for positive levels
of the perturbed density (\ref{eq::modedens}); those for negative levels have
exactly the same pattern, rotated by $90^{\circ}$ and occupy the blank
sectors. The levels of the contours are in steps of 10\% 
of the maximum of $\Sigma _1$ from 10\% to 90\%. The length scale of all 
plots is that of the core radius of the potential, so that $R_{\rm C}=1$. 
 
Solid and dotted circles on the contour plots of a wave pattern mark
the radii $R_{\rm CR}$ and $R_{\rm OLR}$ of circular orbits in 
co-rotation resonance (CR) and outer
Lindblad resonance (OLR) respectively with a neutral $s=0$ mode with
the pattern speed $\Omega_p$ of that mode. All pattern speeds are
too large for there to be an inner Lindblad resonance (ILR). Orbits of
any shape, not just circular ones, may be resonant. For example, the orbits
which are in a CR with a pattern speed $\Omega_p$ are those which 
lie on the horizontal line $\Omega_{\phi}=\Omega_p$ which cuts through 
the lens-shaped region of
Figure \ref{fig1} from the circular orbit boundary on its left to the
radial orbit boundary on its right. These orbits are 
spread out in space and not confined to
one specific circle. They are more concentrated near that
circle when most orbits are near-circular, but not
otherwise. Similarly the orbits in an OLR are those for which
$\Omega_{\phi}=\Omega_p-\Omega_R/2$, and lie on a line through
Figure \ref{fig1} with slope $-1/2$. They too range from circular to
radial. They have lower orbital frequencies than those in the CR
because they lie further out in the disk. An ILR occurs only for
the limited range of $\Omega_p$ values for which the line
$\Omega_{\phi}=\Omega_p+\Omega_R/2$ of slope $1/2$ intersects the
lens-shaped region of Figure \ref{fig1}. When this happens, there are two
circular orbits, as well as a generally wide range of intermediate
orbits, in ILR. For the most part we find unstable modes with growth 
rates $s>0$. They have no resonances, only near-resonances at which the
denominators of the matrix components (\ref{eq::matrix-elements})
are small when $s$ is small.

Below each contour plot is a plot of the radial variations of the
amplitude $P(R)$ of the perturbed density (full curve) and of the
unperturbed density (dashed curve). Below this is a bar chart which
displays the values of the different Fourier components $L_2^l$,
$K_{2,1}^l$ and $W_{2,1}^l$ computed for them. The values of all these
components depend on how the eigenvector ${\bf c}$ is normalized. 
We normalize ${\bf c}$ so as to make the positive and
negative components of $e^{-2st}\Omega _p {\cal L}_2$ sum to $\pm 1$
respectively.
We find that ${\cal K}_{2,1}$ 
is always positive. This does not necessarily imply that all 
modes release gravitational energy and convert it to kinetic energy. 
Such a release occurs only if the sum of the two components 
${\cal K}_{2,1}+{\cal K}_{2,2}$ is positive. Our tables for 
Kuzmin and isochrone disks list values of the ratio 
${\cal K}_{2,2}/{\cal K}_{2,1}$ for each mode. A mode converts 
gravitational energy to kinetic energy if this 
ratio exceeds -1, and vice versa if the ratio is less than -1. 
There is no conversion if the ratio is exactly -1.

\subsection{Modes of Kuzmin Disks}
\label{subsec::example-Kuz-disk}
Miyamoto (1971) models are characterized by the single parameter $n_{\rm M}$.
The orbital population becomes
more nearly circular with increasing $n_{\rm M}$, and ultimately cold
in the limit $n_{\rm M}\rightarrow \infty$. We follow (Hunter 1992) in
working with models for which all orbits circulate in the same direction.
We use units in which $G=M=R_{\rm C}=1$.

\begin{figure*}
\plottwo{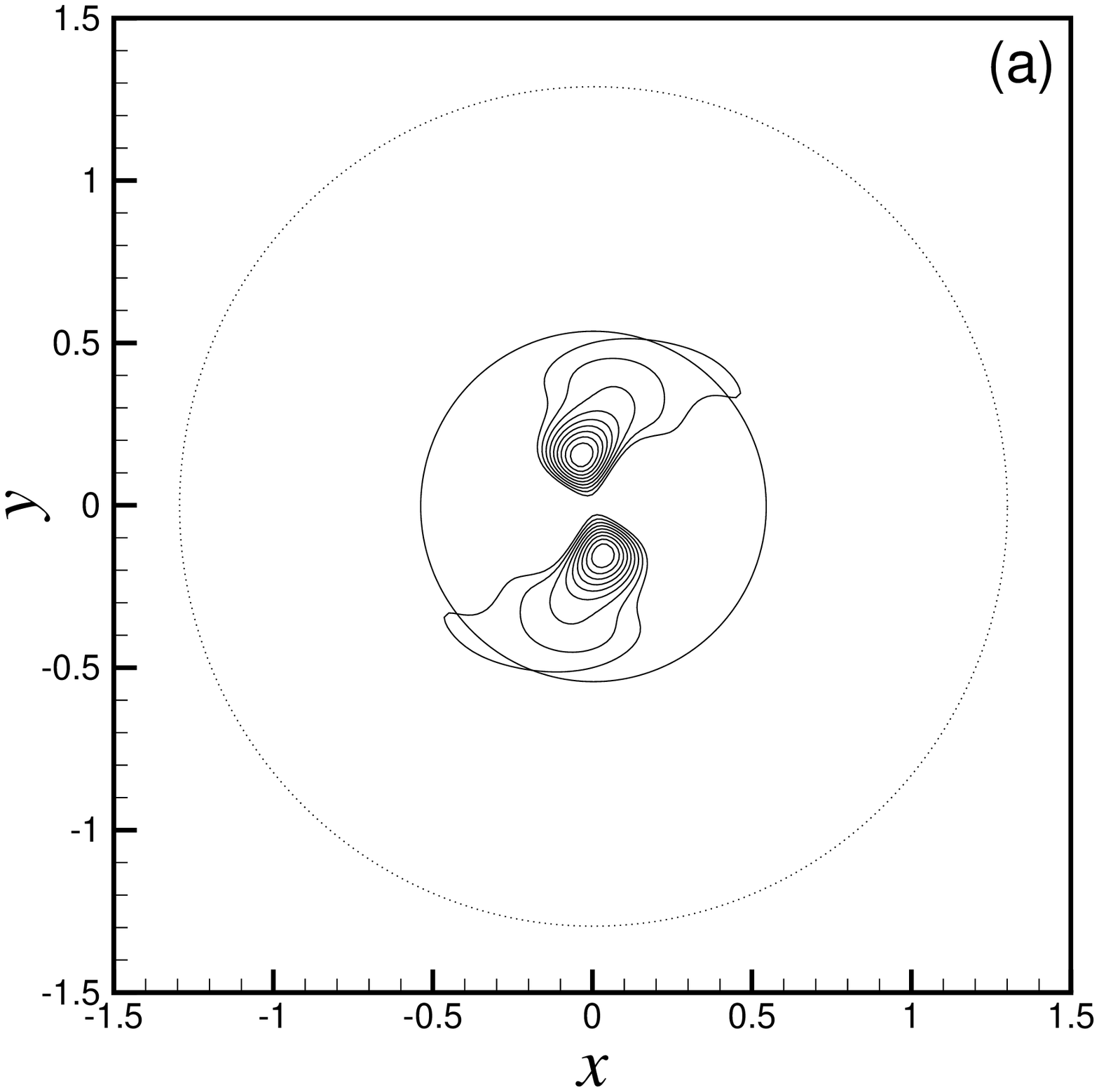}{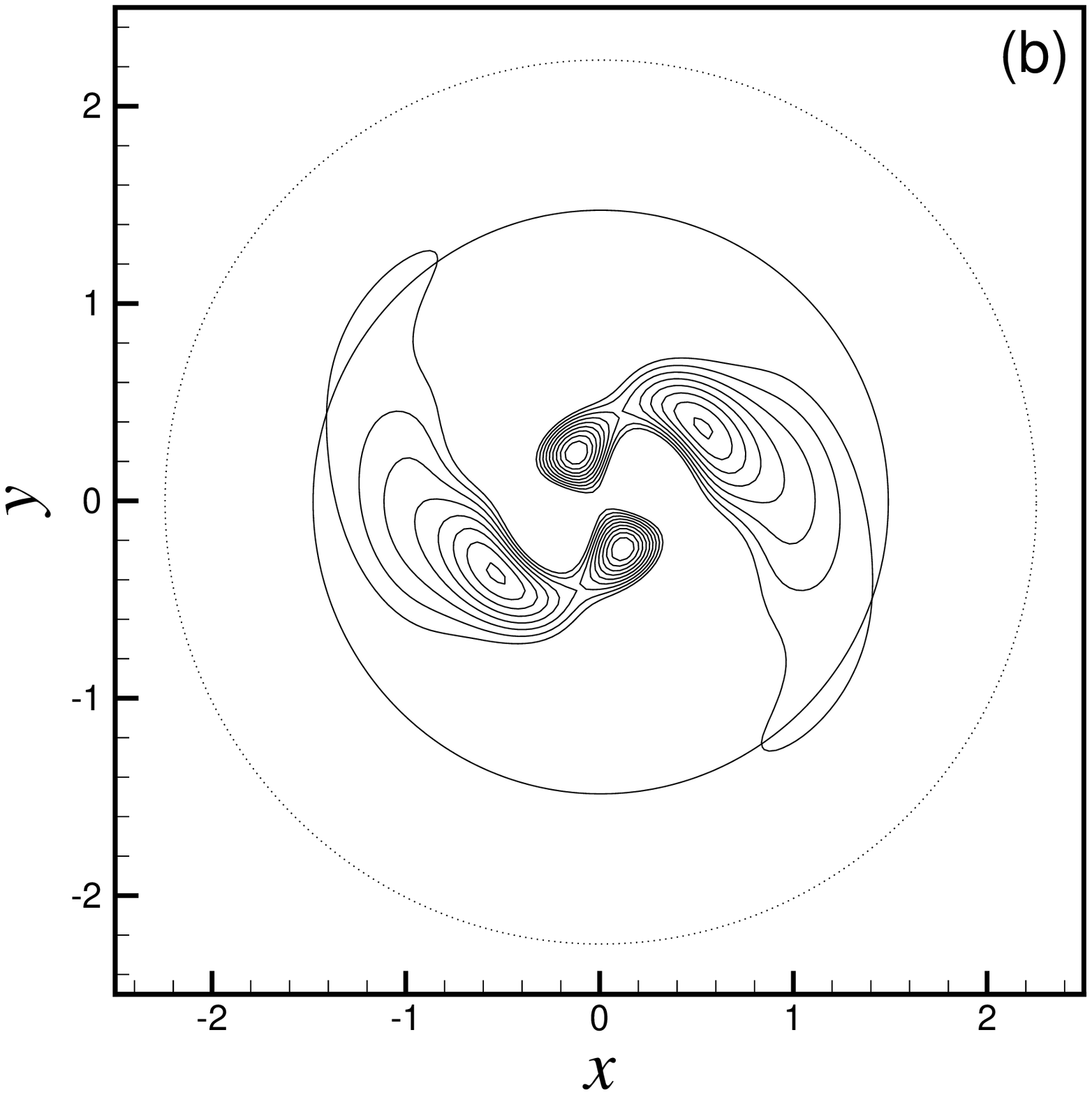}
\plottwo{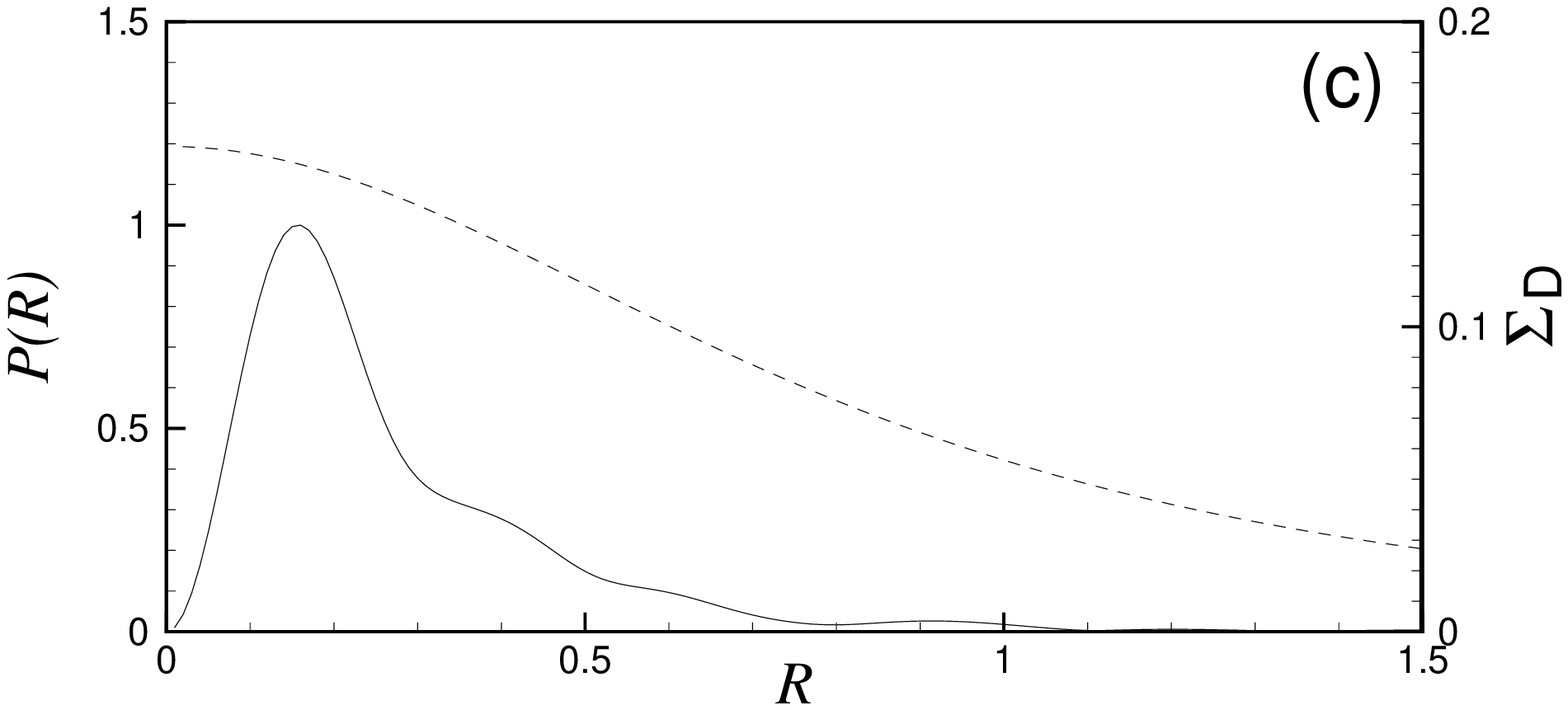}{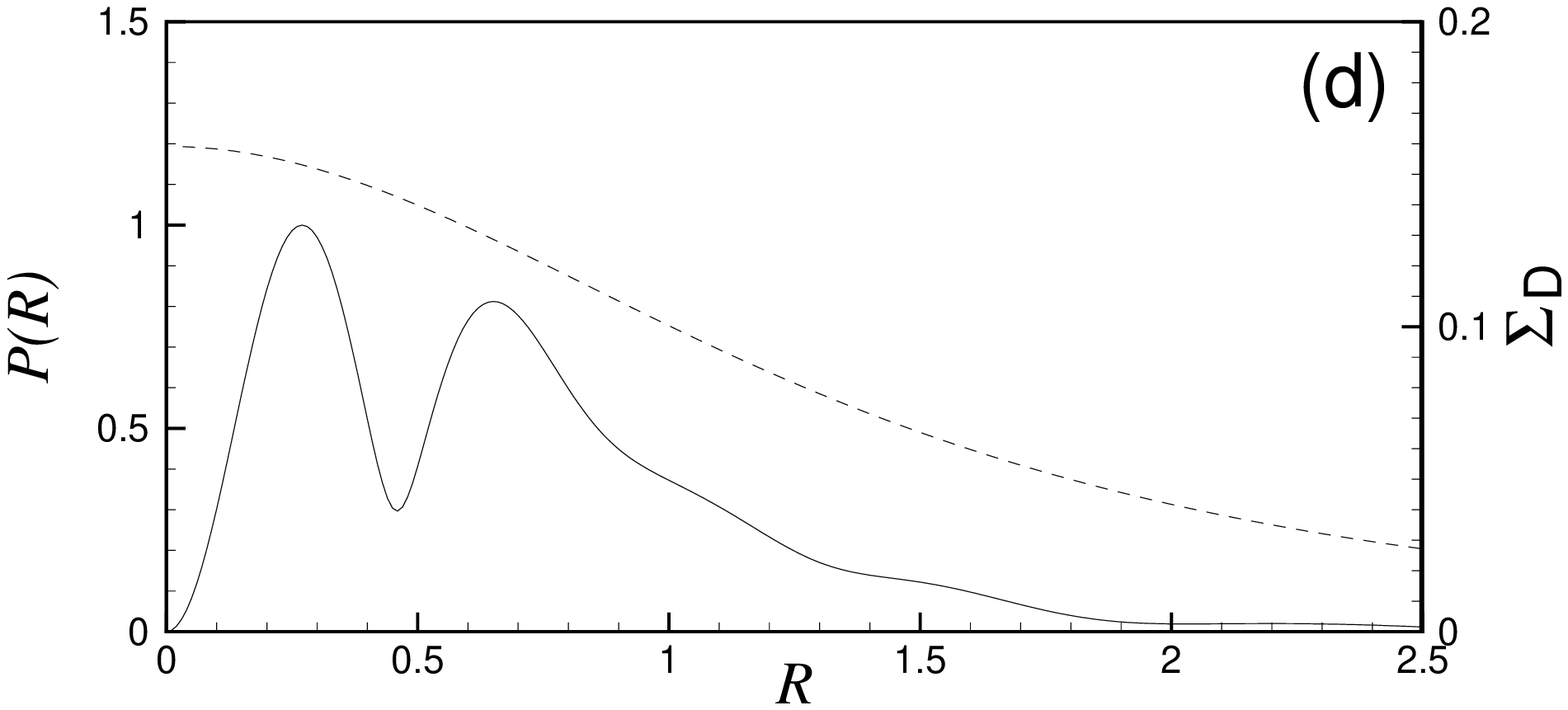}
\plottwo{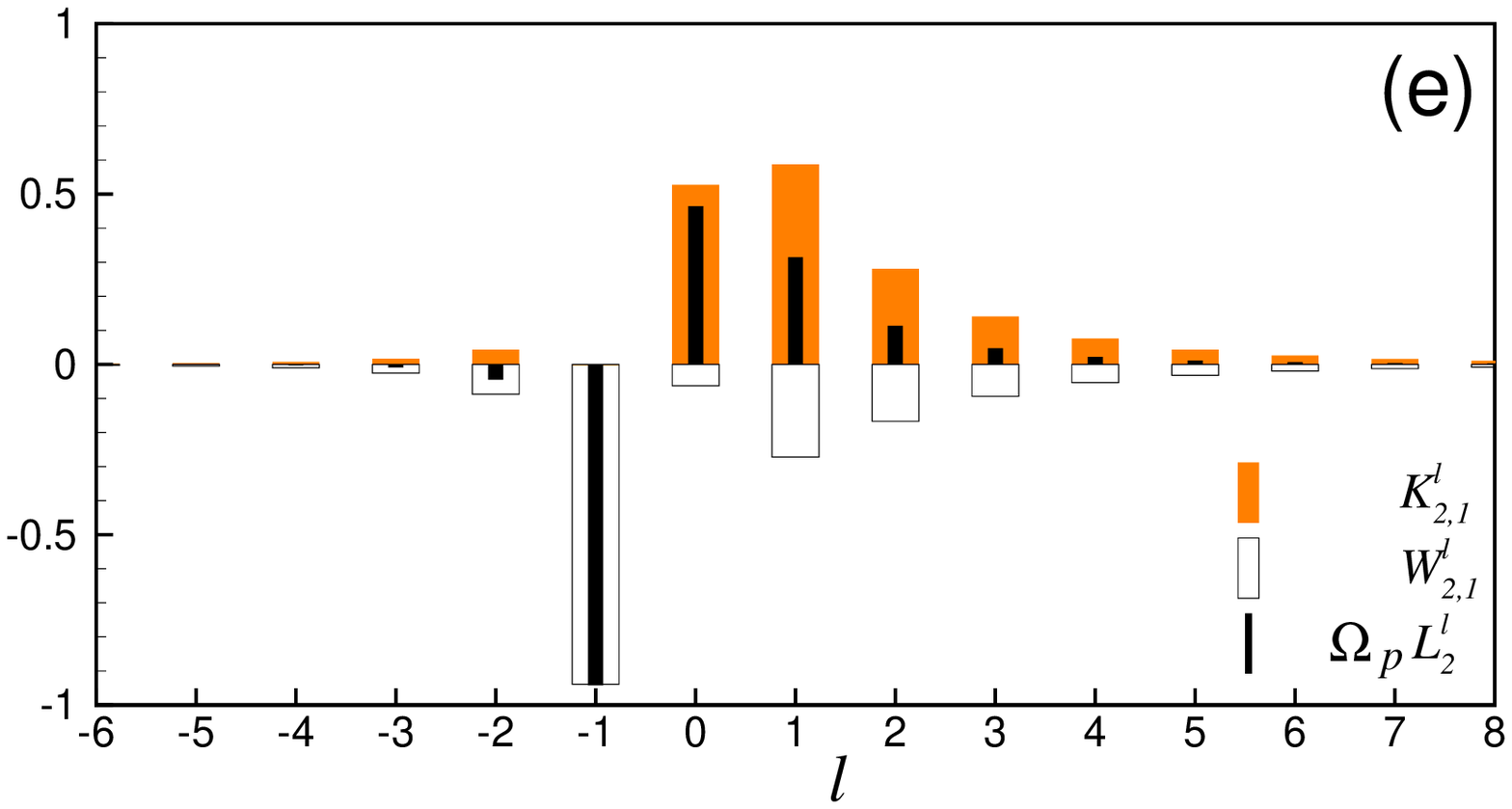}{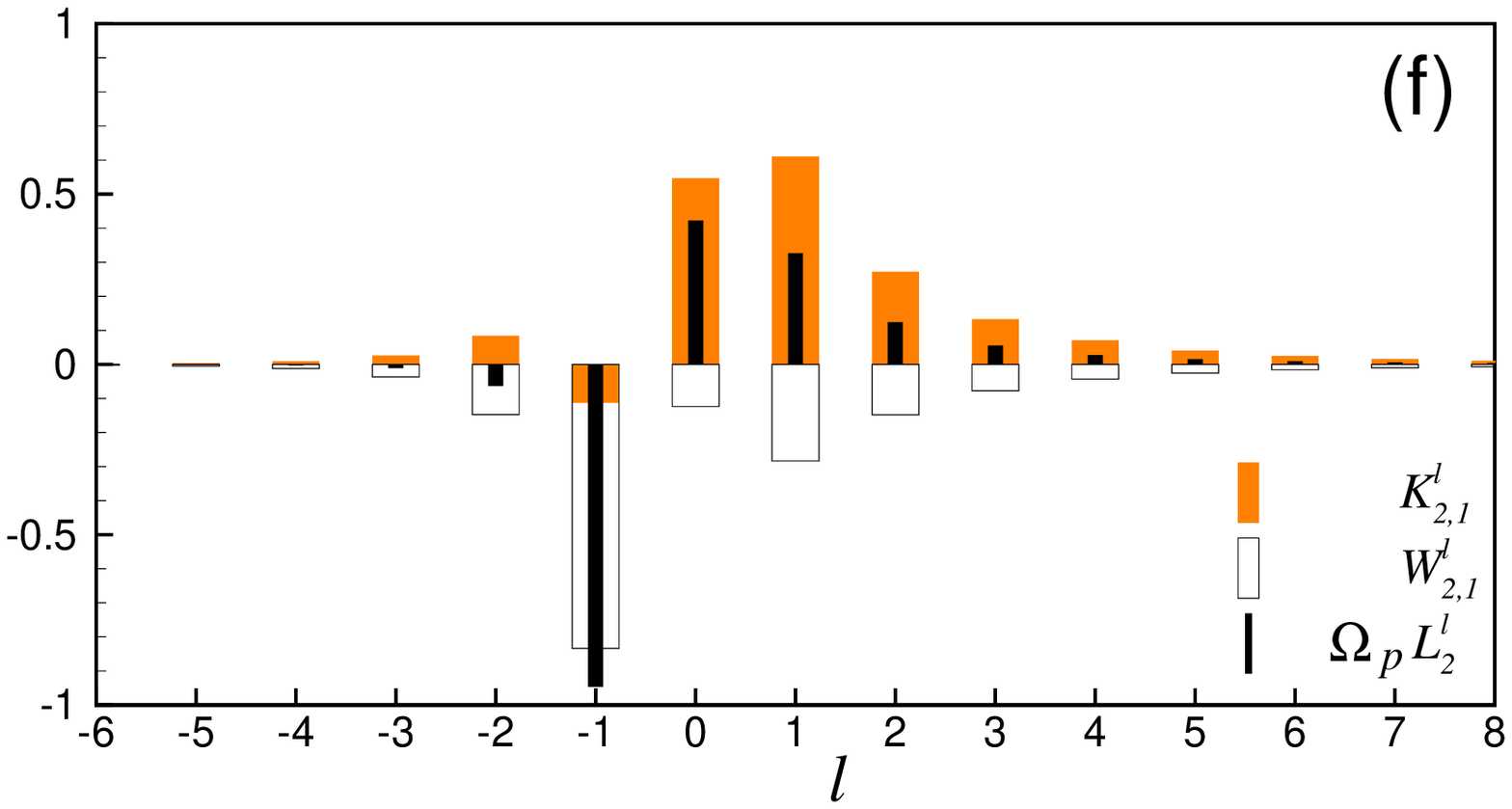}
\caption{The fundamental mode (left panels) and secondary mode
(right panels) of the self-consistent $L_0=0$, $n_{\rm M}=3$, Miyamoto model
for Kuzmin's disk; the first two entries of Table \ref{table1}. 
Here and in all subsequent plots of modes,
the top panels show positive contours of the perturbed density $\Sigma_1$,
in steps of $0.1\Sigma_1$. The solid and dotted circles mark 
the co-rotation and outer Lindblad resonance circles.
The middle panels show the wave amplitude (solid line and left scale) 
and unperturbed density (dashed line and right scale). 
The bottom panels show Fourier components of kinetic energy (grey bars),
gravitational energy (white bars), and angular momentum
(thin bars). 
\label{fig6}}
\end{figure*} 

The frequencies of the fundamental modes for the self-consistent 
($L_0=0$) models listed in Table \ref{table1} differ
substantially from those given earlier by
Hunter (1992) and Pichon \& Cannon (1997). Those results
are incorrect because they omit the boundary integral
${\bf M}^{\rm B}$. The difference is considerable. Neglecting 
${\bf M}^{\rm B}$ gives a pattern speed of $\Omega_p=0.357$ with 
$R_{\rm CR}=1.717$ and a growth rate of $s=0.295$ for $n_{\rm M}=3$.
The true fundamental mode of the $n_{\rm M}=3$ model is the compact and 
rapidly growing central bar shown in Figure \ref{fig6}{\em a}, much
more compact than that shown in Figure 11 of Pichon \& Cannon.
The amplitude $P(R)$ of its perturbed density in Figure \ref{fig6}{\em c}
has a single peak. 
The secondary mode, shown in the right panels of Figure \ref{fig6}, is slower
growing and slower propagating. It is more extensive, has a
double--peaked amplitude, and a more spiral structure which is also
largely confined within the CR circle.
The growth rates and pattern speeds increase with
increasing $n_{\rm M}$. They become increasingly centrally concentrated
as they are largely confined within the CR circle.

The bar charts in Figures \ref{fig6}{\em e} and \ref{fig6}{\em f}
display a standard pattern which is common to all but two of those we
show. Angular momentum is lost by the $l<0$ Fourier components,
primarily $l=-1$, and gained by the $l \geq 0$ components. All Fourier
components lose ${\cal W}_{2,1}$ gravitational energy, much from the
$l=-1$ component, while all components gain ${\cal K}_{2,1}$ kinetic
energy except for $l=-1$.  The positive values for the ratio ${\cal
K}_{2,2}/{\cal K}_{2,1}$ for these modes show that the ${\cal
W}_{2,2}$ and ${\cal K}_{2,2}$ terms reinforce this transfer from
gravitational to kinetic energy.  We show in \S\ref{subsec::barcharts}
how the forms of the bar charts can be understood in terms of the formulae we
derived in \S\ref{sec::angularmomentumenergy}.  Every bar chart shows
that a few Fourier components are significant for most of the transfer of
angular momentum and energy.

\begin{figure*}
\plottwo{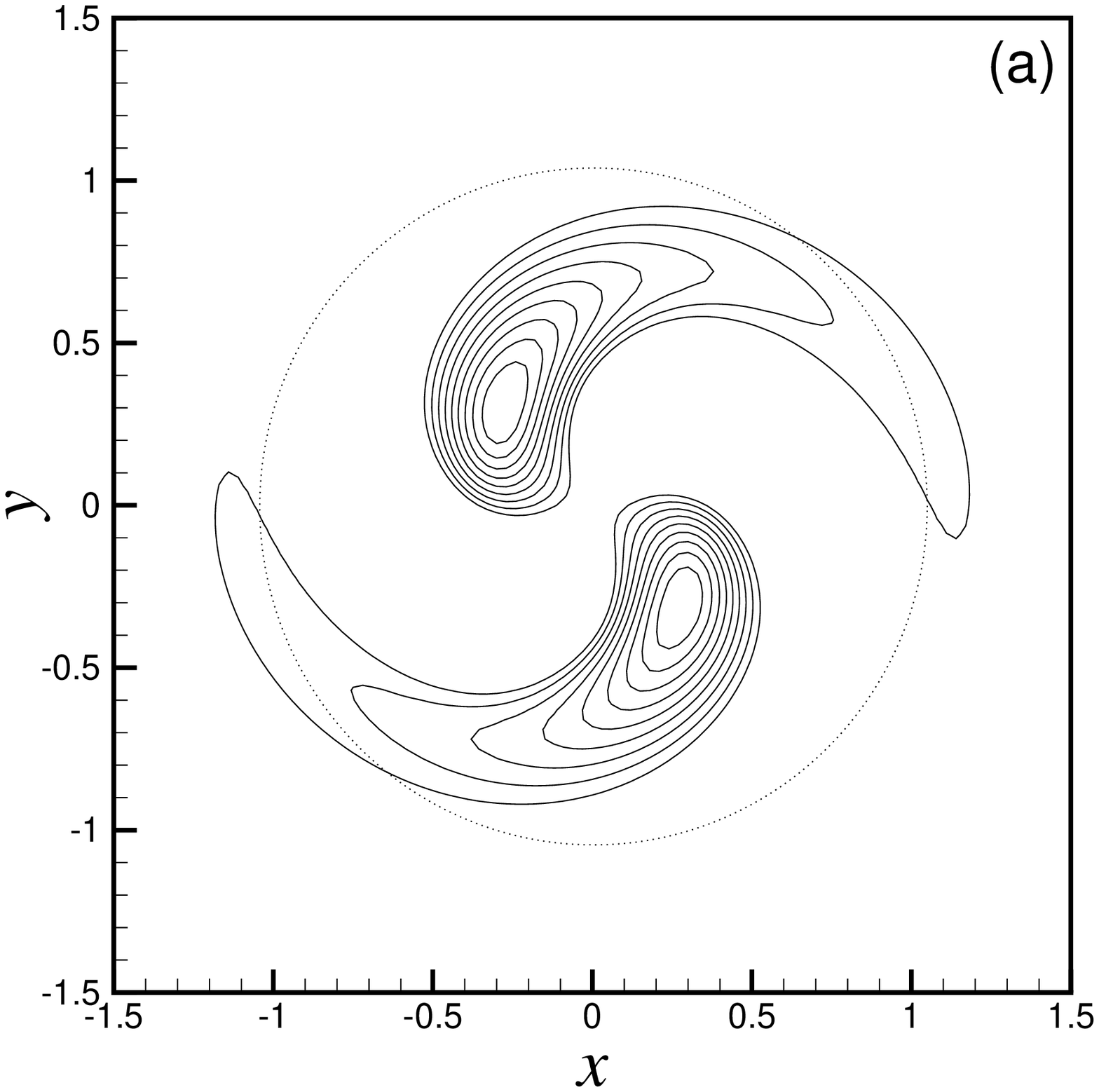}{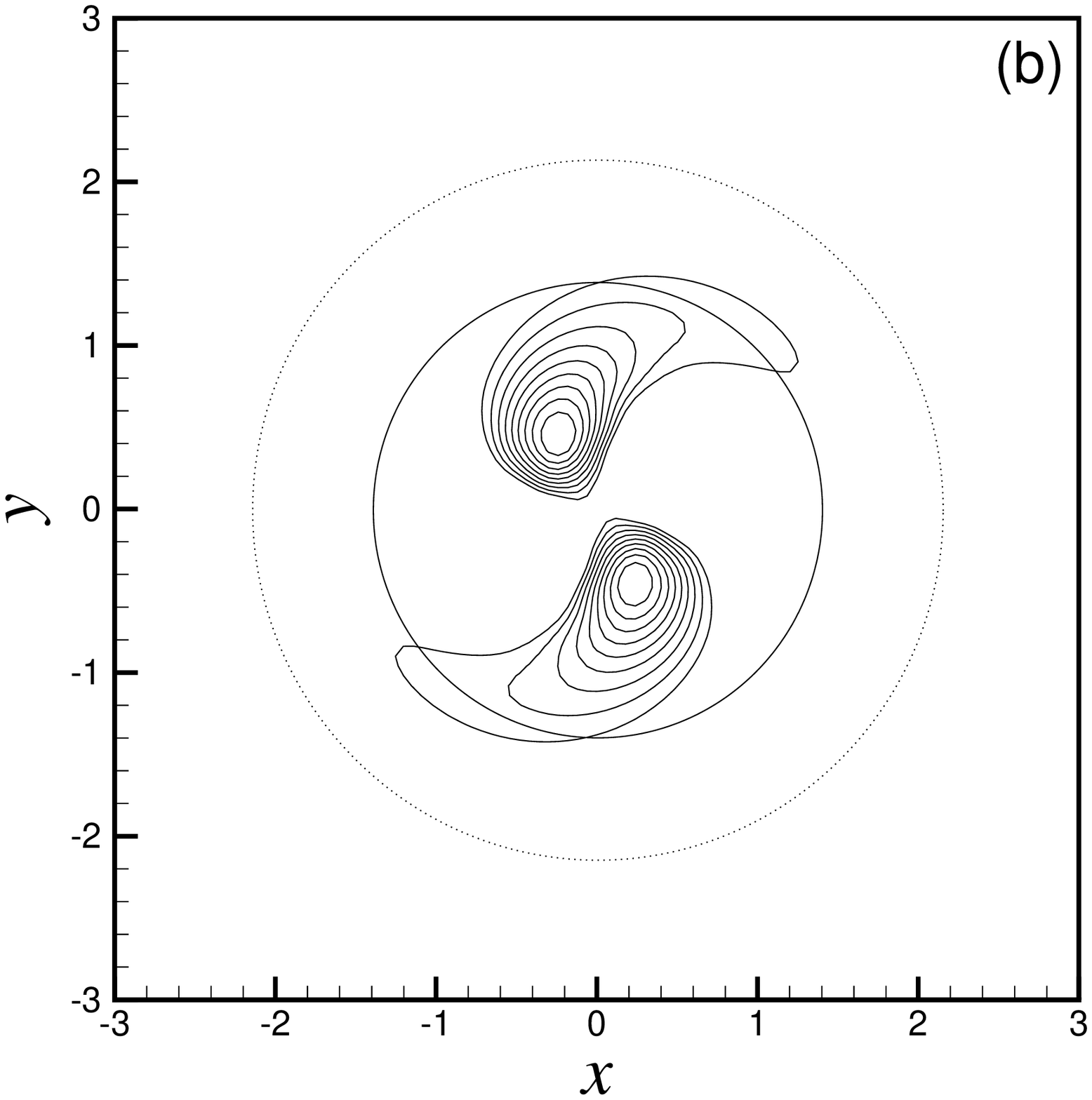}
\plottwo{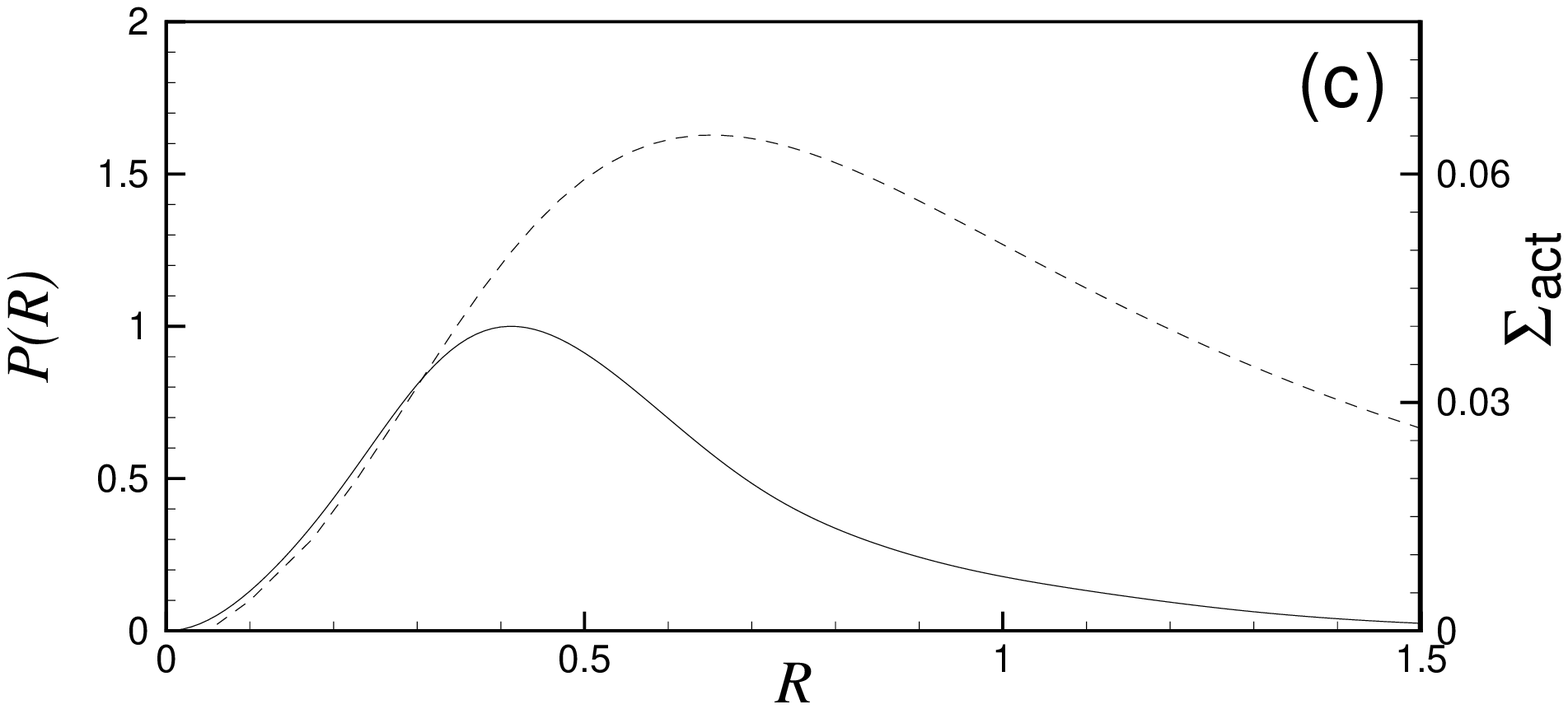}{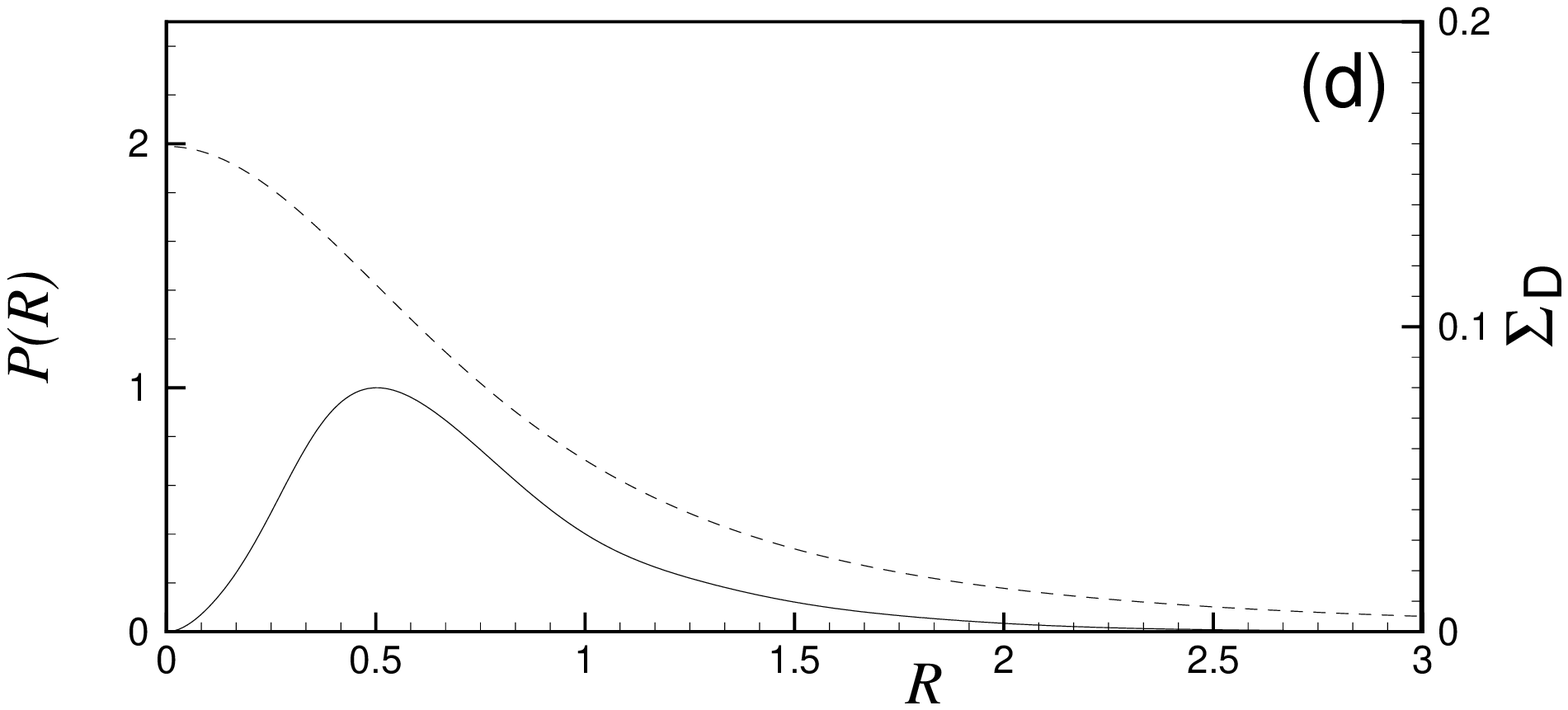}
\plottwo{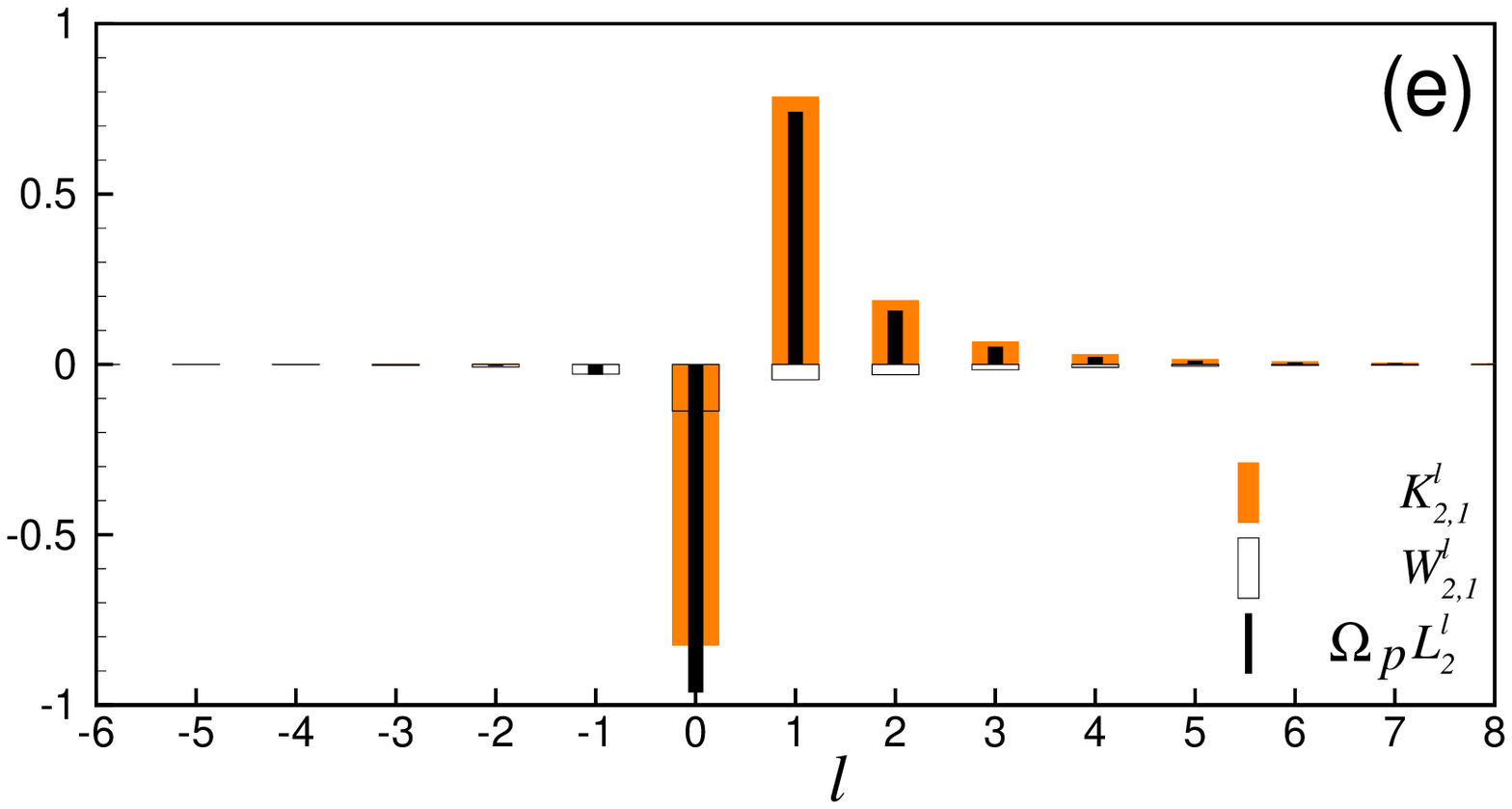}{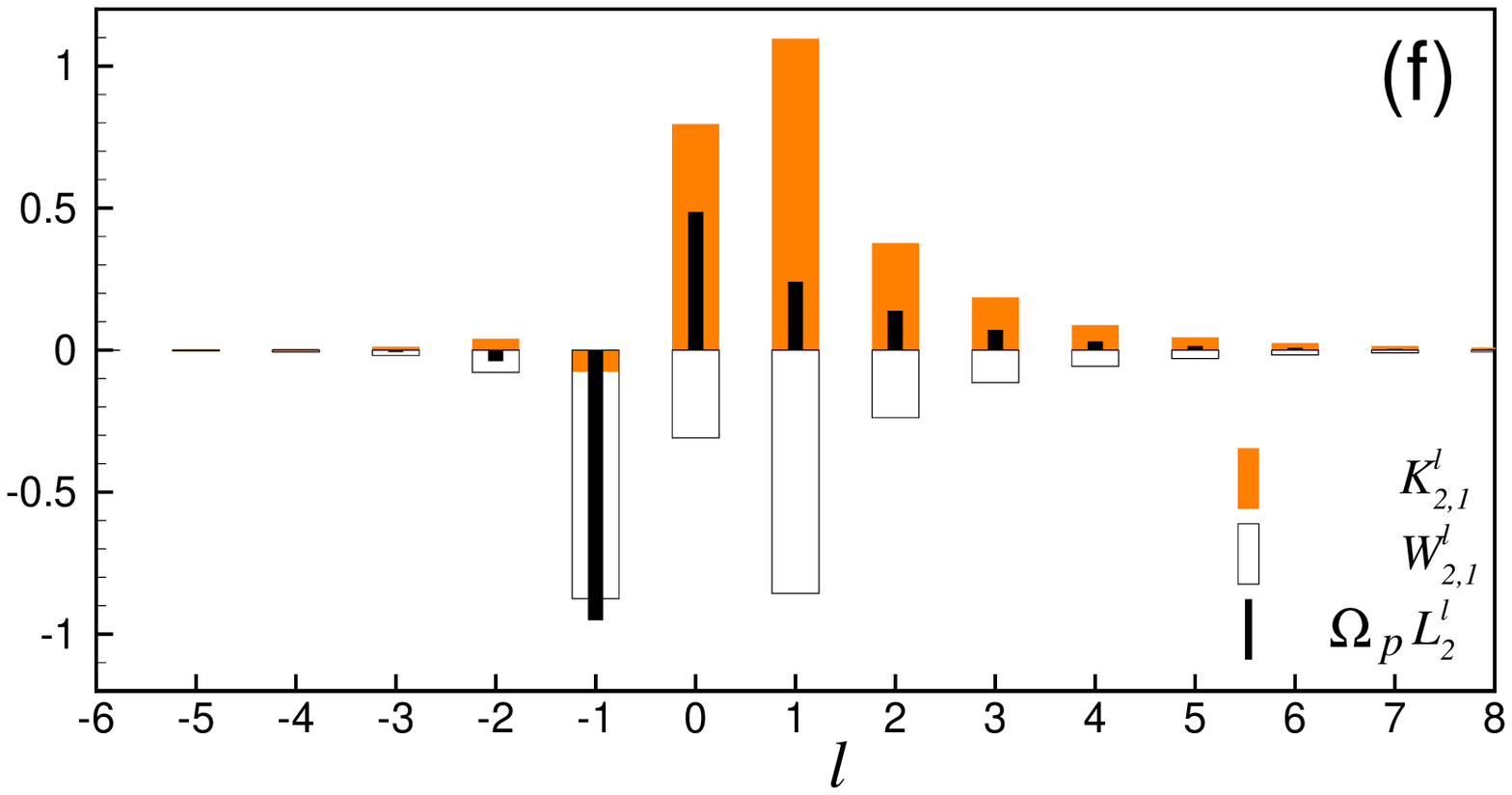}
\caption{The left panels are for the fundamental mode of a cutout
$n_{\rm M}=3$, $L_0=0.2$, unidirectional Miyamoto model. The right
panels are for the fundamental mode of a tapered $m_{\rm K}=6$,
$J_c=0.25$, Athanassoula \& Sellwood model for which 
some 4.6\% of the orbits have been made retrograde.
\label{fig7}} 
\end{figure*}

The second block of results in Table \ref{table1} are for
unidirectional Miyamoto models from which low angular momentum orbits
have been removed by applying the cutout function
(\ref{eq::ourcutout}) with $L_0=0.2$.  This removal reduces the active
surface density $\Sigma_{\rm act}$ so that it tends to zero in the
center, as shown in the dashed curve in Figure \ref{fig7}{\em c}, but
does not introduce a sharp barrier.  It reduces the active mass
$M_{\rm act}$ of the disk by a little more than 10\%, but has a much
larger effect on the modes, and especially the fundamental mode.  The
left panels of Figure \ref{fig7} show the large changes in the
fundamental mode of the $n_{\rm M}=3$ Miyamoto model caused by the
$L_0=0.2$ cutout. The amplitude of the fundamental mode is again
single peaked, though its peak is moved out to $R \approx 0.4$. This
mode rotates so rapidly that no orbits are in CR,
and it now extends out to the OLR circle. Its bar chart is
totally different from that of Figure \ref{fig6}{\em e}, but similar
to that of Figure \ref{fig14}{\em e}, which is for another fundamental
mode of a cutout disk.
The $l=-1$ components are small, and the flow of angular momentum and
${\cal K}_{2,1}$ is from the $l<1$ components, primarily $l=0$, and to
the $l \geq 1$ components, primarily $l=1$.

The cutting-out of low angular momentum orbits and some 10\% of the
mass reduces the growth rate of the fundamental mode by such a large
factor that it is no longer the fastest growing mode. The secondary
mode, which is now the fastest growing, is not shown because
it is changed much less. 
It has a similar but smoother spiral shape than that of
Figure \ref{fig6}{\em b}, and extends just beyond the now--larger
CR circle. The two humps of its amplitude $P(R)$ are
displaced outwards from those of Figure \ref{fig6}{\em d}, and the
inner hump, which lies in a region of large mass reduction, is
diminished. The outer hump, now at $R \approx 0.9$, lies in a region
where the mass reduction is smaller. Hence the reason why the cutout
affects the fundamental mode so much more than the secondary mode is
that it has a much larger effect on the more central orbits which are the major
participants in the fundamental mode than it does on those of the
secondary mode.  Because the ratio ${\cal K}_{2,2}/{\cal K}_{2,1}<-1$
for the both modes of the cutout $n_{\rm M}=3$ model, both induce a
transfer from kinetic to gravitational energy.

The results for higher $n_{\rm M}$ are similar. The cutout increases
the pattern speeds of fundamental modes so much that none have
CRs. It decreases the pattern speeds of secondary
modes, though not enough for there to be ILRs,
which are possible only when $\Omega _p\le 0.130$.  All growth rates are
decreased, though the fundamental mode is still, though barely, 
the faster growing for $L_0=0.2$ and $n_{\rm M}=7$.  As $n_{\rm M}$
increases and orbits become more circular, the relative amount of angular
momentum absorbed by the $l=1$ component of secondary modes increases.

\begin{deluxetable*}{rrrrrrrrrrrrrrrr}
\tabletypesize{\scriptsize}
\tablecaption{Eigenvalues for $m=2$ modes of Athanassoula \& Sellwood  
modified Kalnajs models for Kuzmin's disk \label{table2}}
\tablecolumns{16}
\tablewidth{0pc}
\tablehead{
\colhead{} & 
\colhead{} & 
\colhead{} & 
\colhead{} & 
\colhead{} & 
\multicolumn{3}{c}{Full Model} &   
\colhead{} &
\multicolumn{2}{c}{AS\tablenotemark{a}} &   
\colhead{} &
\multicolumn{2}{c}{P04\tablenotemark{b}} &   
\colhead{} &
\multicolumn{1}{c}{$l=-1$ only} \\
\cline{6-8} 
\cline{10-11} 
\cline{13-14} 
\cline{16-16} \\
\colhead{$m_{\rm K}$} &
\colhead{$\beta$} &
\colhead{$J_c$} &
\colhead{$M_{\rm retro}$} &
\colhead{mode} &
\colhead{$\Omega _p$} &
\colhead{$s$} & 
\colhead{${\cal K}_{2,2}/{\cal K}_{2,1}$} & 
\colhead{} &
\colhead{$\Omega _p$} &
\colhead{$s$} & 
\colhead{} &
\colhead{$\Omega _p$} &
\colhead{$s$} & 
\colhead{} &
\colhead{$\Omega _p$} }
\startdata

4 & 0 & 0.40 & 0.084 & 1 & 0.335 & 0.174 & 1.83 &  & 
               0.168 & 0.020 &  &
               \nodata & \nodata &  &
               0.193  \\ \\

6 & 0 & 0    & 0 & 1 & 0.746 & 0.711 & 1.42 &  & 
               \nodata & \nodata &  &
               \nodata & \nodata &  &
               0.569  \\

6 & 0 & 0.25 & 0.046 & 1 & 0.445 & 0.308 & 0.52 &  & 
               0.233 & 0.066 &  &
               0.22  & 0.114 &  &
               0.264  \\

6 & 0 & 0.25 & 0.046 & 2 & 0.294 & 0.109 & 1.85 &  & 
               0.165 & 0.058 &  &
               0.175 & 0.055 &  &
               0.207  \\ \\

6 & 3 & 0.60 & 0.154 & 2 & 0.158 & 0.027 & -4.00 &  & 
               0.145 & 0.014 &  &
               0.14 & 0.02 &  &
               0.144  \\
8 & 4 & 0.90 & 0.160 & 1 & 0.199 & 0.064 & -1.47 &  & 
               0.173 & 0.035 &  &
               \nodata & \nodata &  &
               0.174  \\
\enddata
\tablenotetext{a}{Athanassoula \& Sellwood (1986)}
\tablenotetext{b}{Polyachenko (2004)}
\end{deluxetable*}

\begin{figure*}
\plottwo{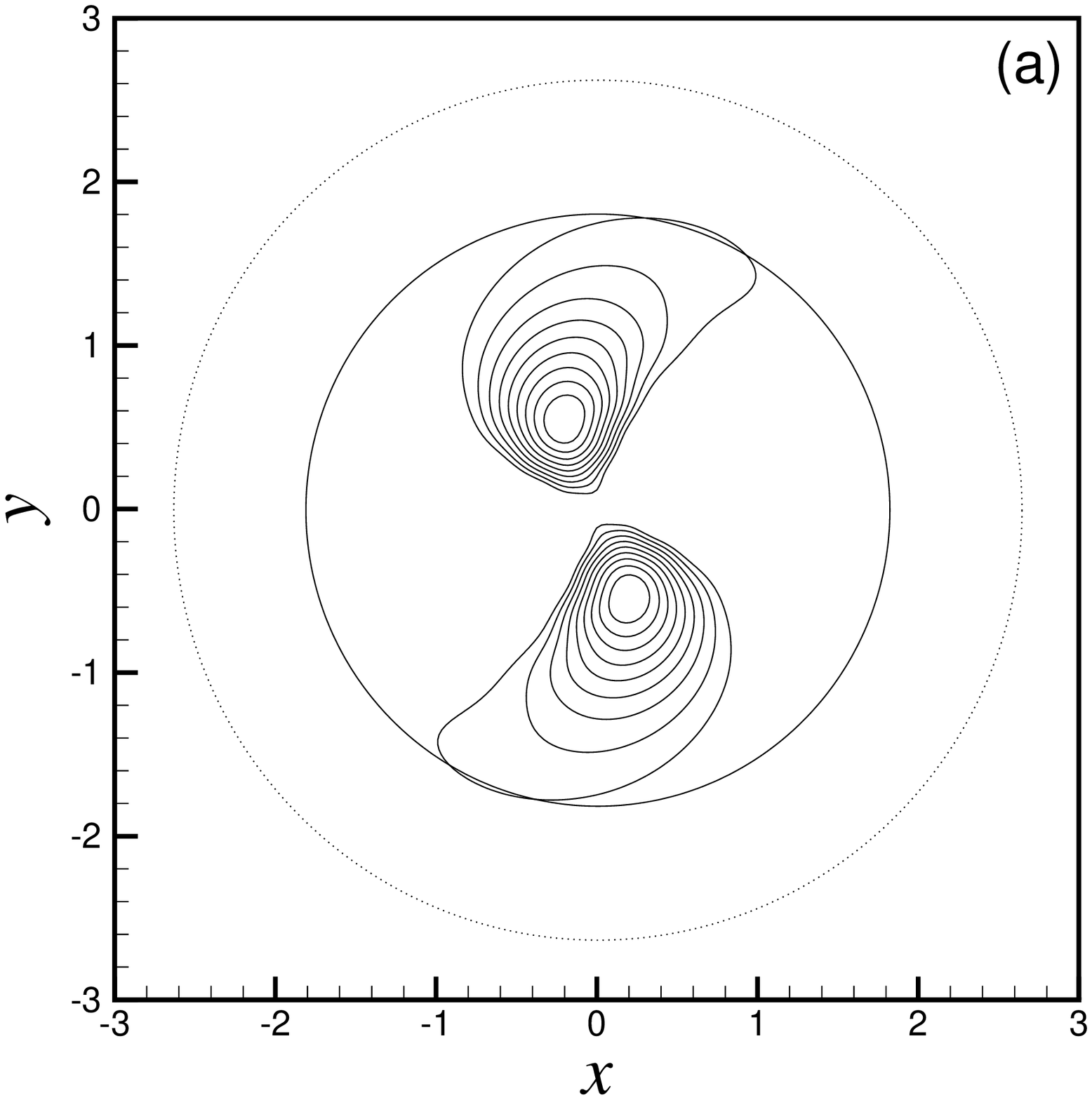}{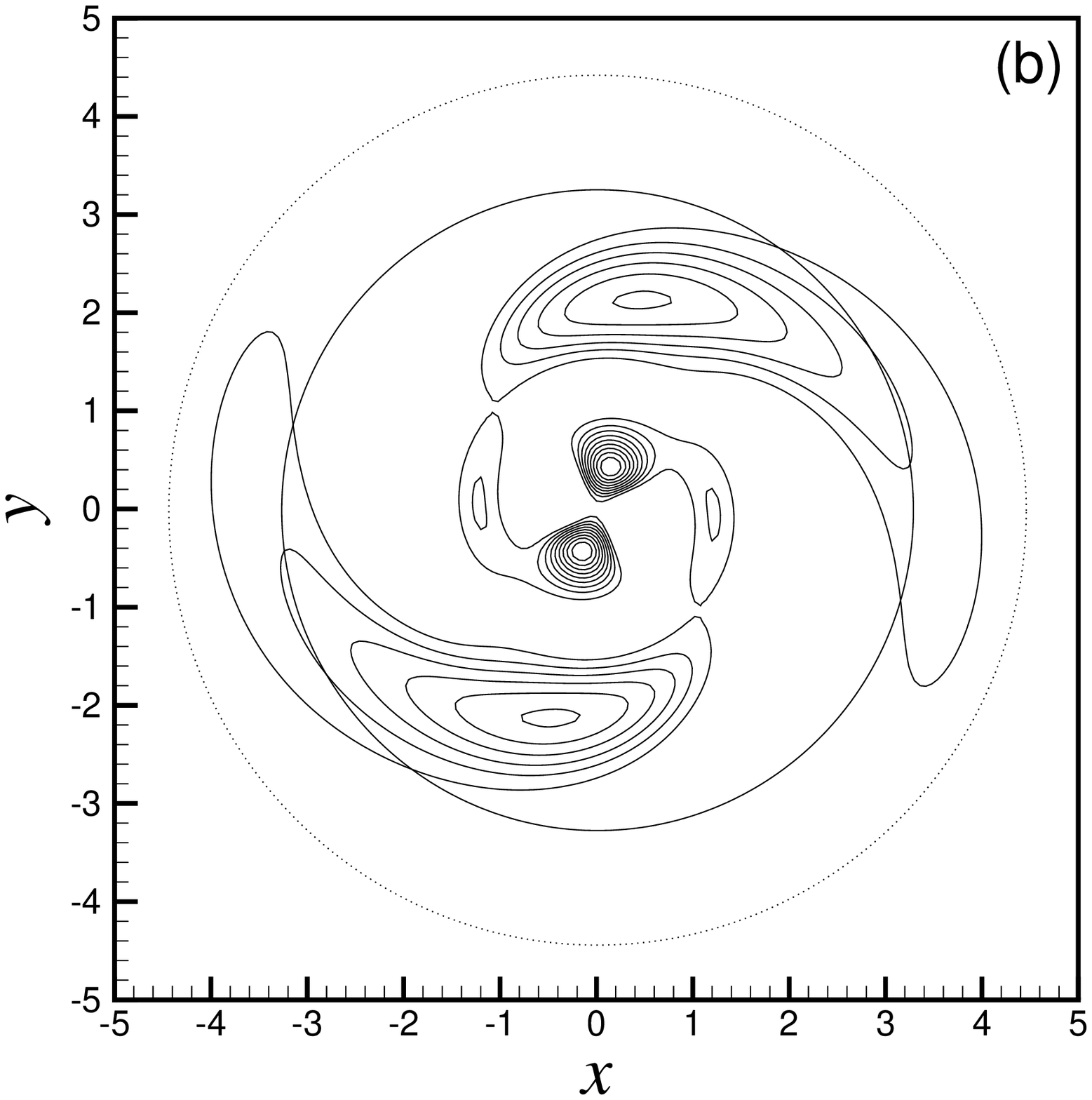}
\plottwo{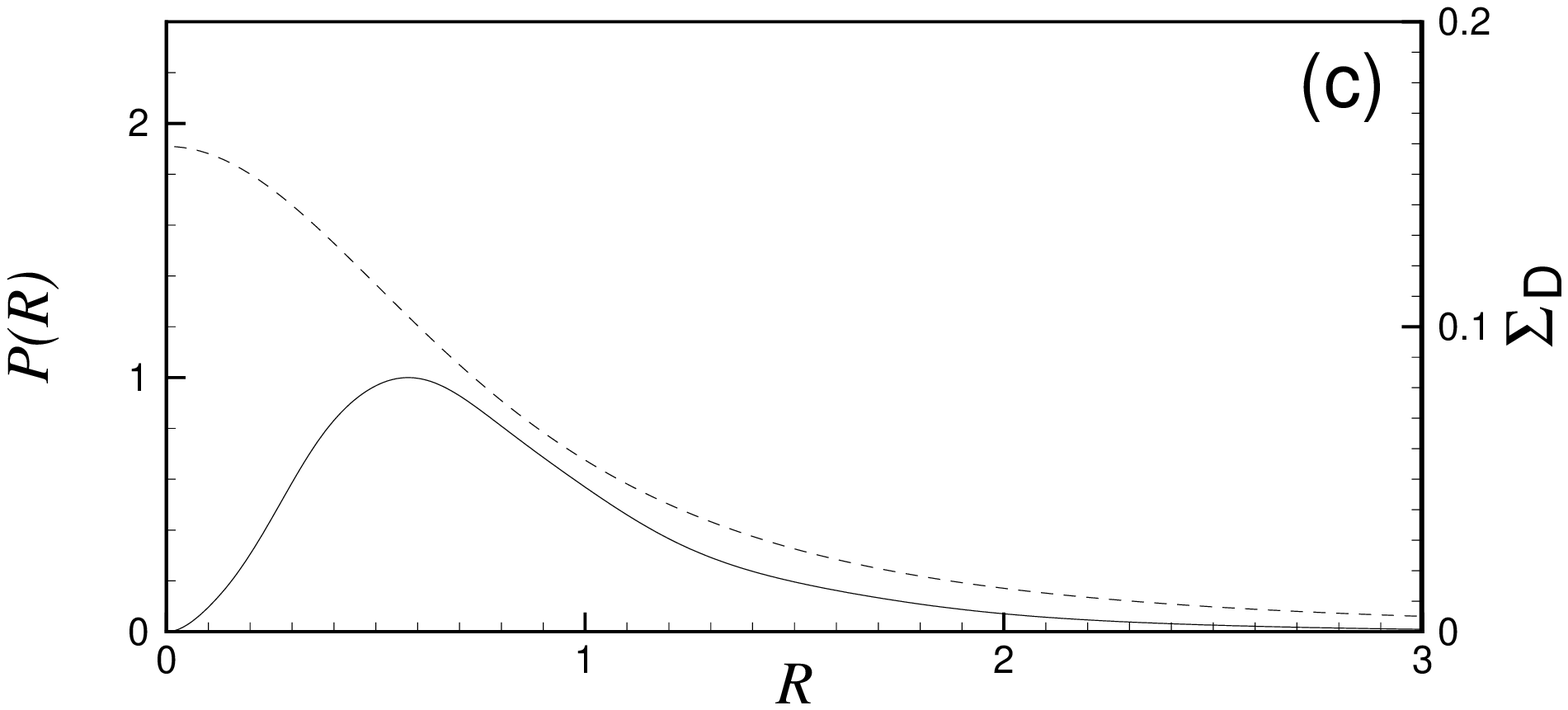}{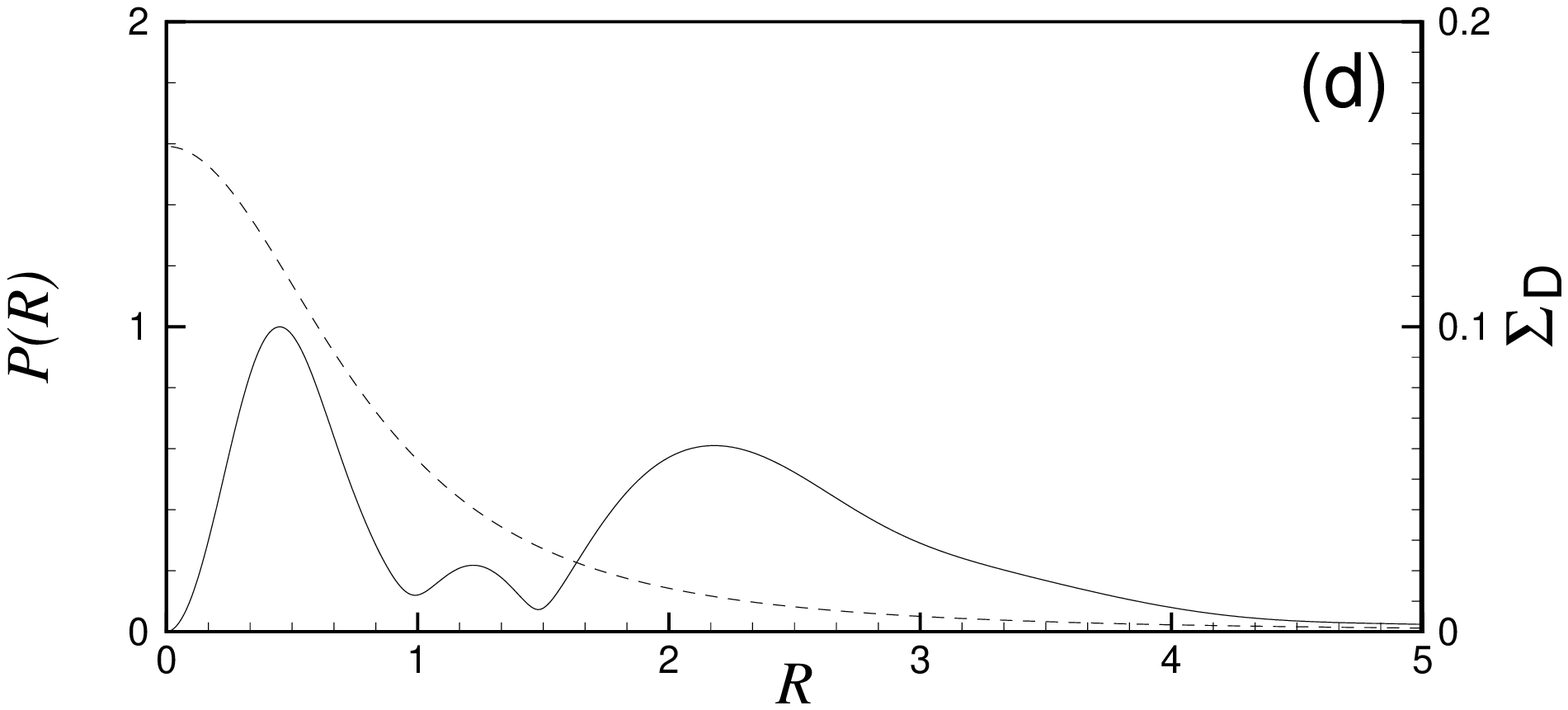}
\plottwo{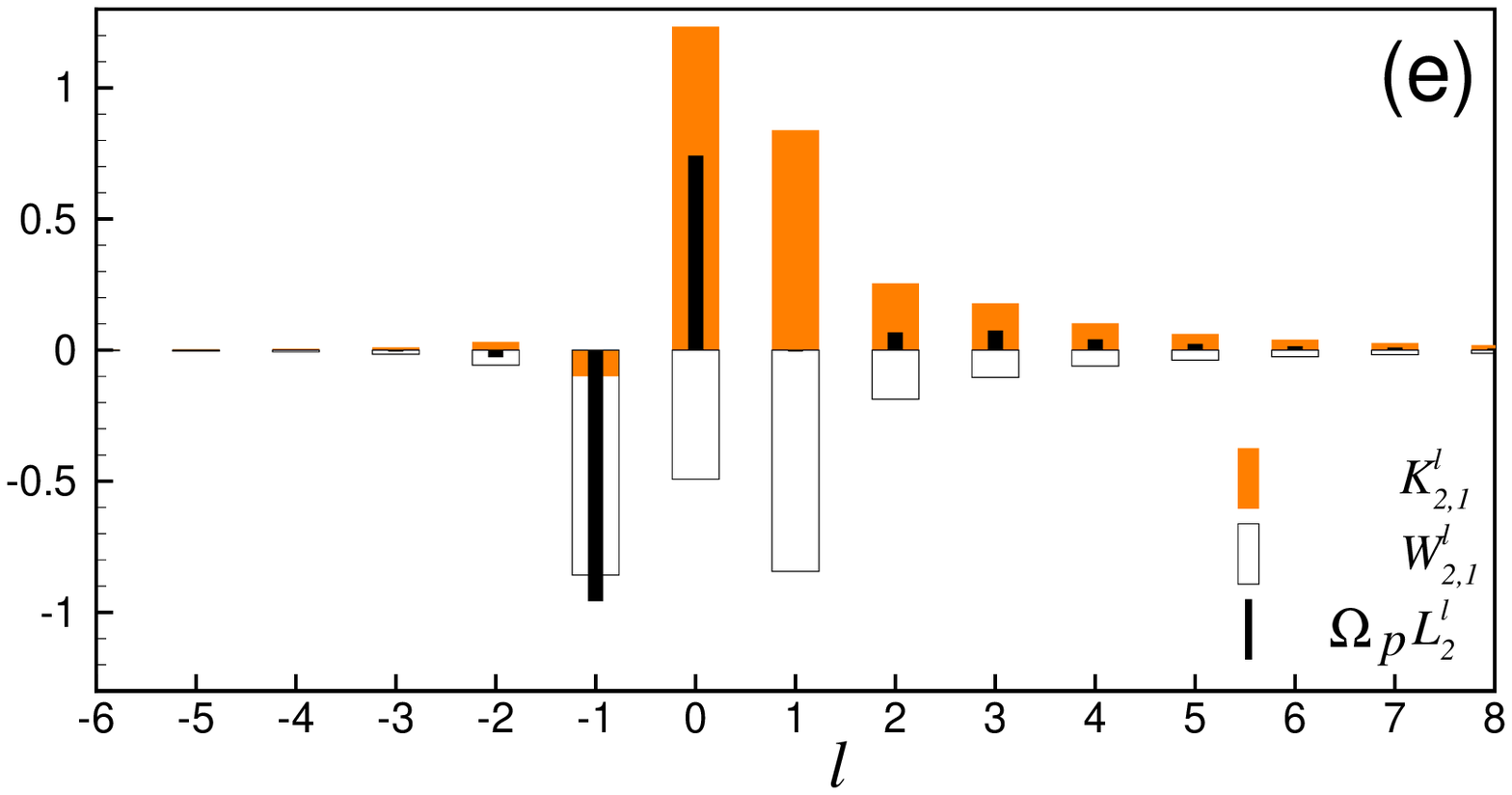}{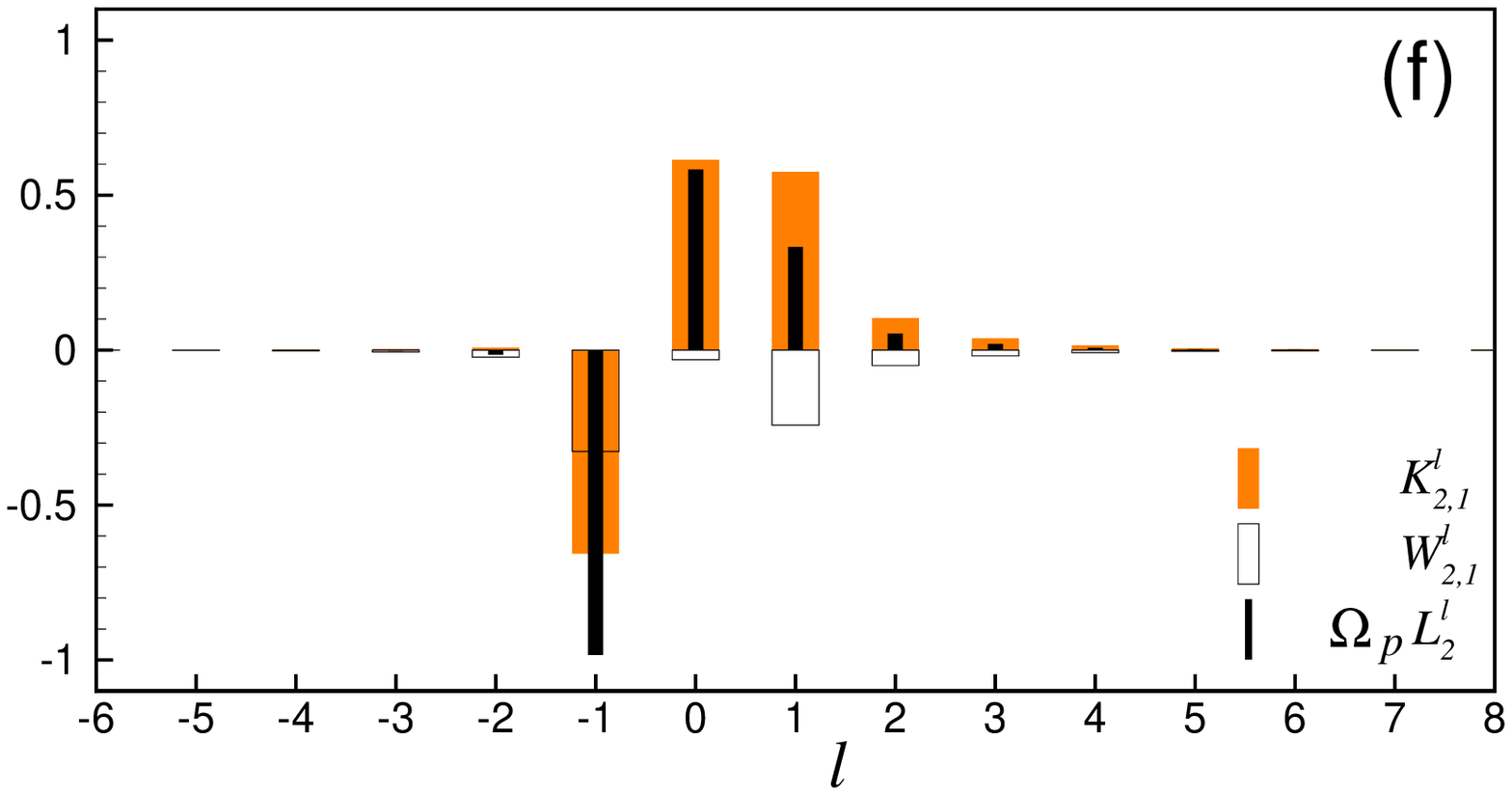}
\caption{The left panels show the fundamental mode of the $m_{\rm
K}=4$ Athanassoula \& Sellwood model, tapered with $J_c=0.4$ and with
$8.4\%$ of its orbits retrograde. The right panels show our computed mode 
for their $\beta=3$, $m_{\rm K}=6$, $J_c=0.6$ model with $15.4\%$
retrograde orbits.
\label{fig8}} 
\end{figure*} 

Toomre's axisymmetric stability parameter $Q$ increases monotonically
outwards from its central value of $2\pi/[3.36(1+n_{\rm M}/2)^{1/2}]$
for Miyamoto models.  Kalnajs (1976b) introduced an alternative set of
models, which also depend on an integer parameter $m_{\rm K}$ which
increases as the models cool, and for which Toomre's $Q$ remains
almost constant. Sellwood \& Athanassoula (1986) modified these models
with the addition of two extra parameters: an angular momentum $J_c$
and $\beta$. They reversed the sense of rotation of a fractional mass
$M_{\rm retro}$ of stars in a unidirectional Kalnajs model with
angular momenta in the range $(0,J_c)$, following equation (5) of Zang
\& Hohl (1978). This gives a smoothly tapered DF, and eliminates the
discontinuity of the unidirectional model.  Our results for some of
these models are given in table \ref{table2}, and displayed in Figures
\ref{fig7} and \ref{fig8}.  The fundamental mode of the unidirectional
$m_{\rm K}=6$ model, listed in Table \ref{table2} has not been plotted
because it is a compact and rapidly rotating bar, a little larger, but
otherwise just like that of the Miyamoto model shown in the left
panels of Figure \ref{fig6}.  The right panels of Figure \ref{fig7}
show how the compact bar is modified by a tapering with $J_c=0.25$
which reverses less than 5\% of the orbits. Like the cutout, it gives
the fundamental mode a smoother, more spiral, and more extensive
pattern, and a slower growth rate. A major difference is that the
taper diminishes the pattern speed, whereas the cutout increases
it. Both the tapering and the cutout cause a large change in the
population of low angular momentum orbits which dominate in the
central part of the disk where the fundamental mode is
concentrated. Whereas the cutout removes many of them, the tapering
merely reverses the rotation of many, and makes the DF isotropic for
$|L| \ll J_c$. Figure \ref{fig7}{\em f} shows a bar chart which is of
the standard pattern, though the tapering has increased the magnitudes
of the $l \geq 0$ energy components.

We computed the fundamental mode for the tapered $m_{\rm K}=4$,
$J_c=0.4$ model because it is one of the few for which Sellwood \&
Athanassoula plot a mode shape. Our results in the left panels of
Figure \ref{fig8} are similar to, though less spiral than, the
fundamental mode of the cooler and more sharply tapered disk
in the right panels of Figure \ref{fig7}, though there is
a remarkable dearth of angular momentum transferred to its $l=1$ component.
The major difference between their Figure 5b and our Figure \ref{fig8}{\em a}
is that theirs, which is less spiral, has some central structure which
is not present in Figure \ref{fig8}{\em a}. 

Table \ref{table2} compares our results with those of Athanassoula \&
Sellwood (1986) for four of the 33 models for which they did $N$-body
simulations. Some discrepancies between the matrix theory and the
simulations with their finite number of particles and gravity
softening are to be expected. Nevertheless, the large discrepancies
between our results and theirs, and that of Polyachenko (2004), for
the $\beta=0$ models are worrisome. Our best explanation for it
is that we have found these disks to be very sensitive to near-radial orbits,
and, as we explain in \S \ref{sec::actionspaceintegrations} and \S
\ref{sec::eigenvaluesearch}, we find eigenvalues to be sensitive to
the accuracy with which the central regions are handled.

In contrast to these discrepant cases, our results for the two
non--zero $\beta$ models listed in Table \ref{table2} agree reasonably
well with those of Athanassoula \& Sellwood.  A positive value of $\beta$
removes orbits with low energies and angular momentum high relative to
$L_c(E)$, and adds to those with higher energies and lower angular momentum.
It is easy to show from their Appendix A that
\begin{equation}
\label{eq::f0AS}
f_0(E,0)=\frac{(m_{\rm K}-2\beta E^2_s)}{2\pi^2}\exp[(\beta(1-E^2_s)],
~~ E_s=\frac{R_{\rm C} E}{GM}.
\end{equation}
The value of $\beta$ can not exceed $m_{\rm K}/2$ because $f_0$ is
then negative at the center of the disk where the scaled energy
$E_s=-1$. Many of Athanassoula \& Sellwood's models have marginal
values $\beta=-m_{\rm K}/2$. Their $f_0$ values peak at intermediate
values of $E_s$, and they have double peaked radial velocity profiles. The
$(m_{\rm K},\beta,J_c)=(6,3,0.6)$ model is one of two to which
Polyachenko (2004) applied his simplified theory to get a value of the
pattern speed $\Omega_p$ which agrees with what we get from $l=-1$
terms only. Our plot of the most unstable mode for this case in Figure
\ref{fig8}{\em b} shows the same concentrated central
structure as in Polyachenko's Figure 6, but his plot has a stronger
second hump around $R=1$ than our weaker one, and lacks the extended
outer spiral that we find. The structure of this mode suggests that it
is a secondary one, though it is the only one we have found. Its low
pattern speed means that it comes closer to having an ILR than any
other mode in our survey.  Its bar chart in Figure \ref{fig8}{\em f}
is noteworthy for its large and negative $l=-1$ component of ${\cal
K}_{2,1}$. This mode is another one for which ${\cal K}_{2,2}/{\cal
K}_{2,1}<-1$, and hence converts kinetic to gravitational energy.  

\begin{deluxetable*}{rrrrrrrrrrrrrr}
\tabletypesize{\scriptsize}
\tablecaption{Eigenvalues of Kalnajs models for the isochrone disk.
                                       \label{table3}}
\tablecolumns{14} 
\tablewidth{0pc}
\tablehead{
\colhead{} & 
\colhead{} & 
\colhead{} & 
\multicolumn{3}{c}{Full Model} &   
\colhead{} &
\multicolumn{2}{c}{K78\tablenotemark{a}} &   
\colhead{} &
\multicolumn{2}{c}{PC97\tablenotemark{b}} &   
\colhead{} &
\multicolumn{1}{c}{$l=-1$ only} \\
\cline{4-6} 
\cline{8-9} 
\cline{11-12} 
\cline{14-14} \\
\colhead{$m_{\rm K}$} &
\colhead{$M_{\rm retro}$} &
\colhead{mode} &
\colhead{$\Omega _p$} &
\colhead{$s$} & 
\colhead{${\cal K}_{2,2}/{\cal K}_{2,1}$} & 
\colhead{} &
\colhead{$\Omega _p$} &
\colhead{$s$} & 
\colhead{} &
\colhead{$\Omega _p$} &
\colhead{$s$} & 
\colhead{} &
\colhead{$\Omega _p$} }
\startdata
6 & 0.056 & 1 & 0.169 & 0.080 & -0.11 &  & 
               0.170 & 0.075 &  &
               0.170 & 0.075 &  &
               0.094  \\
6 & 0.056 & 2 & 0.121 & 0.035 & -1.58 &  & 
               \nodata & \nodata &  &
               \nodata & \nodata &  &
               0.079  \\ \\
9 & 0.029 & 1 & 0.235 & 0.149 & -0.05 &  & 
               0.235 & 0.145 &  &
               0.235 & 0.145 &  &
               0.127  \\
9 & 0.029 & 2 & 0.183 & 0.089 & -0.92 &  &
               \nodata & \nodata &  &
               \nodata & \nodata &  &
               0.109  \\ \\
12 & 0.019 & 1 & 0.292 & 0.217 & 0.07 &  & 
               0.295 & 0.210 &  &
               0.295 & 0.210 &  &
               0.150  \\
12 & 0.019 & 2 & 0.234 & 0.148 & -0.60 &  &
               \nodata & \nodata &  &
               0.230 & 0.145 &  &
               0.133  \\
\enddata
\tablenotetext{a}{Kalnajs (1978).}
\tablenotetext{b}{Pichon \& Cannon (1997).} 
\end{deluxetable*}

\begin{figure*}
\plottwo{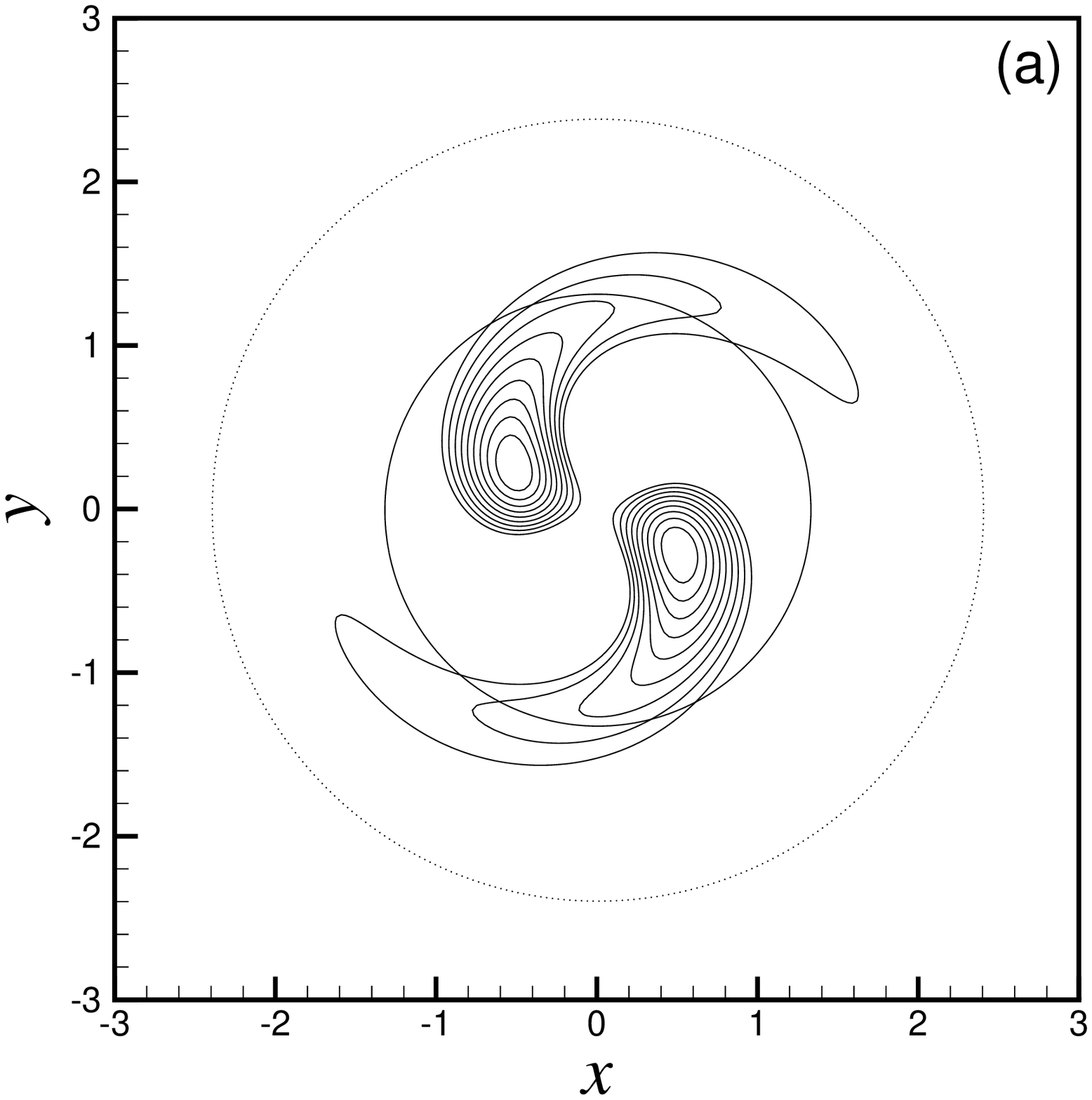}{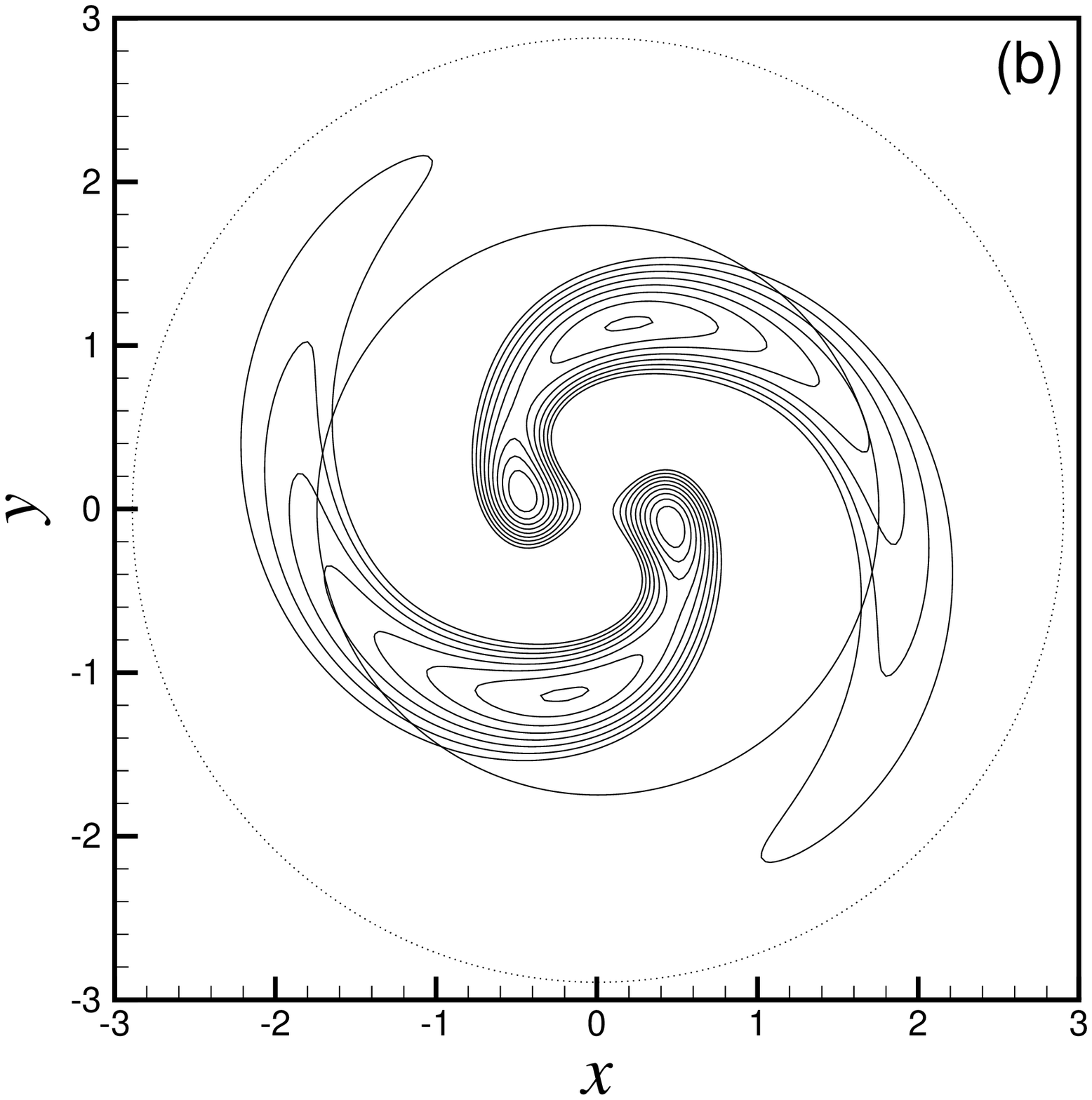}
\plottwo{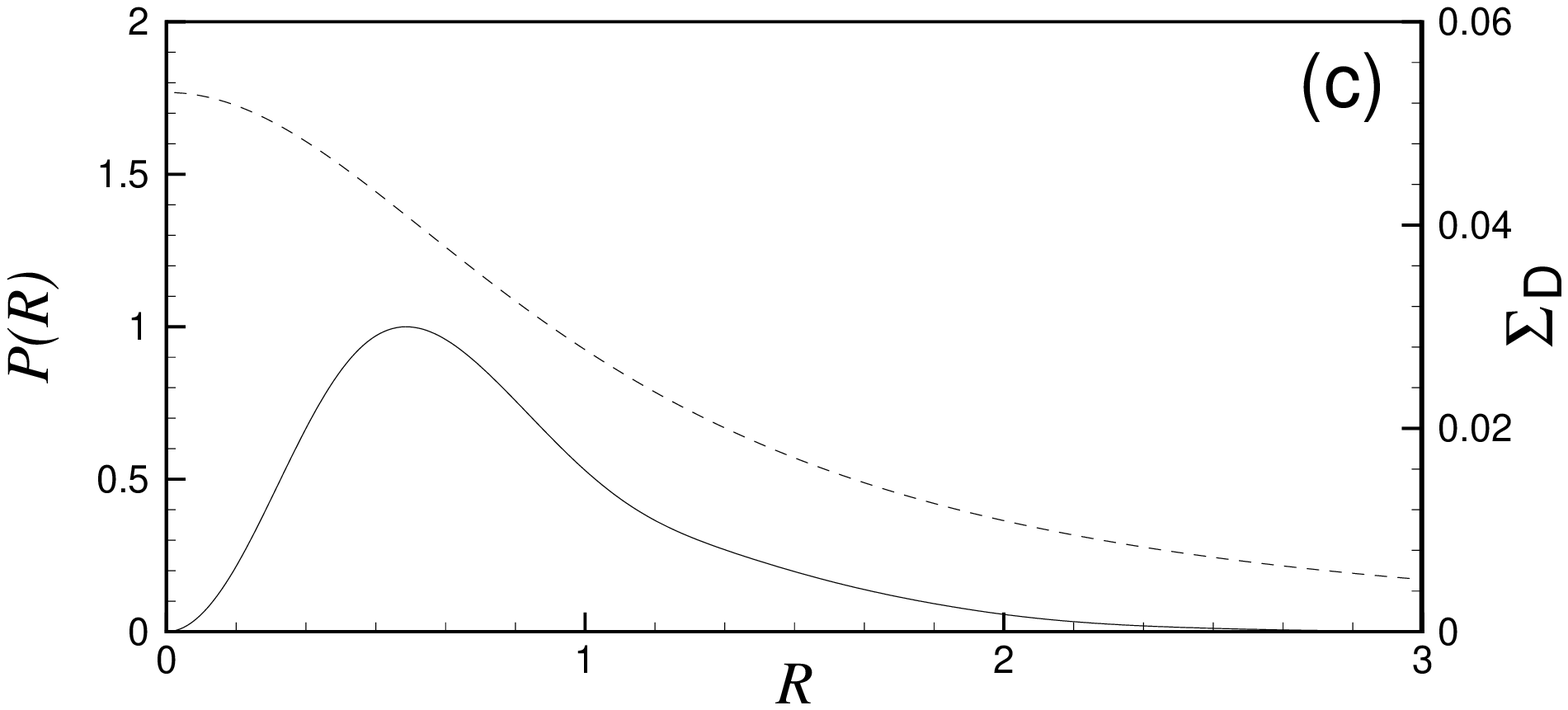}{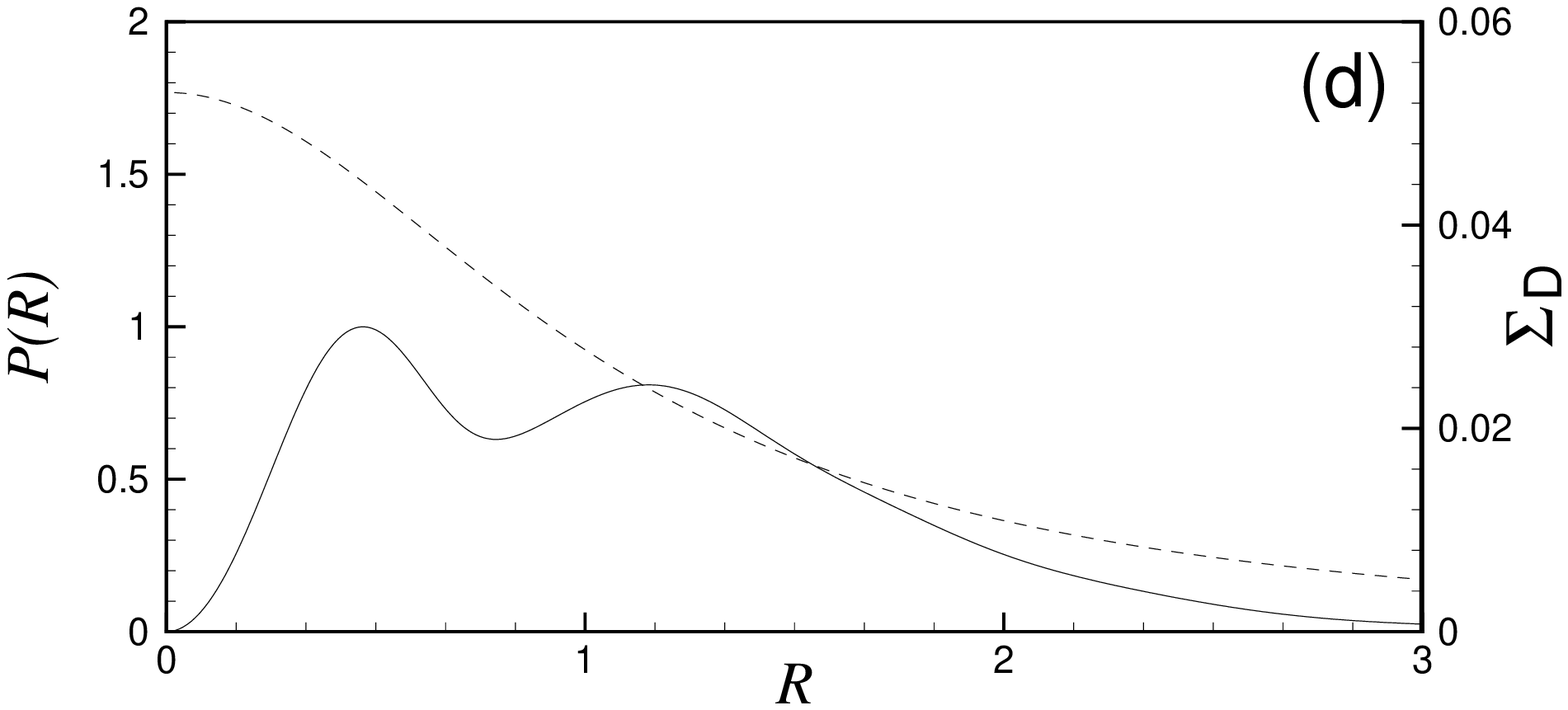}
\plottwo{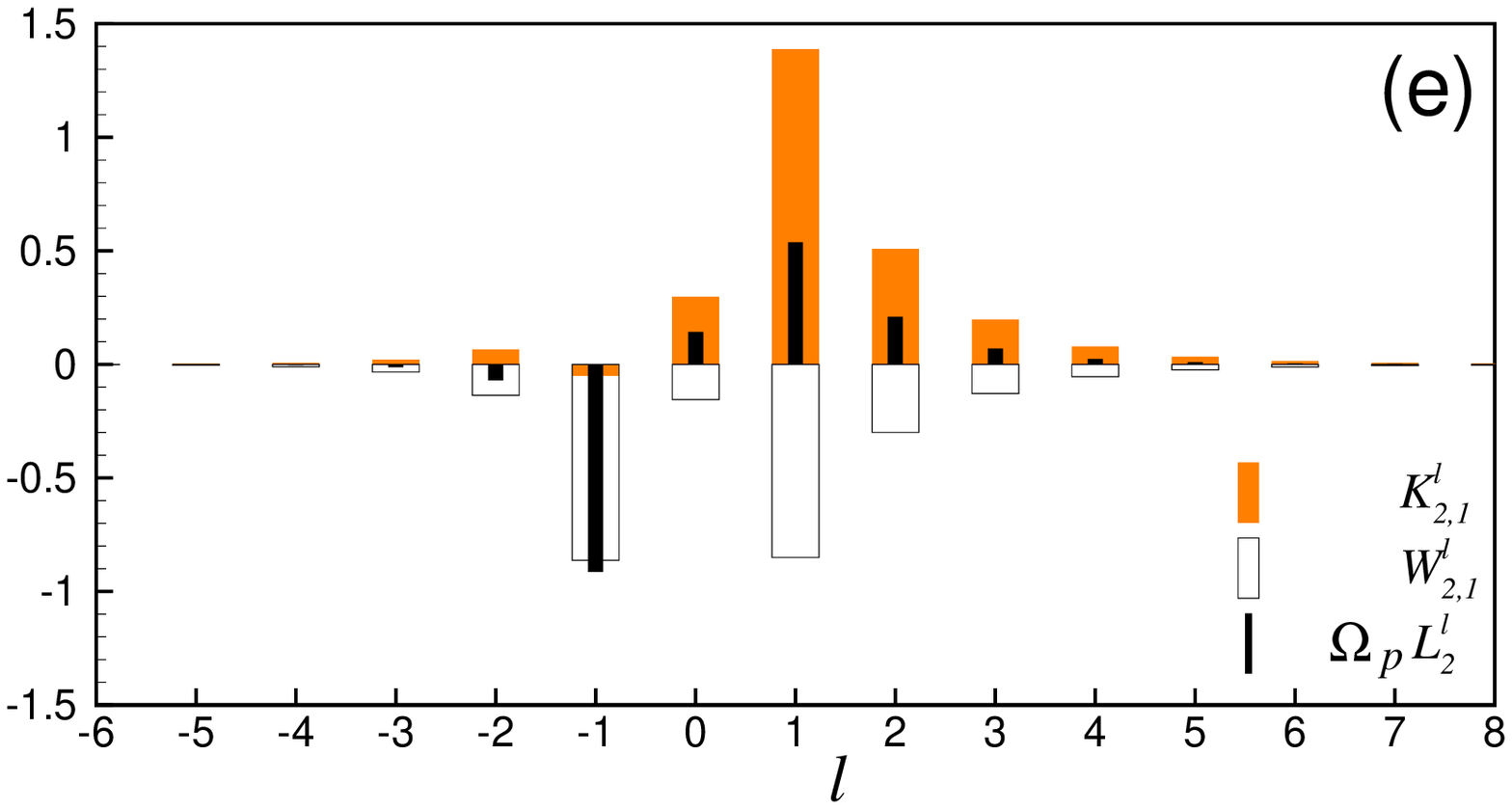}{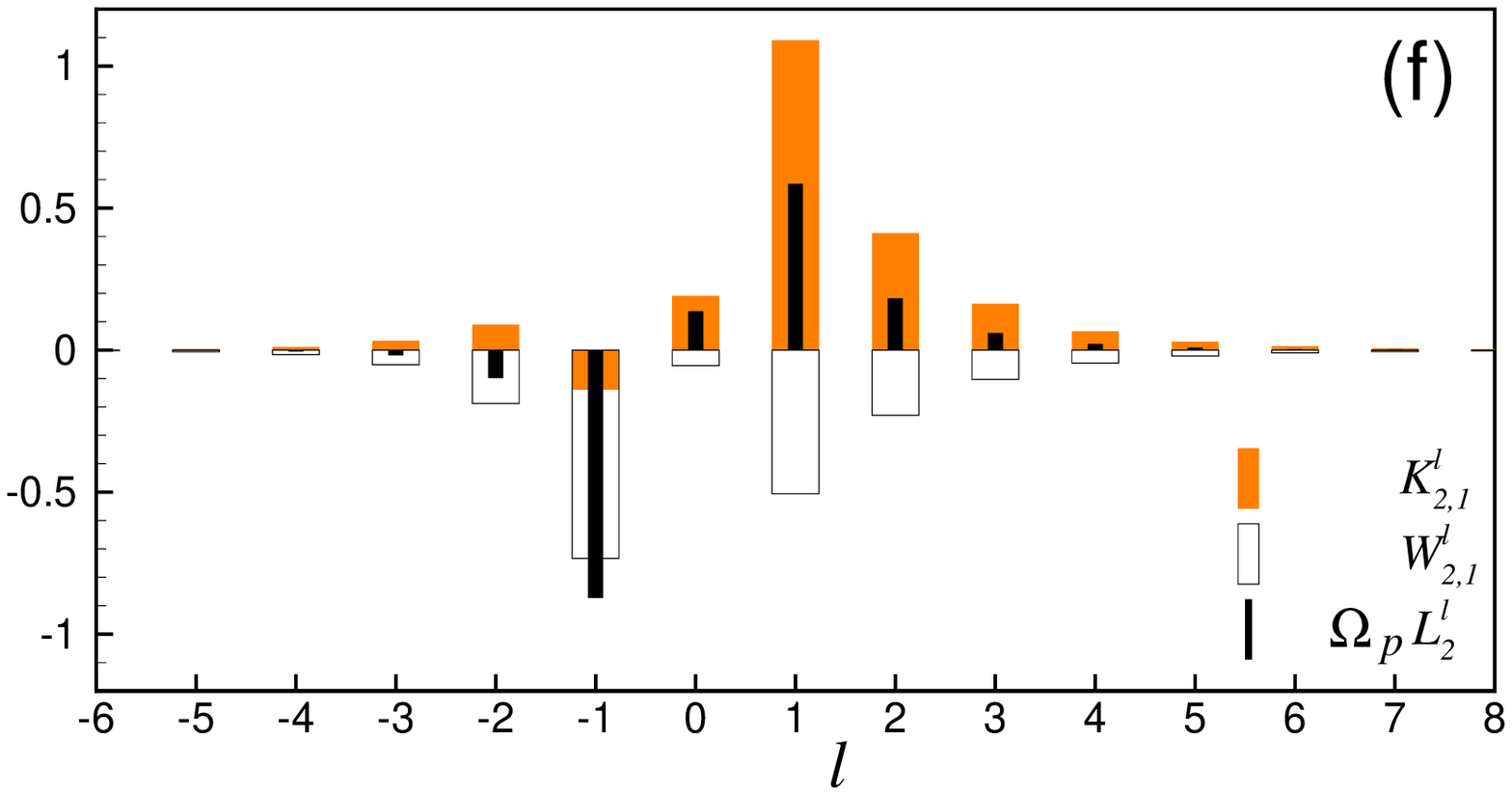}
\caption{Fundamental mode (left panels) and secondary mode
(right panels) of the isochrone disk with $m_{\rm K}=12$. 
\label{fig9}}
\end{figure*}

\subsection{Modes of Isochrone Disks}
\label{subsec::example-Kiso-disk}
Kalnajs (1976b) gives unidirectional models for the isochrone disk.
They contain an integer parameter $m_{\rm K}$
which increases as the models cool, and have fairly uniform 
values of the Toomre parameter $Q$. 
Kalnajs calculated modes for modified versions of these models
in which the sense of rotation of some stars is reversed to make them
retrograde. Kalnajs's formula for retrograde stars is given in equation (13)
of Earn \& Sellwood (1995). Combining that with the $x \to 0$ limit of
equation (26) of Kalnajs (1976b) and using GR formula (7.126.1) gives
\begin{equation}
\label{eq::isoretrodf}
f_0={m_{\rm K} \over 6\pi^2\left[1+J_R+ \vert J_{\phi} \vert \right]
^{2m_{\rm K}-2}}, ~~ J_{\phi}<0,
\end{equation}
for the DF of retrograde stars in units in which $G=M=R_{\rm C}=1$. There are
now retrograde stars of all angular momenta, not the limited ranges of the 
models of \S \ref{subsec::example-Kuz-disk}. Integration over phase space gives
\begin{equation}
\label{eq::isoretromass}
M_{\rm retro}={m_{\rm K} \over 3(m_{\rm K}-2)(2m_{\rm K}-3)},
\end{equation}
for the fractional mass in retrograde stars.

Table \ref{table3} gives our results and those of
others in units in which $G=M=R_{\rm C}=1$ for two modes of these models.
The mutual agreement is now gratifyingly close.
Both pattern speeds and growth rates increase with increasing $m_{\rm K}$,
as they do with Miyamoto models with increasing $n_{\rm M}$.
Pichon \& Cannon (1997) found secondary modes for some
values of $m_{\rm K}$ other than those we have listed in our table.
Again there are no ILRs. ILRs occur
only when $\Omega_p<0.0593$ in the units used here, for which the
ranges of both $\Omega_R$ and $\Omega_{\phi}$ are only
a half of what they are in Figure \ref{fig1}.

Figure \ref{fig9} displays two modes of the isochrone disk with
$m_{\rm K}=12$. The wave patterns match those of Earn \& Sellwood's
Figure 1. Spirality increases with $m_{\rm K}$. 
The secondary mode is more extended than the fundamental,
and the amplitude of its spiral arm is two--peaked, versus the single
peak of the fundamental mode.  Both bar charts show that the $l=1$
component absorbs much more angular momentum  and kinetic
energy than the $l=0$. This
is different from Figure \ref{fig6}, but is in line with the trend 
we noted with Miyamoto models with increasing $n_{\rm M}$, 
though note that $m_{\rm K}=12$ corresponds to
a much larger $n_{\rm M}$ value than the $n_{\rm M}=3$ of
Figure \ref{fig6}. The secondary mode of the warmest $m_{\rm
K}=6$ model is the only one for which kinetic is converted to
gravitational energy.

We studied some other models for the isochrone potential, 
and found, as did Pichon \& Cannon (1997), 
that their modes are qualitatively similar to those 
for Kuzmin's disk. Only their scales differ because the isochrone 
disk is the more spread out.   
   
\subsection{Modes of the Exponential Disk}
\label{sec::example-exponential}

\begin{deluxetable*}{rrrrrrrrrrrr}
\tablecaption{Eigenvalues for $m=2$ modes of exponential disks with $N=6$.
              \label{table4}}
\tablecolumns{11}
\tablewidth{0pt}
\tablehead{ \colhead{} & \colhead{} & \colhead{} &
            \colhead{} & \colhead{} &
\multicolumn{4}{c}{Full Model} &
\colhead{} &
\multicolumn{1}{c}{$l=-1$ only} \\
\cline{6-9} 
\cline{11-11} \\
\colhead{mode} & 
\colhead{$R_{\rm D}$} &
\colhead{$\Sigma _s R_{\rm D}$} &
\colhead{$L_0$} &
\colhead{$M_{\rm act}$} &
\colhead{$\Omega _p$} & 
\colhead{$s$} & 
\colhead{$R_{\rm CR}$} &
\colhead{$R_{\rm OLR}$} &
\colhead{} &
\colhead{$\Omega _p$} }
\startdata

1 & 1 & 0.42 & 0 & 1.000  & 0.914 & 1.151 & 
                      0.444 & 1.688 &  & 0.698 \\

2 & 1 & 0.42 & 0 & 1.000  & 0.693 & 0.324 & 
                      1.040 & 2.334 &  & 0.364 \\ \\

1 & 1.2 & 0.38 & 0 & 1.000 & 0.840 & 0.935 & 
                      0.646 & 1.871 &  & 0.591 \\ 

2 & 1.2 & 0.38 & 0 & 1.000 & 0.607 & 0.059 & 
                      1.309 & 2.701 &  & 0.330 \\ \\

1 & 1.4 & 0.36 & 0 & 1.000 & 0.805 & 0.794 & 
                      0.737 & 1.967 &  & 0.540 \\ 

2 & 1.4 & 0.36 & 0 & 1.000 & 0.467 & 0.179 & 
                      1.893 & 3.572 &  & 0.294 \\ \\

1 & 1.6 & 0.34 & 0 & 1.000 & 0.768 & 0.642 & 
                      0.834 & 2.077 &  & 0.488 \\ 

2 & 1.6 & 0.34 & 0 & 1.000 & 0.443 & 0.119 & 
                      2.024 & 3.775 &  & 0.277 \\ \\

1 & 1 & 0.42 & 0.1 & 0.967 & 1.170 & 0.259 & 
                     \nodata & 1.212 &  & \nodata \\ 

2 & 1 & 0.42 & 0.1 & 0.967 & 0.537 & 0.288 & 
                     1.571 & 3.082 &  & 0.331 \\ \\

1 & 1 & 0.42 & 0.3 & 0.913 & 1.044 & 0.228 & 
                     \nodata & 1.423 &  & \nodata \\ 

2 & 1 & 0.42 & 0.3 & 0.913 & 0.439 & 0.274 & 
                     2.047 & 3.811 &  & 0.302 \\ \\
\enddata
\end{deluxetable*}  

\begin{figure*}
\plottwo{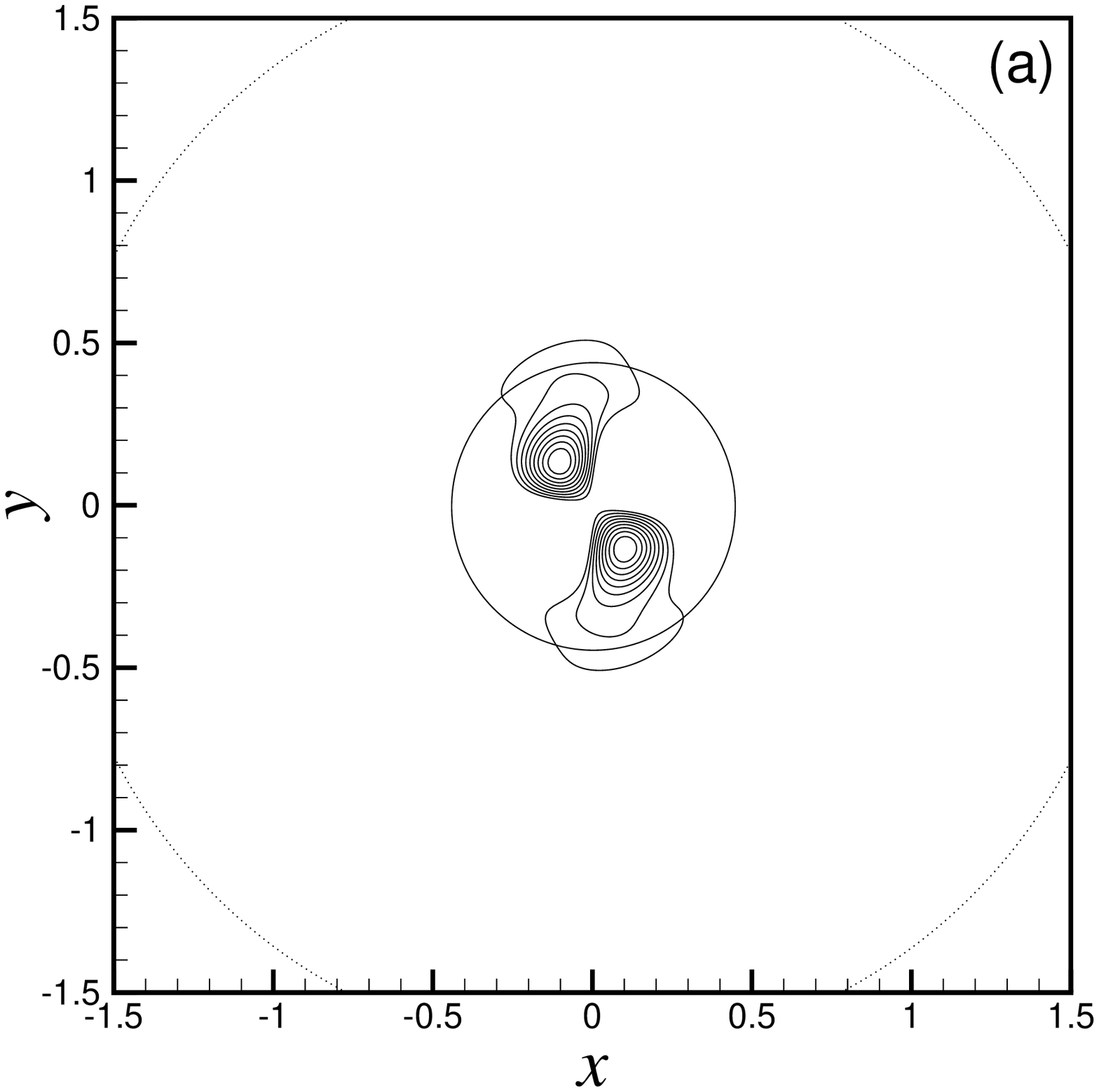}{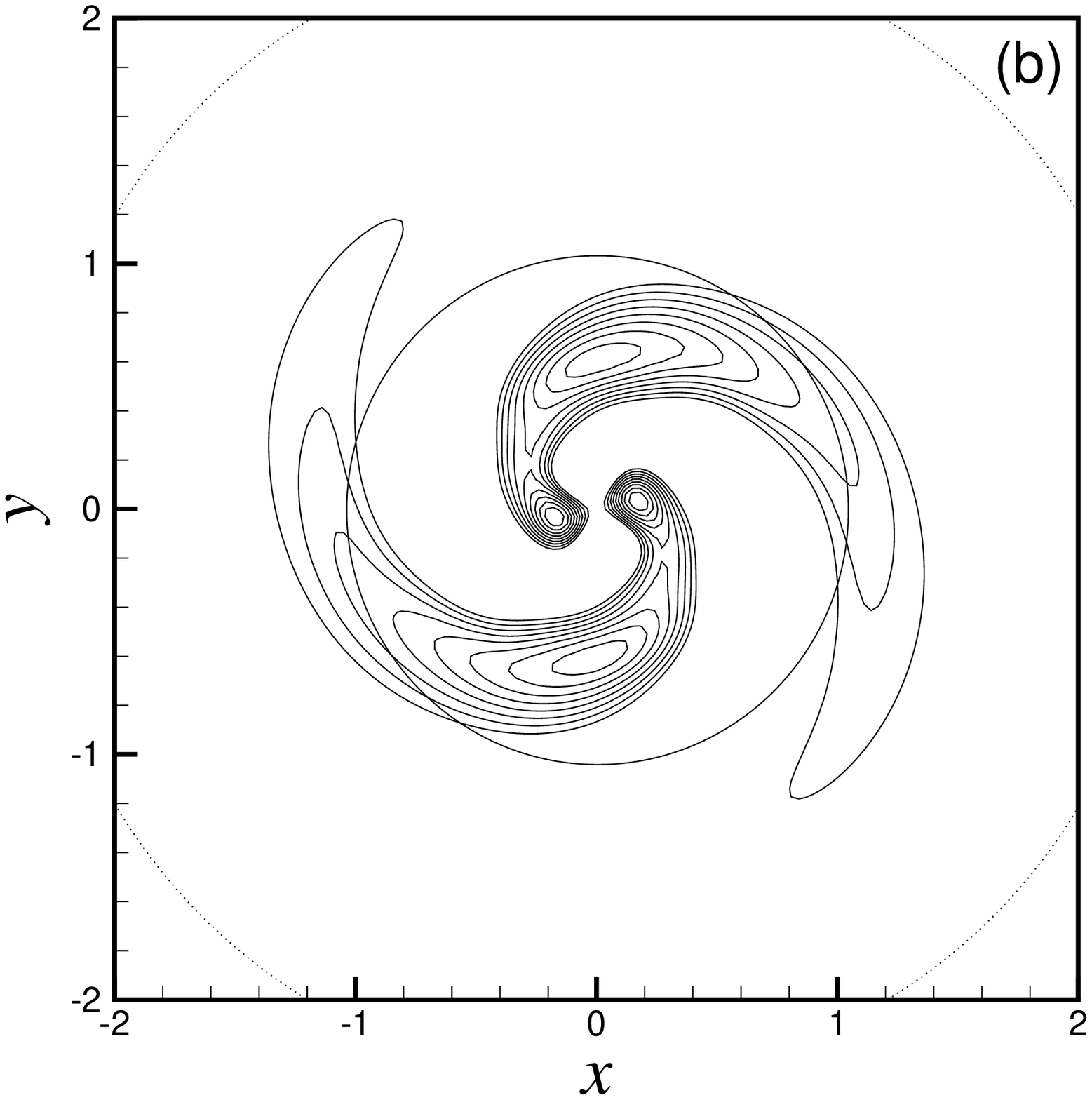}
\plottwo{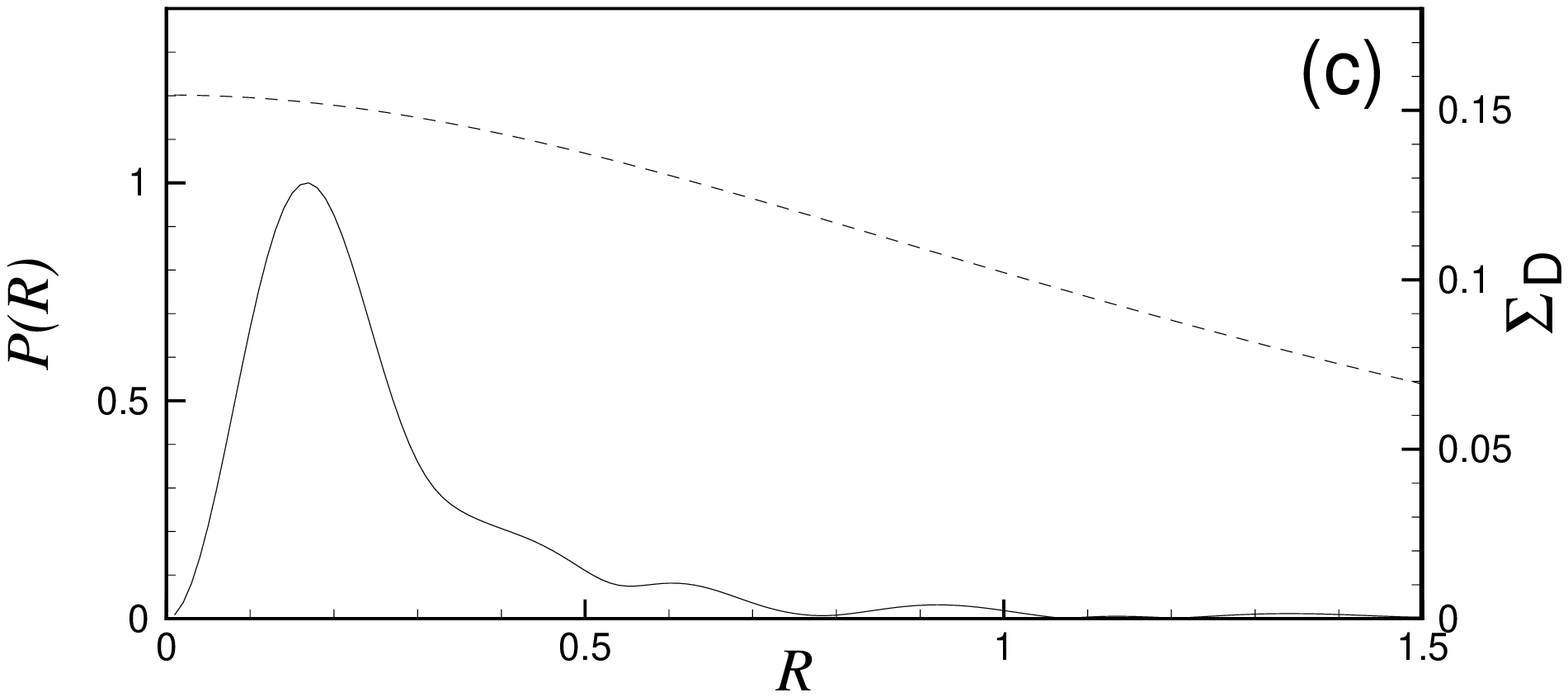}{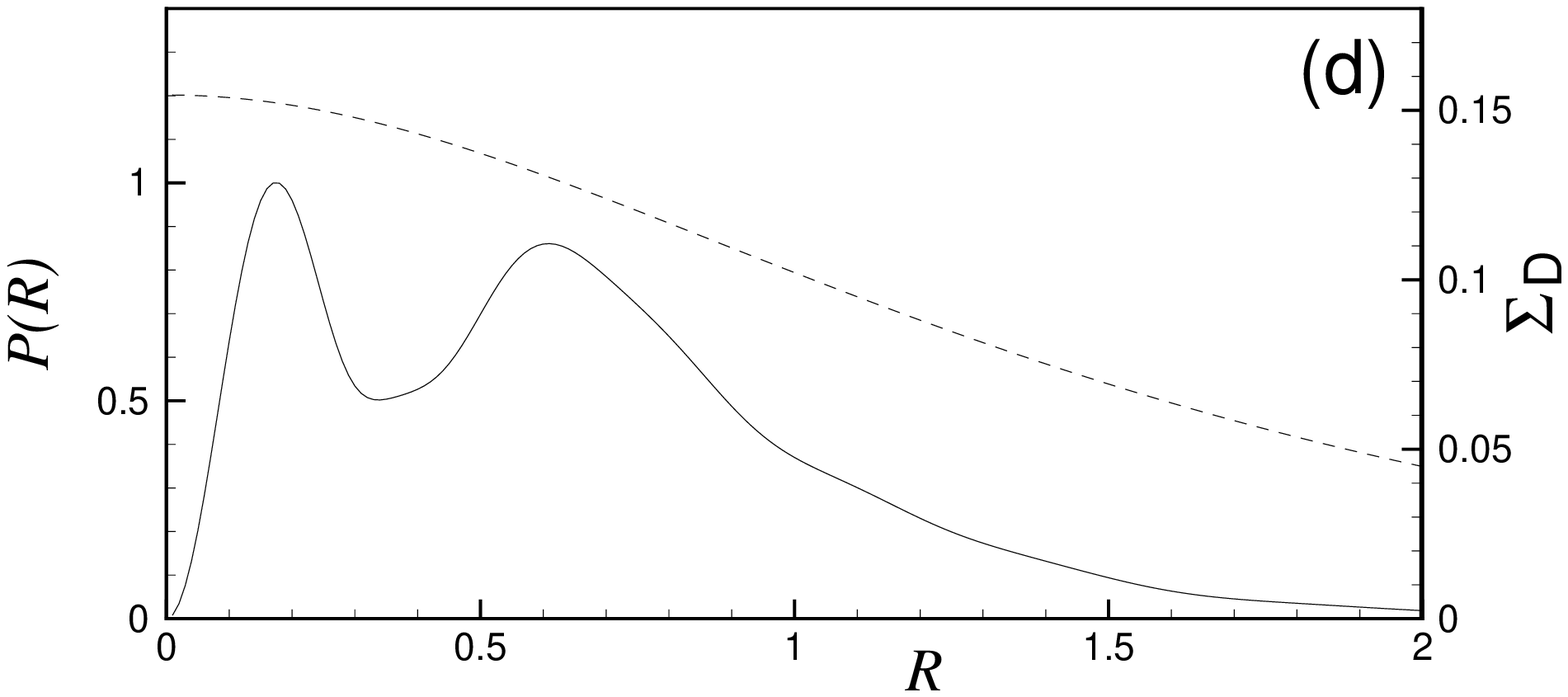}
\plottwo{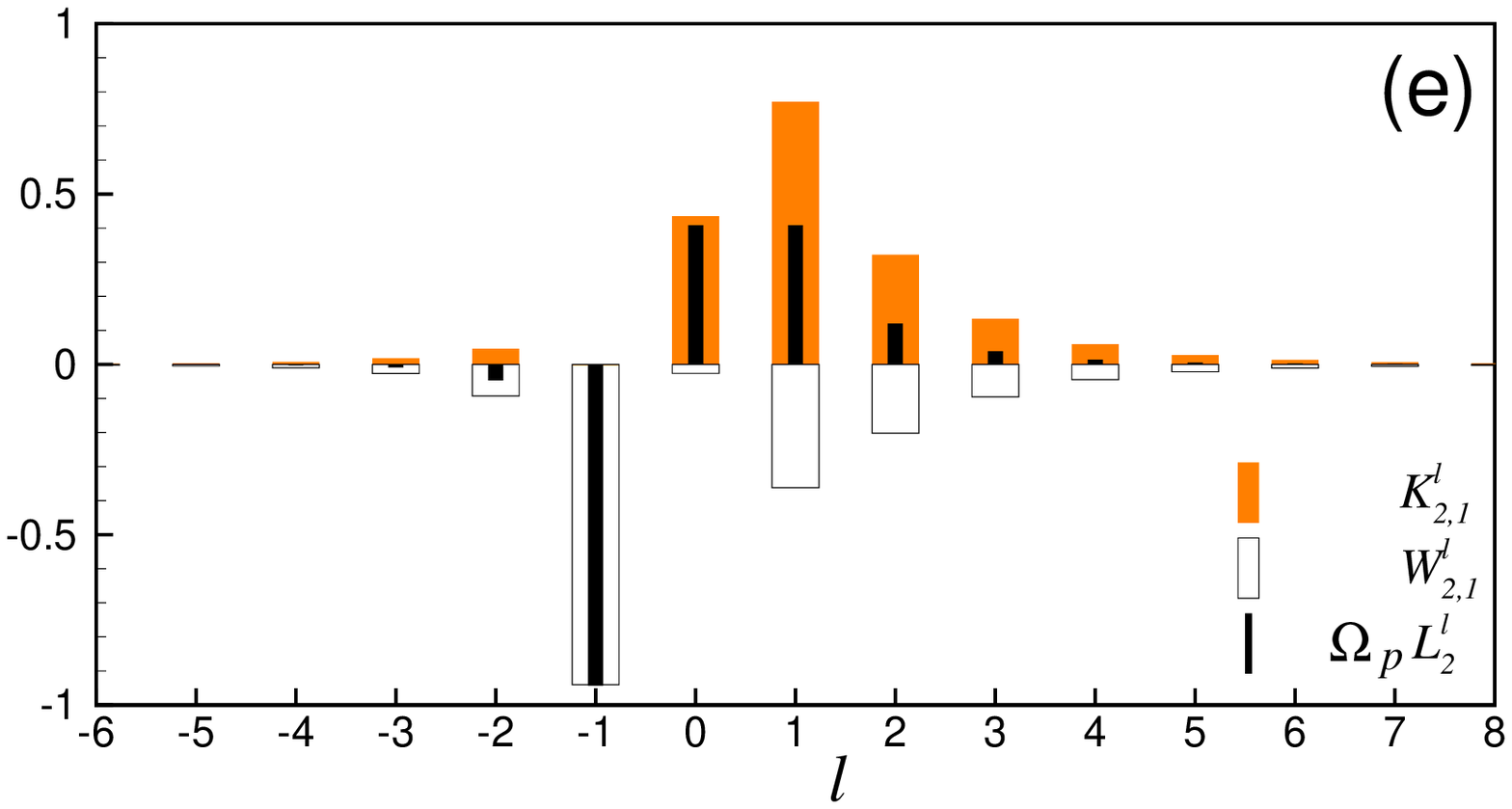}{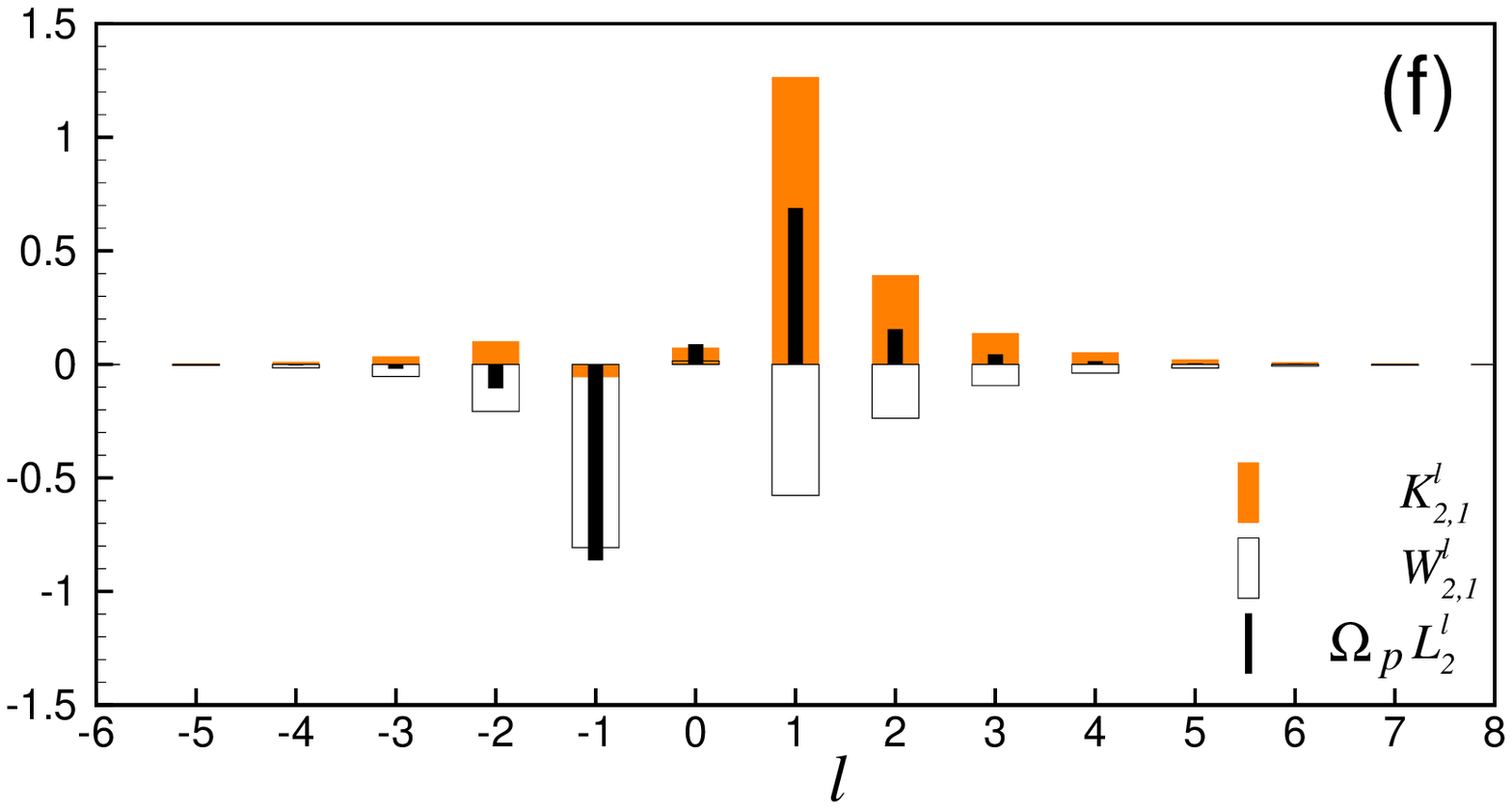}
\caption{Modes of the exponential disk for $N=6$, $R_{\rm D}=1$, 
and $\Sigma _s R_{\rm D}=0.42$. The left panels are for the fundamental mode,
and the right panels are for the secondary mode. 
\label{fig10}} 
\end{figure*}

\begin{figure}
\plotone{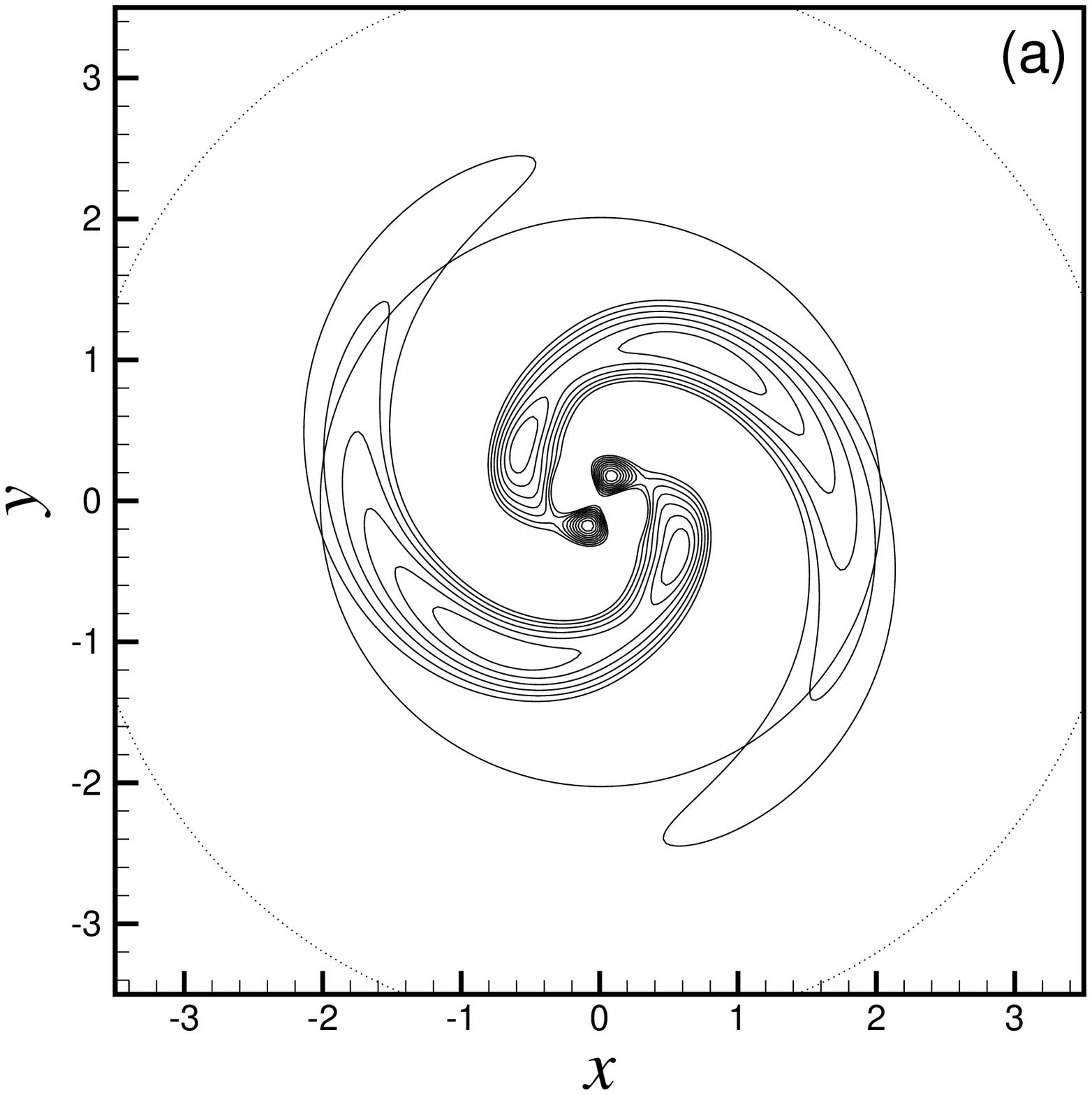}
\plotone{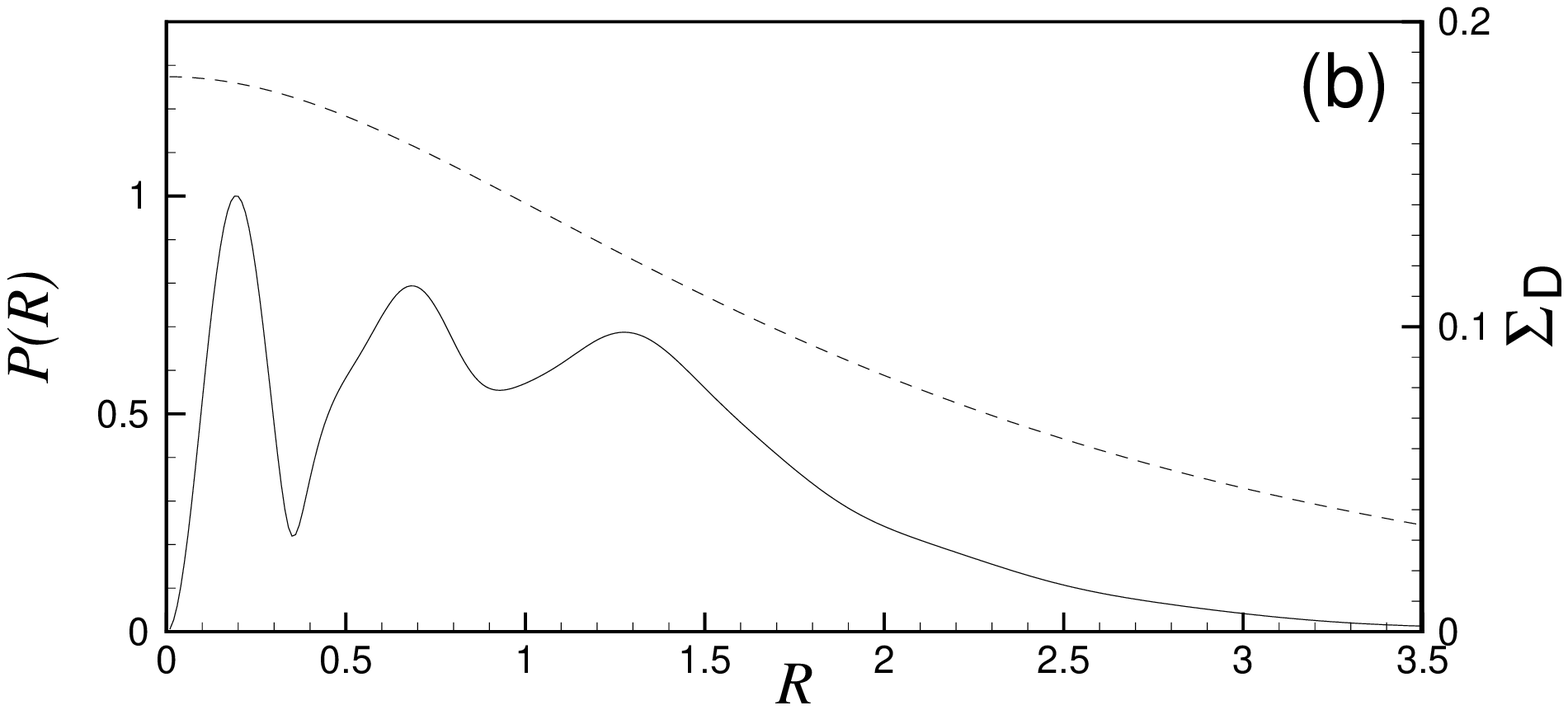}
\plotone{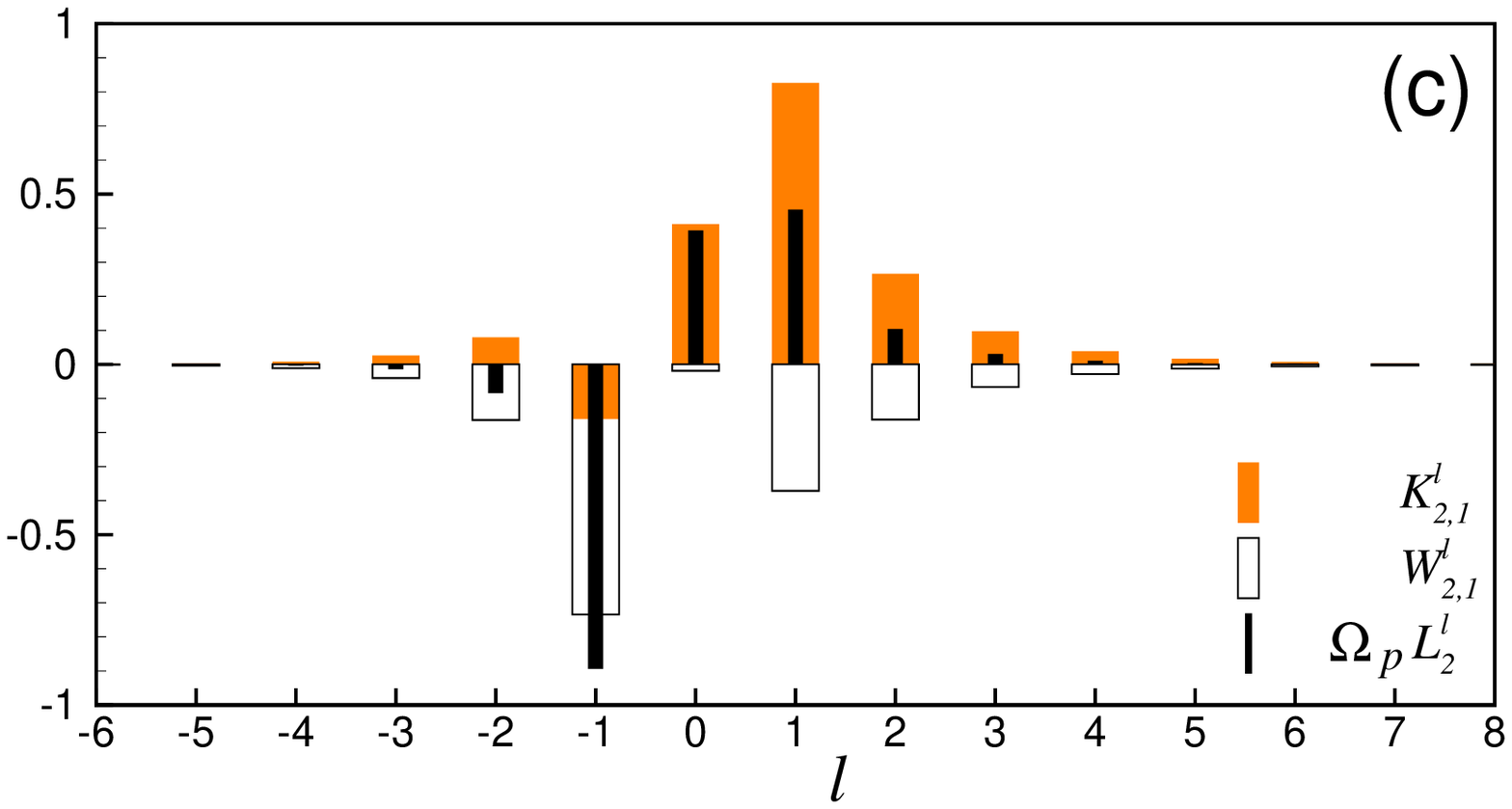}
\caption{The secondary mode of an $N=6$ exponential disk of larger
extent with $R_{\rm D}=1.6$ and $\Sigma _s R_{\rm D}=0.34$.  
\label{fig11}}  
\end{figure}  

\begin{figure}
\plotone{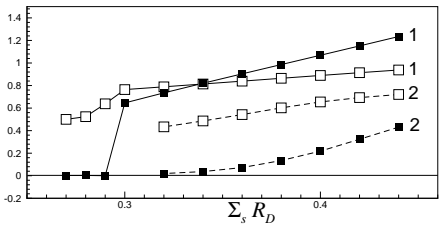}
\caption{The variation of the pattern speed $\Omega_p$ (open squares) 
and growth rate $s$ (filled squares) of the $N=6$, $R_{\rm D}=1$ exponential 
disk as the mass of the disk is varied. Data for the fundamental modes (1)
and secondary modes (2) are connected by solid and dashed lines, 
respectively. The secondary modes are double peaked. 
\label{fig12}}    
\end{figure}

\begin{figure}
\plotone{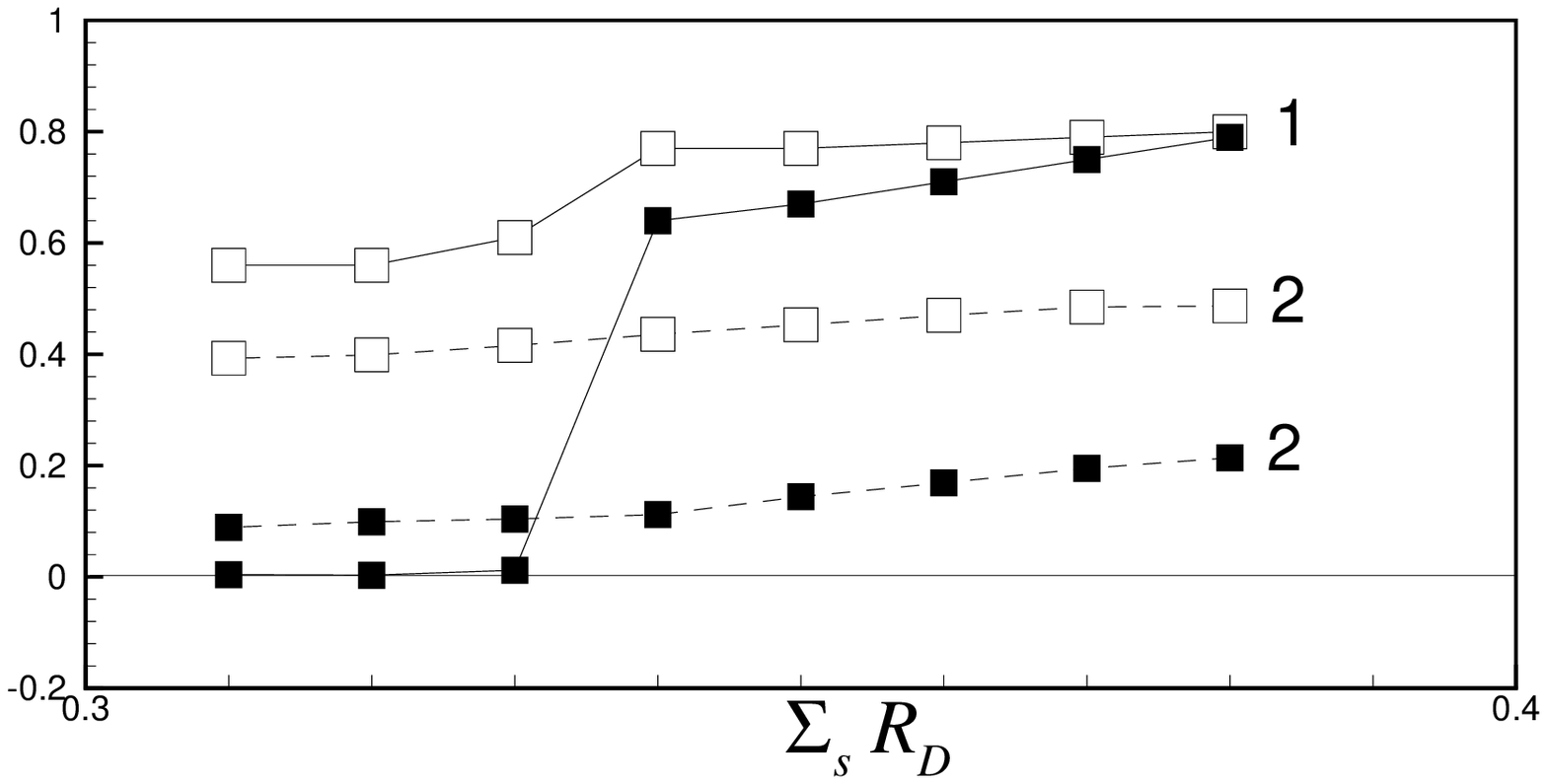}
\caption{Same as Figure \ref{fig12} but for $N=6$, $R_{\rm D}=1.6$.
The secondary modes now have triple peaks.  
\label{fig13}}  
\end{figure}

Our exponential disk models are the most varied. After normalizing with
units in which $G=v_0=R_{\rm C}=1$, we are still left with three parameters;
$N$ which measures the tendency of the orbits to circularity
(cf Figure \ref{fig5}), $R_{\rm D}$ which measures the length scale of
the exponential disk relative to the core radius of the potential,
and the density scale $\Sigma _s$.
A cutout introduces a further parameter $L_0$.
The central density of the disk is $\Sigma_se^{-\lambda}$ and its 
total mass $M_{\rm D}=2\pi\Sigma_s R_{\rm D}(R_{\rm C}+R_{\rm D})e^{-\lambda}$.
As Figure \ref{fig5}{\em a} shows, disks become increasingly
centrifugally supported with increasing $N$ and hence this parameter
is similar to the parameters $n_{\rm M}$ and $m_{\rm K}$ which we varied
for previous models. Here we restrict our exponential disk
models to the single case of $N=6$ to allow us to study the consequences
of varying the other parameters. As Figure \ref{fig5}{\em b} shows,
$N=6$ models have Toomre's $Q$ parameter close to unity over a
substantial central part of the disk, whose size increases with $R_{\rm D}$ 
when $R_{\rm D} \ge 1$.

Table \ref{table4} lists the fundamental and secondary modes for six
different $N=6$ models.  It lacks values of the ratio ${\cal
K}_{2,2}/{\cal K}_{2,1}$ because we were unable to compute accurate
values of ${\cal K}_{2,2}=-{\cal W}_{2,2}$ with our chosen basis
functions, for the reasons discussed in \S \ref{sec::basisfunctions}.
ILRs occur in the cored logarithmic potential only for
$\Omega_p<0.106$, which is much smaller than any of the pattern
speeds.  Figure \ref{fig10} shows the two modes for the first $R_{\rm
D}=1$, $\Sigma _s R_{\rm D}=0.42$, model.  The fundamental mode is a
rapidly rotating bar confined within the CR circle, and similar in all
respects to that of the Kuzmin disk in Figure \ref{fig6}{\em a}.  The
secondary mode has a slightly lower pattern speed and growth rate, and
has a spiral form. Its amplitude has the usual two peaks within the CR
circle, and its pattern extends a little beyond the CR circle but lies
within the OLR circle. Its bar chart is remarkable for its small $l=0$
components.

The first four models of Table \ref{table4} are chosen to study the effect of
varying $R_{\rm D}$. The parameter $\Sigma _s R_{\rm D}$ has to be changed too because
it is necessary to remain within the physically allowed region of
Figure \ref{fig2}. Its values are near 
marginal in that they are approximately 90\% of their allowed maximum.
Despite the decrease of $\Sigma _s R_{\rm D}$, the total disk mass grows 
as its scale length $R_{\rm D}$ increases, though, as Figure \ref{fig3} shows,
the halo/bulge grows in importance as $R_{\rm D}$ decreases, and the 
disk becomes progressively less maximal.
The structure of the fundamental bar mode does not
change along this sequence; it remains a compact and rapidly rotating
central bar. Figure \ref{fig11}{\em b} for $R_{\rm D}=1.6$ shows that the
amplitude of the secondary mode has developed a third hump. This
development occurs around $R_{\rm D} \approx 1.25$, and seems to be related
to the anomalously low growth rate at $R_{\rm D}=1.2$ 
(See Table \ref{table4}). The biggest difference between the bar chart
Figure \ref{fig11}{\em c} for the larger $R_{\rm D}=1.6$ disk and that of
Figure \ref{fig10}{\em f} for $R_{\rm D}=1$ is the greatly increased 
significance of the $l=0$ component.

Figures \ref{fig12} and \ref{fig13} illustrate how decreasing $\Sigma_s$ 
and hence the mass of the disk, whilst keeping its length scale
$R_{\rm D}$ fixed, stabilizes the disk. The transition of the fundamental
mode from stability to instability appears to take place through a
pitchfork bifurcation. That of the secondary mode appears to take
place through a tangent bifurcation. The order in which the two modes
are stabilized is different for the two different values of $R_{\rm D}$.
When $\Sigma _s R_{\rm D}$ is decreased at a fixed $R_{\rm D}$, then the part of
the rotational velocity due to the halo/bulge components increase from
those shown in Figure \ref{fig3} which are drawn for values of
$\Sigma _s R_{\rm D}$ which are 90\% of the allowed maximum. The stabilization
shown in figures \ref{fig12} and \ref{fig13} is therefore similar to
that which is achieved by sufficiently massive halos
(Kalnajs 1972, Ostriker \& Peebles 1973, Hohl 1975).

The stabilization of both modes in the neighborhood of 
$\Sigma _s R_{\rm D} \approx 0.3$ for both $R_{\rm D}=1$ and $R_{\rm D}=1.6$ suggests
that the stability boundary approaches the boundary of the physically
feasible region plotted in Figure \ref{fig2} as $R_{\rm D}$ increases,
i.e. as $\lambda$ decreases. It raises the possibility that the two boundaries
intersect before the $\lambda \to 0$ limit of an exponential disk in a
singular logarithmic potential, for which $\Sigma _s R_{\rm D}=0.304$
is reached. Our current computer algorithms are not capable of
approaching that limit, but they do show the $R_{\rm D}=2$ disk to be quite stable. 
This suggests that the classical
exponential disk ($\lambda =0$) with a completely flat rotation curve
($R_{\rm C}=0$) may be stable against bisymmetric excitations. This disk
is much less than maximal, and requires a substantial central
bulge/halo to maintain its equilibrium.

\begin{figure*}
\plottwo{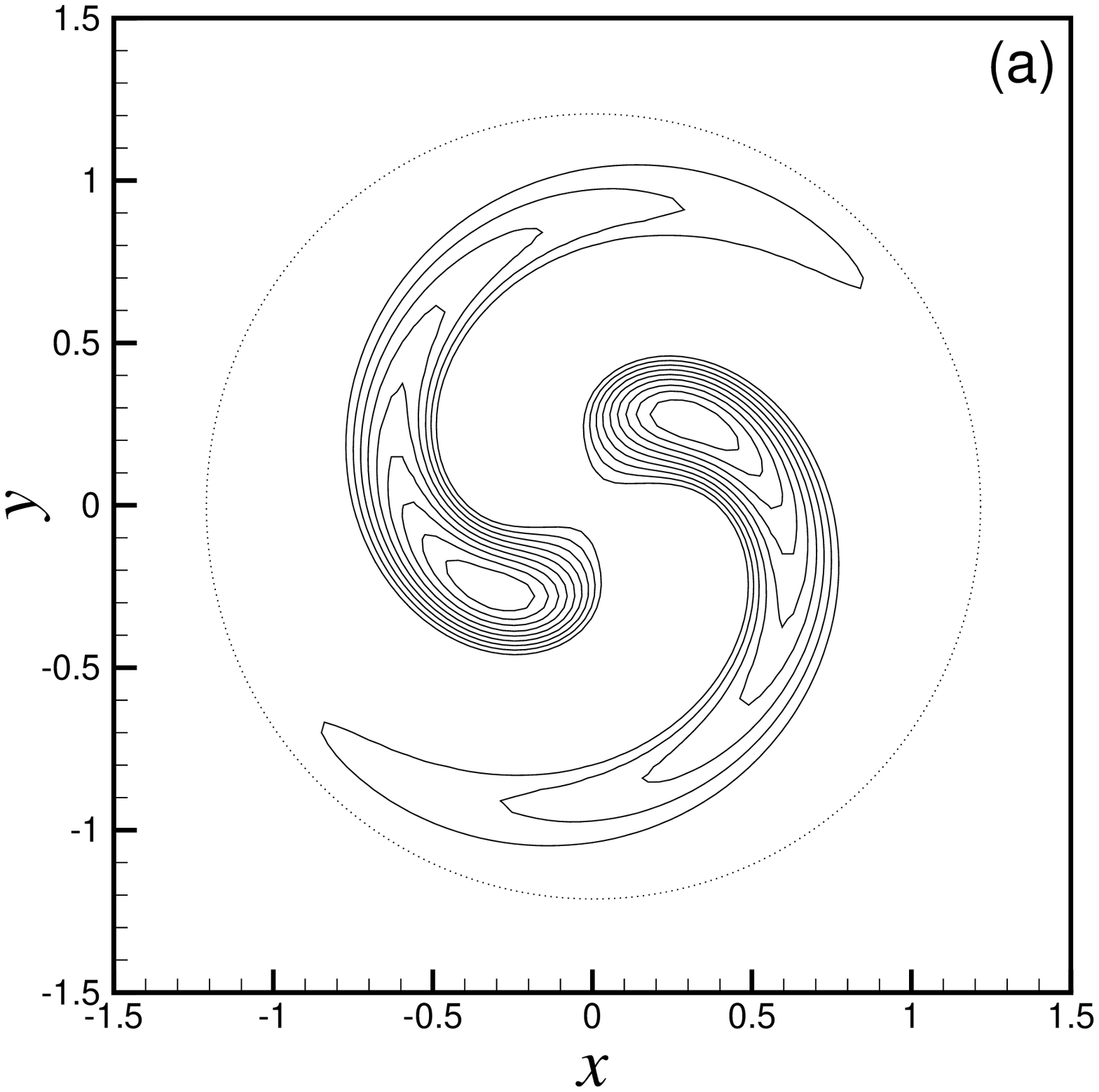}{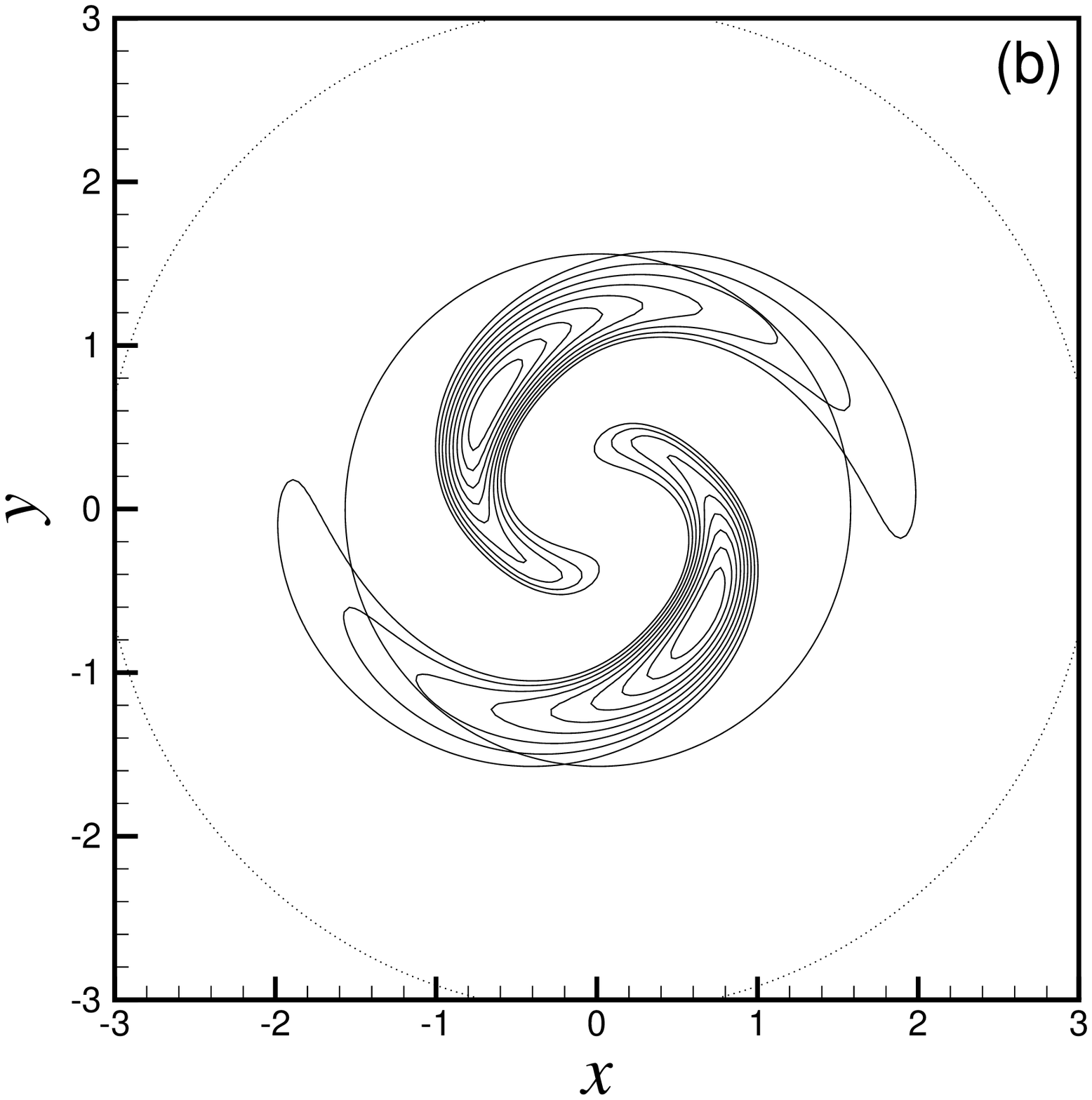}
\plottwo{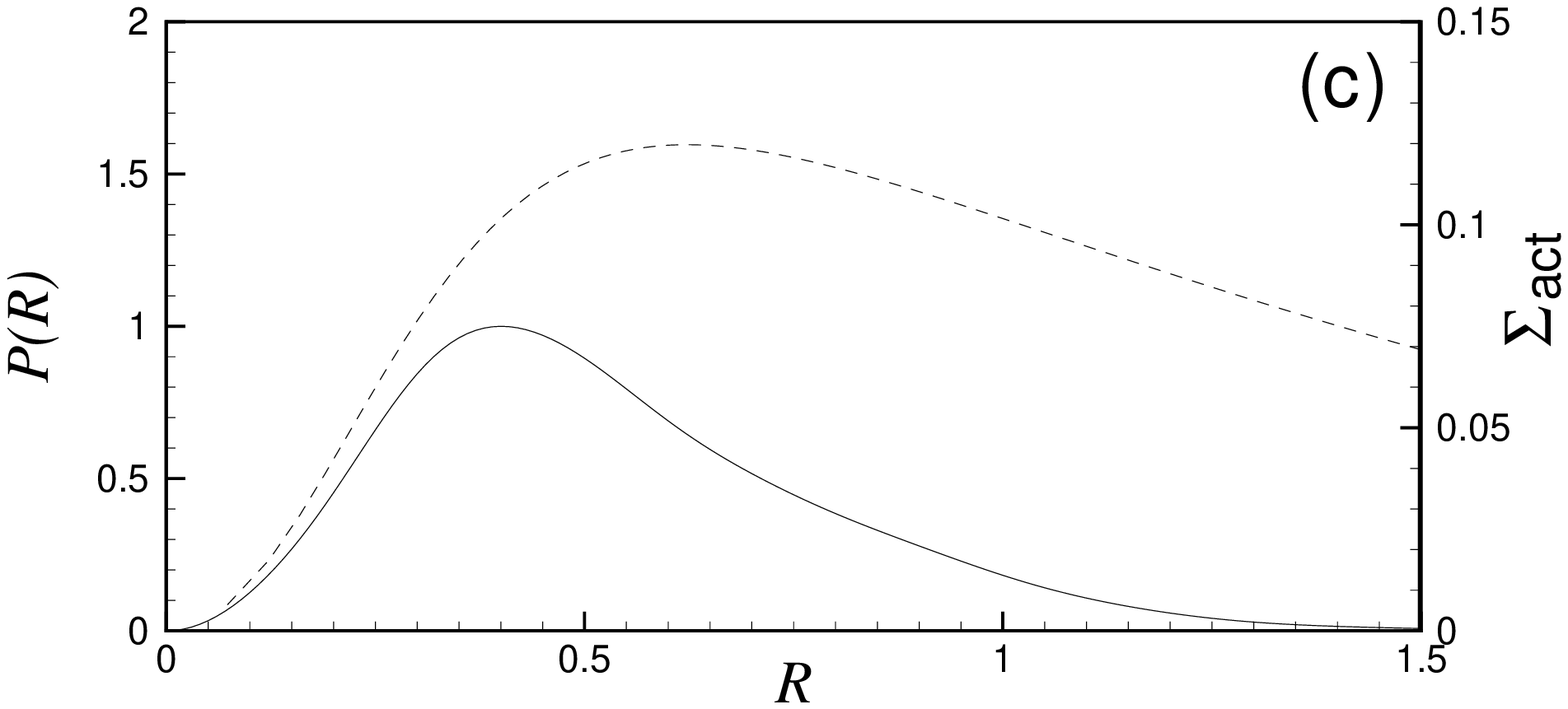}{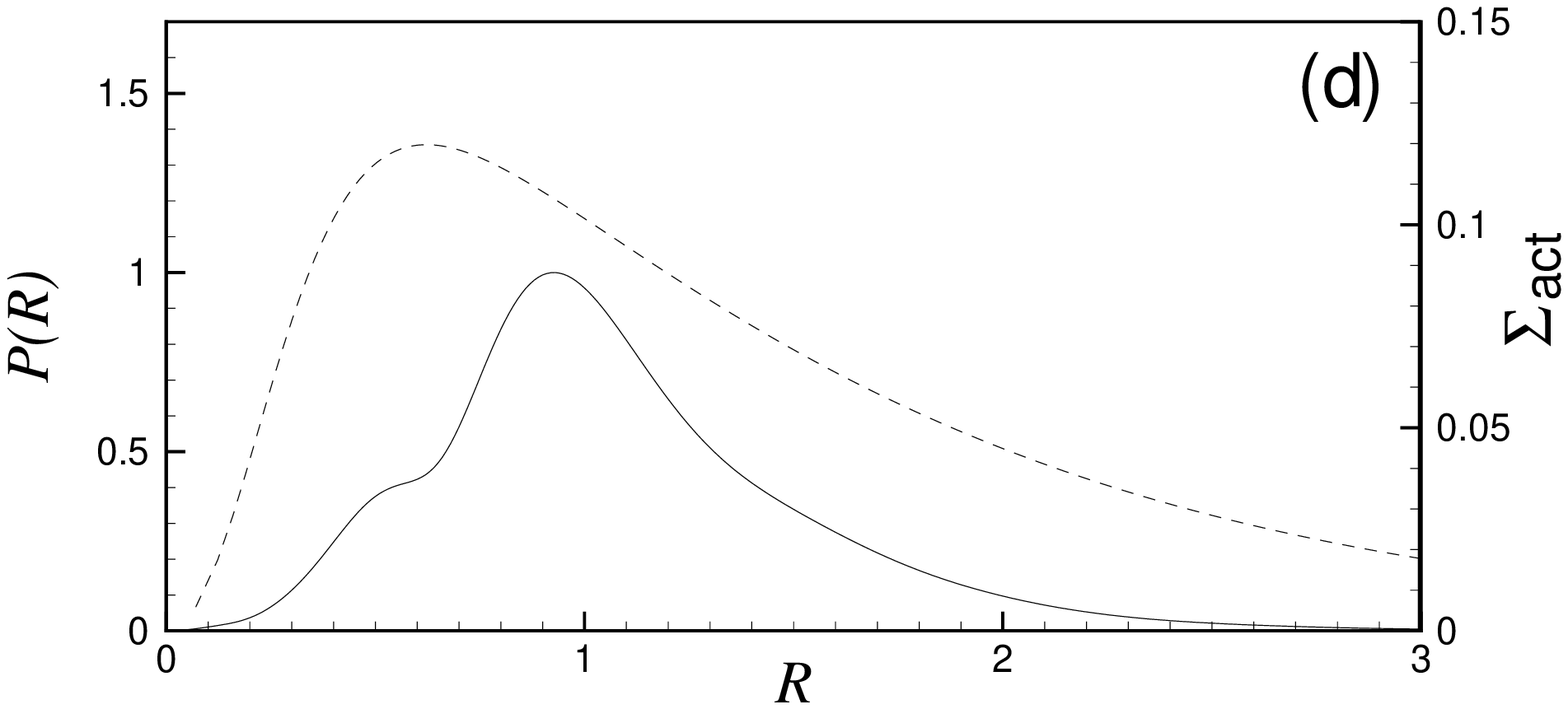}
\plottwo{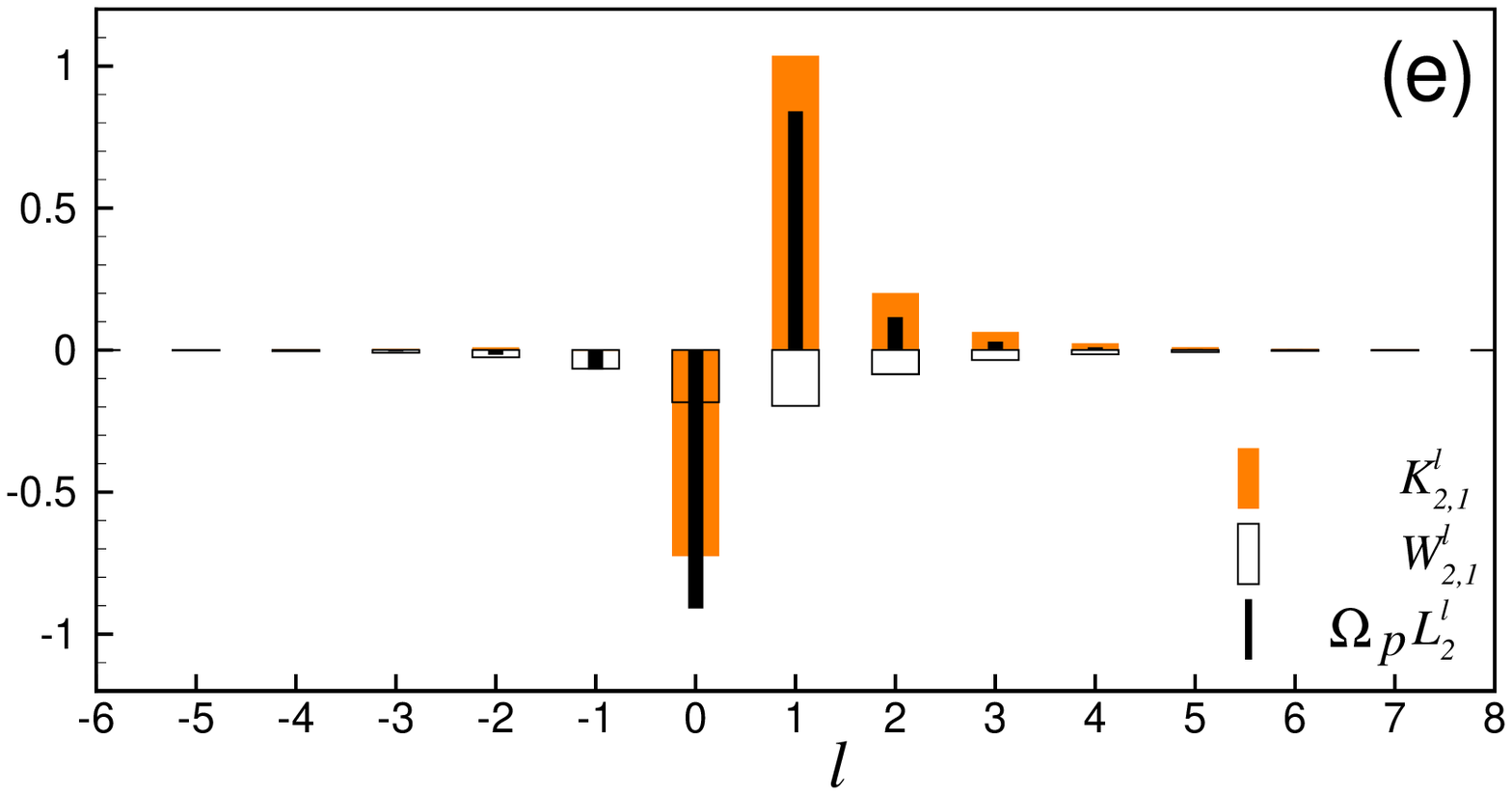}{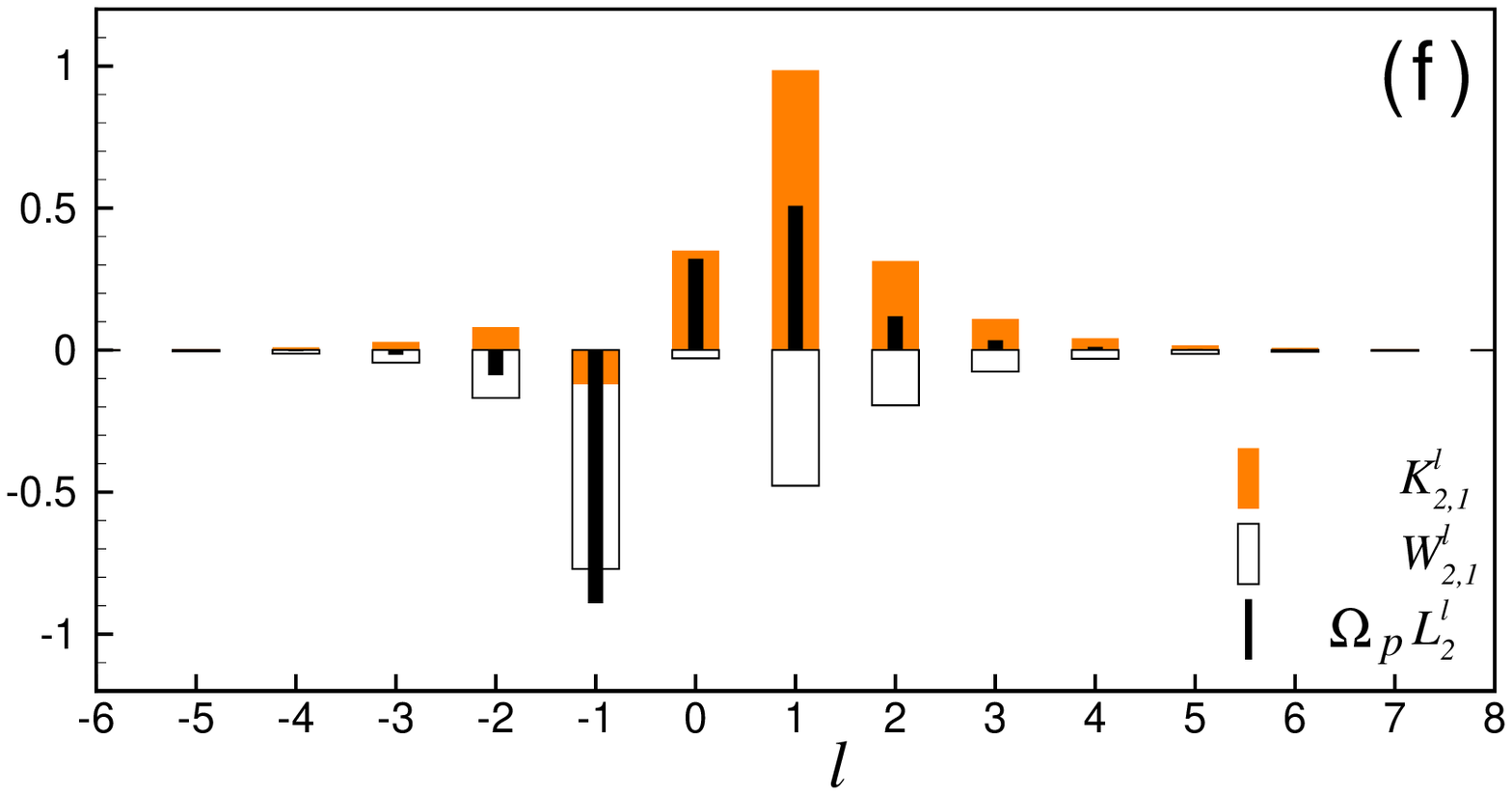}
\figcaption{Modes of the $N=6$, $R_{\rm D}=1$, and $\Sigma _s R_{\rm D}=0.42$ 
exponential disk with an $L_0=0.1$ cutout. The active surface density
is shown by the dashed lines in the central panels. The left panels 
show the fundamental mode and the right panels the secondary mode. 
\label{fig14}} 
\end{figure*}

The final Figure \ref{fig14} and the last two models of Table
\ref{table4} show the effect of cutting out low angular momentum
orbits via the cutout function (\ref{eq::ourcutout}). There is a major
change between $L_0=0$ and $L_0=0.1$. Comparison with Figure
\ref{fig10} shows that the fundamental mode is changed much more than
the secondary one, as we found with the Kuzmin disk.  Its growth rate
is diminished substantially, and its pattern speed is increased so
much that there is no longer a CR circle.  It is more spiral and
extensive as the peak of its amplitude has moved outwards from the
region in which the density has been diminished substantially.  Its
bar chart has undergone a large change and now resembles that 
of Figure \ref{fig7}{\em e}; the $l=-1$ components are
insignificant and the flow of both angular
momentum and $K_{2,1}$ is primarily from $l=0$ to $l=1$.
The growth rate and pattern speed of the more
extensive secondary mode is changed much less by the cutout. The
decrease in the inner density of the disk leads to the near total
disappearance of the inner hump of its amplitude. The $l=0$ components
are again significant in the bar chart of Figure \ref{fig14}{\em f},
unlike the $R_{\rm D}=1$ case of Figure \ref{fig10}{\em f}, but like that
for the $R_{\rm D}=1.6$ case of Figure \ref{fig11}{\em c}. The modes for
$L_0=0.3$ are similar to those for $L_0=0.1$.  The singular behavior
of $e^{-L^2/L_0^2}$ makes the study of small values of $L_0$ and 
the approach to the limit $L_0 \to 0$ computationally difficult.

We can not compare our results directly with those of Sellwood (1989)
which are for uncored exponential disks in the cored logarithmic
potential (\ref{eq::pot-coredlog}). Not only are the disks different,
but his are for the larger range $ 2 \le R_{\rm D}/R_{\rm C} \le 8.33$
of disk radii than ours. His $m=2$ modes have pattern speeds that are also
too fast for ILRs, but the spiral mode for $R_{\rm D}/R_{\rm C}=5$
shown in the left of his Figure 3 extends out to the OLR circle, and
so is more extensive than any of ours for exponential disks without
cutouts. As noted in \S \ref{subsec::expdisk_in_coredlog}, Sellwood's
value of $0.6$ for the critical parameter $v_0(R_{\rm D}/GM_{\rm D})^{1/2}$, 
which corresponds to $[e^{\lambda}/2\pi\Sigma_s R_{\rm D}(1+\lambda)]^{1/2}$ 
in our scalings, is smaller than any of
ours. That critical parameter is $0.86$ when stability is achieved for
$\Sigma_s R_{\rm D}=0.29$ in Figure \ref{fig12} for example. That
value is still significantly less than the $1.1$ which Efstathiou,
Lake \& Negroponte (1982) found to be necessary for their $N$--body
experiments, but that larger value may well be necessary for nonlinear
stability.

\section{DISCUSSION}
\label{sec::discussion}
We analyze and explain the properties of the bar charts in
\S\ref{subsec::barcharts}, and discuss the implications of our
findings for Polyachenko's (2004) simplified theory in
\S\ref{subsec::P04results}.

\subsection{Transfer of Angular Momentum and Energy}
\label{subsec::barcharts}
The rate of change of the perturbed angular momentum
(\ref{eq::dLtwodt}) contains a factor $s/\vert l\Omega _R+m\Omega
_{\phi} -\omega|^2 \vert^2$, which tends to
$\pi\left[m(\Omega_{\phi}-\Omega_p)+l\Omega_R\right]$ as $s \to 0$. As
Lynden-Bell \& Kalnajs (1972) note, this implies that a neutrally
stable wave emits and absorbs angular momentum only at resonances.
For the unstable modes we discuss, the factor $s/\vert l\Omega
_R+m\Omega _{\phi} -\omega|^2 \vert^2$ peaks at resonances, sharply so
when $s$ is small. This peaking implies that emission and absorption
occurs mainly near resonances.  Figure \ref{fig1} shows that, for
given pattern speed $\Omega_p>0$, the orbits associated with $l \geq
0$ resonances lie at successively smaller values of the orbital
frequencies $\Omega_R$ and $\Omega_{\phi}$, and hence successively
further out in the disk. That is the reason for interpreting the flows
of angular momentum, and also kinetic energy, which are proportional
to the quantities shown in the bar charts, as outward flows.

It is evident from equations (\ref{eq::W_2_1_ell}),
(\ref{eq::L_2_ell}), and (\ref{eq::K_2_1_ell}) that the partial
derivatives $\partial f_0/\partial J_R$ and $\partial f_0/\partial
J_{\phi}$ play a major role in determining the signs of the Fourier
components shown in the bar charts. All the other terms in $L_2^l$ are
either magnitudes or constants.  Lynden-Bell \& Kalnajs argued
that both derivatives are negative for a physically reasonable DF. If
this is so, then equation (\ref{eq::L_2_ell}) shows that $L_2^l > 0$
for $l \geq 0$.  The compensating negative values of $L_2^l$,
necessary because there is no net change of angular momentum, can and
must occur for $l<0$.  The fact that $L_2^{-1}$ is negative, as we
find, can be accounted for by $|\partial f_0/\partial J_R|$ being much
larger than $|\partial f_0/\partial J_{\phi}|$. The prediction matches
the standard pattern we find in \S \ref{sec::numresults}. The
assumptions that $\partial f_0/\partial J_R<0$ and $|\partial
f_0/\partial J_R| \gg |\partial f_0/\partial J_{\phi}|$ are generally
valid.  Figure 6 of Kalnajs (1976b) gives examples; the contours of
the two unidirectional DFs shown there decrease much more rapidly with
increasing $J_R$ than with $J_{\phi}$. However the modified
$\beta=m_{\rm K}/2$ models of \S \ref{subsec::example-Kuz-disk}
are exceptional because they have regions in which $\partial
f_0(E,L)/\partial E$, and hence $\partial f_0(J_R,J_{\phi})/\partial
J_R$, are positive.

Though positive values of $\partial f_0/\partial J_R$ are unusual,
positive values of $\partial f_0/\partial J_{\phi}$ are not. They are
unavoidable with disks which are totally, or mostly, unidirectional. 
If $f_0^P(J_R,0)=0$, as it is for the unidirectional 
models of Zang (1976) and Evans \& Read (1998a) and our cutout
disks, then $\partial f_0/\partial J_{\phi}$ must be positive for some
positive values of $J_{\phi}$ because otherwise $f_0(J_R,J_{\phi})$
could never become positive for $J_{\phi}>0$. Similarly $\partial
f_0/\partial J_{\phi}$ must be positive for some $J_{\phi}>-J_c$ for
the tapered models of Table \ref{table2}. Furthermore the explicit
formula (\ref{eq::isoretrodf}) shows that $\partial f_0/\partial
J_{\phi}>0$ for $J_{\phi}<0$ for the isochrone models of \S
\ref{subsec::example-Kiso-disk}. Although no positive values of
$\partial f_0/\partial J_{\phi}$ are visible in Figure 6 of Kalnajs
(1976b), these are disks for which the boundary integral
\begin{equation}
\label{eq::L_2_ellB}
-m^2\pi^2\int {\rm d} J_R \left[{f_0^P(J_R,0)|\tilde V_{l}|^2 \over
|l\Omega _R+m\Omega _{\phi} -\omega|^2} \right]_{J_{\phi}=0},
\end{equation}
must then be included in the expression for $L_2^l$. It is negative for
every $l$, as are the contributions to the integral (\ref{eq::L_2_ell})
from positive values of $\partial f_0/\partial J_{\phi}$. Unidirectional
disks for which $f_0^P(J_R,0)>0$ can be regarded as $J_c \to 0$ limits
of tapered disks; the boundary integral (\ref{eq::L_2_ellB}) then
accounts for the effect of the large positive values of $\partial
f_0/\partial J_{\phi}$ which occur in the taper. 

Despite the occurrence of regions of positive values of
$\partial f_0/\partial J_{\phi}$, they are generally either too small, or
confined to too limited regions, to modify the standard bar chart pattern.
However, cutout disks have central regions in which 
$\partial f_0/\partial J_{\phi}$ is large and positive. That is the reason for
the negative values of $L_2^0$ in the barcharts of Figures \ref{fig7}{\em e}
and \ref{fig14}{\em e} for fundamental modes of cutout disks.
The bar charts for the secondary modes of these cutout disks have
the standard form, as is seen in Figure \ref{fig14}{\em f}. This is
because $\partial f_0/\partial J_{\phi}<0$ for the more distant orbits
which are the more important for the secondary mode. 

Integral (\ref{eq::K_2_1_ell}) for $K_{2,1}^l$ differs from
integral (\ref{eq::L_2_ell}) for $L_2^l$ by a factor 
$(\Omega _{\phi}+l\Omega _R/2)$. This is positive for $l \ge -1$
for all direct orbits, but negative for $l \le -2$ (See Figure \ref{fig1}),
and is the reason why the signs of $K_{2,1}^l$ match those of $L_2^l$
for $l \ge -1$, but are their opposites for $l \le -2$. This pattern holds
even for the exceptional cases, with the result that $K_{2,1}^l$ is
positive for all $l$, except for $l=-1$ in standard bar charts and
$l=0$ in our exceptional ones.

Integral (\ref{eq::W_2_1_ell}) for $W_{2,1}^l$ differs from that for $L_2^l$
by a factor $(\Omega_p-\Omega _{\phi}-l\Omega _R/2)$. This factor is
positive for all orbits for $l=0$ modes when the pattern speed $\Omega_p >1$,
which is why $W_{2,1}^0$ is negative like $L_2^0$ for the exceptional
fundamental modes. $W_{2,1}^0$ and $L_2^0$ otherwise have opposite signs
for other modes with slower pattern speeds and $L_2^0>0$, and which
are dominated by orbits within CR. In fact $W_{2,1}^l$ is negative
for all $l$ in all bar charts. This is explained by noting that
$(\Omega_p-\Omega _{\phi}-l\Omega _R/2)$ is negative for modes which 
lie principally within OLR for $l \geq 1$, and positive for 
$l \leq -1$, so that the signs of $W_{2,1}^l$ are respectively the 
opposite and the same as those of $L_2^l$.

A striking feature of the bar charts is the rapidity with which the
quantities displayed there decrease with increasing $|l|$.
There are two reasons for this. One is the increase in the
denominators of their integrands in the regions where most orbits lie.
The $l=-1$ terms are generally important, even though there is no ILR,
because the denominator term $2(\Omega _{\phi}-\Omega_p)-\Omega _R$ is
never large for any orbit. Another reason is the decay of the
Fourier coefficients $\tilde V_{l}$ with increasing $|l|$.

\subsection{Abnormal Orbits and Polyachenko's Theory}
\label{subsec::P04results}
Tables \ref{table1} through \ref{table4} show consistently
that eigenvalues estimated using Polyachenko's
simplified theory are lower than those
calculated from the full matrix theory. Pattern speeds $\Omega_p$ well
exceed the narrow range of $\Omega_i$ values which the cored
potentials considered here allow, and so his basic assumption of small
$|\Omega_p-\Omega_i|$ is then not valid.  Our bar charts show clearly
that a few Fourier components other than $l=-1$ are always
significant.  The orbits which participate in the unstable modes that we
find are predominantly abnormal in the sense defined by Lynden-Bell
(1979) because all sufficiently central, as well as more radial,
orbits are abnormal. In fact even circular
orbits are normal only for $R/R_{\rm C}>2.44$ for Kuzmin's disk,
$R/R_{\rm C}>3.73$ for the isochrone, and $R/R_{\rm C}>25.3$ for the cored
logarithmic potential (See Figure \ref{fig4}). Our modes
lie primarily within these limits.

A quite different situation arises with the scale-free potentials
$V_0(R)={\rm sgn}(\alpha)R^{\alpha}$ for $\alpha \ne 0$, and
$V_0(R)=\ln R$ for $\alpha =0$, studied by Evans \& Read (1998a,b).
Even central orbits can then be normal because the normal/abnormal
classification then depends solely on the ratio $y=L/L_c(E)$, and is
independent of energy. Lynden-Bell's criterion (\ref{eq::abnormalcondition}) is
satisfied, and an orbit abnormal, if
\begin{eqnarray}
\label{eq::scalefreeabnormalcondition}
&-&\left[f(y)+\frac{\pi y}{2}\right]^2 f^{\prime\prime}(y)- \nonumber \\
&{}& \left[\frac{2-\alpha}{2+\alpha}\right]
\left[f^{\prime}(y)+\frac{\pi}{2}\right]^2 
\left[f(y)-yf^{\prime}(y)\right]>0,
\end{eqnarray}
where the function $f(y)=\pi J_R/L_c(E)$ and depends also on $\alpha$.
Its derivative
$f^{\prime}(y)=-\pi\Omega_{\phi}/\Omega_R=\frac{1}{2}|g(\alpha,y)|$,
where $g(\alpha,y)$ is the function defined and analyzed in Touma \&
Tremaine (1997).  The criterion (\ref{eq::scalefreeabnormalcondition})
gives $y<0.723$ for the scalefree logarithmic potential, so that all
the more circular orbits with large $L/L_c(E)$ are normal.  For
$\alpha=-0.25$ and a falling rotation curve, there is a wide range of
normal orbits, and only orbits with $y<0.496$ are abnormal.  For
$\alpha=0.25$ and a rising rotation curve, only nearly circular orbits
are normal, and all those with $y<0.973$ are abnormal.  All orbits are
abnormal for $\alpha>0.275$.  Interestingly Evans \& Read, who looked
specifically at the cases of $\alpha=\pm 0.25$, found that unstable
modes grow more vigorously for the rising rotation curve case of
$\alpha=0.25$ with many abnormal orbits. ILRs occur with
their modes because the range of $\Omega_i$ is unbounded for
scale-free potentials.

\section{SUMMARY}
\label{sec::summary}
This paper develops the theory of modes in thin stellar disks. It then
implements that theory for a selection of disks. The theory basically
is that pioneered by Kalnajs (1971,1977). Our development of its
Eulerian form to second order is new. The expressions for angular
momentum and total energy were given earlier by Kalnajs (1971) and
Lynden-Bell \& Kalnajs (1972), but the expressions for the two
separate components are new. The boundary integral terms which must be
included in the method are also new. These integrals are avoided when
a Lagrangian form of the theory is used. The Lagrangian form leads to
more complicated computations, which is why prior computational work,
except for that of Vandervoort (1999), has all been done for using the
Eulerian form.  We show in Appendix \ref{app::lagrangian} how the
passage from the Lagrangian to the Eulerian form allows us to
interpret the boundary integrals as boundary fluxes. This analysis
does not apply to the quite different version of Lagrangian theory
which Vandervoort uses.

Our applications of the theory are to $m=2$ modes in potentials with
smooth cores. In all but one case (see Table \ref{table2}), our search
has found two modes, a centrally concentrated fundamental mode with a
single peak in amplitude, and a more extensive and more spiral
secondary mode with at least two peaks. There may be others.  We find
that the shape, pattern speed, and growth rate of the fundamental mode
are especially sensitive to the relatively small proportion of orbits
which provide the density in central regions because they have low
angular momentum.  The growth rate of the fundamental mode is reduced
substantially, and its shape becomes more extensive and spiral, if
those orbits are either removed or reversed. Removing them increases
the pattern speed, while reversing them decreases it.  Removing them
also makes the more extensive and more spiral secondary mode the
faster growing in all but one of the cases in Tables \ref{table1} and
\ref{table4}. This sensitivity to orbits, many of which are near
radial, might suggest a connection with the phenomenon of {\it radial
orbit instability}. However that phenomenon, reviewed recently in
Merritt (1999), occurs in hot anisotropic spherical stellar systems in
which radial orbits are sufficiently predominant. The sensitivity we
find here arises even with cool stellar systems with few near radial
orbits.

With two exceptions, modes are largely confined within the CR circle,
but are too fast for there to be any orbits in ILR. The lack on an ILR
means that modes can propagate into, and be reflected from, the center
of the disk, even in cases in which our densities drop to zero there.
The modes are unstable, sometimes rapidly so, as swing-amplifier
theory (Toomre 1981) predicts. We have achieved stability only by
decreasing the active mass of the disk. In the exceptional cases, the
modes lie within the OLR circle, and the pattern speeds are then too
fast for any orbits to be in CR. All the unstable modes transfer
angular momentum and kinetic energy outwards, and most release
gravitational energy and convert it to kinetic energy. This flow of
angular momentum and energy is derived from the second order extension
of a linear theory, and so can describe only the early stages of the
growth of an instability, and not its later fully nonlinear
development. Only a few Fourier components account for almost all of
the angular momentum and energy.  Polyachenko's (2004) theory, which
is equivalent to ignoring all but one term of our Fourier development,
seems to be an oversimplification.




\acknowledgments
This work has been supported in part by the National Science Foundation
through grant DMS-0104751.


\appendix

\section{ANGULAR MOMENTUM AND ENERGY}
\label{app::angmomenergy}

Both the perturbed angular momentum ${\cal L}_2$  and the first component
${\cal K}_{2,1}$ of the kinetic energy contain integrals of the form
\begin{equation}
\label{eq::Iintegral}
{\cal I}_2(t)=\int\!\!\!\int S(J_R,J_{\phi})f_2{\rm d}{\bf J}d{\Theta},
\end{equation}
for different functions $S$. Such integrals can be evaluated as
follows
\begin{equation}
{{\rm d}{\cal I}_2(t) \over {\rm d}t}=
  \int\!\!\!\int S(J_R,J_{\phi}){\partial f_2 \over
                 \partial t}{\rm d}{\bf J}{\rm d}{\Theta}
          = \int\!\!\!\int S(J_R,J_{\phi})\left({\partial f_2 \over
                 \partial t}+[f_2,{\cal H}_0]\right)
             {\rm d}{\bf J}{\rm d}{\Theta},
\end{equation}
because the terms added are angle derivatives of periodic functions 
which integrate to zero over the angles. 
Equation (\ref{eq::Boltzmann-equationtwo}) gives the term in parentheses
as the sum of two terms. The first contributes
\begin{equation}
\int\!\!\!\int -S[f_0 , V_2]{\rm d}{\bf J}{\rm d}{\Theta} = \int\!\!\!\int 
         S {\partial f_0 \over \partial J_{\nu}}
    {\partial V_2 \over \partial \theta_{\nu}} 
              {\rm d}{\bf J}{\rm d}{\Theta} 
    = \int\!\!\!\int {\partial \over \partial \theta_{\nu}}
    \left( SV_2 {\partial f_0 \over \partial J_{\nu}} \right)
                     {\rm d}{\bf J}{\rm d}{\Theta}=0.
\end{equation}
Here the subscript $\nu$ represents the pair of subscripts $(R,\phi)$, and
we suppose the summation convention to apply to it. The last step again
uses the fact that any integral of an integrand which is
a derivative with respect to an angle vanishes.
The second term contributes
\begin{equation}
\label{eq::Sintegral}
\int\!\!\!\int -S[f_1,V_1]
     {\rm d}{\bf J}{\rm d}{\Theta}  = \int\!\!\!\int S \left[
     {\partial f_1 \over \partial J_{\nu}}
     {\partial V_1 \over \partial \theta_{\nu}}
    -{\partial f_1 \over \partial \theta_{\nu}}
     {\partial V_1 \over \partial J_{\nu}}\right] 
      {\rm d}{\bf J}{\rm d}{\Theta}
   = \int\!\!\!\int \left[ -f_1 {\partial V_1 \over \partial \theta_{\nu}}
     {\partial S \over \partial J_{\nu}}
    +{\partial V_1 \over \partial \theta_{\nu}}
     {\partial (Sf_1) \over \partial J_{\nu}}
    -{\partial (Sf_1)\over \partial \theta_{\nu}}
     {\partial V_1 \over \partial J_{\nu}}\right]
       {\rm d}{\bf J}{\rm d}{\Theta}.
\end{equation}
The combination of the second and third components vanishes because it
can be combined to an integral of a divergence:
\begin{equation}
\int\!\!\!\int\left[{\partial \over \partial J_{\nu}}
 \left(Sf_1{\partial V_1 \over \partial \theta_{\nu}}\right)-
 {\partial \over \partial \theta_{\nu}}
 \left(Sf_1 {\partial V_1 \over \partial J_{\nu}}\right)\right]
 {\rm d}{\bf J}{\rm d}{\Theta}.
\end{equation}
The second set of component with angle derivatives integrate to zero,
but so too do the integrals of derivatives with respect to the actions. 
That is because the differentiated quantities vanish 
at the limits in action space,
as $J_{\phi}\to \pm \infty$ and $J_R \to \infty$ where the perturbation
tends to zero, and at $J_R=0$ where $V_1$ is independent of $\theta_R$ 
because $\Psi^m_{l,j}(0,J_{\phi})=0$ for $l \neq 0$
(cf \S\ref{sec::boundary-integrals}).
The remaining first component of (\ref{eq::Sintegral}) 
can be evaluated for $S=J_{\phi}$ as in (\ref{eq::Wtwoone}), to
give
\begin{equation}
\label{eq::dLtwodt}
{{\rm d} {\cal L}_2(t) \over {\rm d}t} =
    -2ms\pi^2e^{2st} \sum_{l=-\infty}^{\infty}\int {\rm d}{\bf J}
 \left(l{\partial f_0\over \partial J_R}
       +m{\partial f_0\over \partial J_{\phi}} \right) 
    {|\tilde V_{l}|^2 \over |l\Omega _R+m\Omega _{\phi} -\omega|^2}.
\end{equation}
This result agrees with that of Lynden-Bell \& Kalnajs (1972) when account
is taken of the fact that their Fourier coefficients are larger than ours
by a factor of $4\pi^2$, and their $\omega$ has the opposite sign.
For $S={\cal H}_0$, we get
\begin{equation}
{{\rm d}\over {\rm d}t}\int\!\!\!\int {\cal H}_0f_2
          {\rm d}{\bf J}{\rm d}{\Theta} =
    -2s\pi^2e^{2st} \sum_{l=-\infty}^{\infty}\int {\rm d}{\bf J}
 \left(l{\partial f_0\over \partial J_R}
       +m{\partial f_0\over \partial J_{\phi}} \right) 
     {(l\Omega _R+m\Omega _{\phi})|\tilde V_{l}|^2
        \over |l\Omega _R+m\Omega _{\phi} -\omega|^2},
\end{equation}
A simple consequence is that
\begin{equation}
{{\rm d} {\cal E}_2 \over {\rm d}t}=
{{\rm d} ({\cal K}_2+{\cal W}_2) \over {\rm d}t}=
\Omega_p {{\rm d} {\cal L}_2 \over {\rm d}t},
\end{equation}
where ${\cal E}$ is the total energy. The undifferentiated values
which are quoted in \S\ref{sec::angularmomentumenergy}
follow because of the simple time dependence on $e^{2st}$.

A deeper analysis of the second order equation
(\ref{eq::Boltzmann-equationtwo}), though not its full solution, 
is needed to evaluate the integral
\begin{equation}
\label{eq::Wtwotwo}
{\cal W}_{2,2}=-{\cal K}_{2,2}=\int\!\!\!\int V_0f_2 
                {\rm d}{\bf J}{\rm d}{\Theta}.
\end{equation}
We rewrite equation (\ref{eq::Boltzmann-equationtwo}) as
\begin{equation}
\label{eq::newBoltzmann-equationtwo}
{\partial f_2\over \partial t}+[f_2,{\cal H}_0] +[f_0,V_2]
    = -[f_1,V_1]. 
\end{equation}
The left hand side of (\ref{eq::newBoltzmann-equationtwo}), which 
is homogeneous in subscript 2 quantities, has
the same form as the first order problem for which we derived the
homogeneous linear equations (\ref{eq::dispersion-relation}).
Equation (\ref{eq::newBoltzmann-equationtwo}) leads in a similar way 
to inhomogeneous linear equations. Its right hand side 
contains both axisymmetric terms
and non-axisymmetric ones with wavenumber $2m$. We need consider only
the axisymmetric terms and the part of the solution for $f_2$ which
they cause, because only they will contribute to the integral
(\ref{eq::Wtwotwo}) for ${\cal W}_{2,2}$. They have a Fourier expansion
\begin{equation}
e^{2st}\sum_{l=-\infty}^{\infty}\tilde N_l(J_R,J_{\phi}) e^{il\theta _R}, 
~~ {\rm where} ~~
e^{2st}\tilde N_l=\frac{1}{(2\pi)^2}\int {\rm d}{\Theta}e^{-il\theta _R}
\left(-\frac{1}{4}[f_1,\bar V_1]-\frac{1}{4}[\bar f_1,V_1]\right).
\end{equation}
We represent the potential and density of the axisymmetric  part of $f_2$ by
series
\begin{equation}
V_2 = e^{2st}\sum_{j=0}^{\infty} a_j \psi^0_j(R), ~~
\Sigma _2 = e^{2st}\sum_{j=0}^{\infty} a_j \sigma^0_j(R),
\end{equation}
like those of equations (\ref {eq::pot-exp}) and (\ref {eq::density-exp})
but now in a complete set of axisymmetric basis functions.
We use Fourier expansions
\begin{equation}
f_2= e^{2st} \sum_{l=-\infty}^{\infty}
      \tilde g_{l}(J_R,J_{\phi})e^{il\theta _R},~~
V_2=e^{2st} \sum_{l=-\infty}^{\infty}
      \tilde U_{l}(J_R,J_{\phi})e^{il\theta _R},
\end{equation}
like those of equations (\ref{eq::expansion-f})
and (\ref{eq::expansion-V}), with
\begin{equation}
\tilde U_{l}= \sum_{j=0}^{\infty}a_j \Psi^0_{l,j},
\end{equation}
and with Fourier coefficients $\Psi^0_{l,j}$ defined as in equation 
(\ref{eq::fourier-coeffs}) for $m=0$. Matching Fourier coefficients 
in equation (\ref{eq::newBoltzmann-equationtwo}) gives
\begin{equation}
\label{eq::fcoeffeqtwo}
(2s+il\Omega_R)\tilde g_{l}-il{\partial f_0\over \partial J_R}\tilde U_{l}
 =\tilde  N_l.
\end{equation}
Then, following the same procedure as used in \S\ref{sec::linear-pert-theory},
we obtain the matrix equation
\begin{equation}
\label{eq::matrix-equationtwo}
[{\bf M}(0,2is)-{\bf D}(0)]{\bf a}={\bf h},
\end{equation}
where the components of the column vector ${\bf h}$ are given by
\begin{equation}
\label{eq::hjintegral}
h_j=4\pi^2 \sum_{l=-\infty}^{\infty}\int {\rm d}{\bf J}
            {i\tilde  N_l \Psi^0_{l,j} \over (l\Omega _R-2is)}.
\end{equation}
The matrix ${\bf M}(0,2is)$ is real because each $\pm l$ pair in the
sum (\ref{eq::matrix-elements}) combines two complex conjugate
quantities, because $\Psi^0_{l,j}$ is even in $l$ [cf
eq. (\ref{eq::fourier-coeffs})].  The right hand side ${\bf h}$ of
equation (\ref{eq::matrix-equationtwo}) is real because the $\pm l$
pairs in the sum (\ref{eq::hjintegral}) also combine two complex
conjugate quantities, due also to the fact that $\tilde
N_l=\bar{\tilde N_{-l}}$ because the $\tilde N_l$ are the Fourier
coefficients of a real function.  Hence equation
(\ref{eq::matrix-equationtwo}) is a real matrix equation, and its
solution for the unknown vector ${\bf a}$ is real.  Knowing ${\bf a}$,
we can evaluate
\begin{equation}
\label{eq::W22-equation}
{\cal W}_{2,2}=\int V_0\Sigma _2 {\rm d}{\bf x}
= 2\pi e^{2st}\sum_{j=0}^{\infty} a_j \int\limits_{0}^{\infty}
V_0(R) \sigma^0_j(R)R{\rm d}R.
\end{equation}
The reason why it is so much easier to compute ${\cal W}_{2,1}$,
${\cal K}_{2,1}$,  and 
${\cal L}_2$ is that they need only the single Fourier coefficient
$\tilde g_0$. Equation (\ref{eq::fcoeffeqtwo}) gives $\tilde g_0$ simply
as $\tilde N_0/2s$ and the matrix equation is not needed.

The computation of ${\bf a}$ can be checked by verifying that the total
mass due to the axisymmetric density $\Sigma_2$ vanishes. This is
\begin{equation}
\label{eq::Sigma2mass}
{\cal M}_2(t)=2\pi e^{2st}\sum_{j=0}^{\infty} a_j \int\limits_{0}^{\infty}
\sigma^0_j(R)R{\rm d}R=0.
\end{equation}
Formally, the constancy of ${\cal M}_2$ follows from the analysis of
Appendix \ref{app::angmomenergy}; it is the $S=1$ case of integral 
(\ref{eq::Iintegral}).
With the Clutton-Brock functions (\ref{eq::CBfunctions-den}), the sum
(\ref{eq::Sigma2mass}) can be evaluated as 
$(4be^{2st}/G)\sum_{j=0}^{\infty} a_j$ using GR formula (7.225.3).

Substituting the Fourier series (\ref{eq::expansion-f}) and
(\ref{eq::expansion-V}) for $f_1$ and $V_1$ and carrying out the
angle integrations gives
\begin{equation}
\label{eq::iNl-equation}
i\tilde  N_l =\frac{1}{4}\sum_{k=-\infty}^{\infty}\left[
\left(k{\partial \over \partial J_R}+m{\partial \over \partial J_{\phi}}\right)
\left(\tilde f_k \bar{\tilde V}_{k-l}
 -\bar{\tilde f_k} \tilde V_{k+l}\right)
-l\left(\bar{\tilde V}_{k-l}{\partial \tilde f_k \over \partial J_R}
 +\tilde V_{k+l}{\partial \bar{\tilde f_k} \over \partial J_R}\right)
\right]
\end{equation}
Derivatives of $\tilde f_k$, and hence second order derivatives of $f_0$, 
are avoided by integrating equation (\ref{eq::hjintegral}) by parts 
with respect to the actions. This gives the following integral over
the whole action space:
\begin{eqnarray}
\label{eq::hj-equation}
h_j = \pi^2 \sum_{l=-\infty}^{\infty}\sum_{k=-\infty}^{\infty}
\int {\rm d}{\bf J} 
 \Biggl[ &-& \left(\tilde f_k \bar{\tilde V}_{k-l}
-\bar{\tilde f_k} \tilde V_{k+l}\right)
\left(k{\partial \over \partial J_R}+m{\partial \over \partial J_{\phi}}\right)
\left({\Psi^0_{l,j} \over l\Omega _R-2is} \right) \nonumber \\
 &+& l\tilde f_k{\partial \over \partial J_R}
\left({\bar{\tilde V}_{k-l}\Psi^0_{l,j} \over l\Omega _R-2is} \right)
+l\bar{\tilde f_k}{\partial \over \partial J_R}
\left({\tilde V_{k+l}\Psi^0_{l,j} \over l\Omega _R-2is} \right) 
\Biggr].
\end{eqnarray}
For the unidirectional disk with DF (\ref{eq::sharp-cut-DF}), 
$h_j$ is given by the integral (\ref{eq::hj-equation}) 
over $J_R \geq 0$, $J_{\phi} \geq 0$ with 
$f_0=f_0^P$, plus the following boundary integral:
\begin{eqnarray}
\label{eq::hj-boundaryintegral}
h^B_j  =  \pi^2 \sum_{l=-\infty}^{\infty}\sum_{k=-\infty}^{\infty}
\int_0^{\infty} {\rm d} J_R mf_0^P (J_R,0)  
 \Biggl[ &-& \left(\tilde U_k \bar{\tilde V}_{k-l}
-\bar{\tilde U_k} \tilde V_{k+l}\right)
\left(k{\partial \over \partial J_R}+m{\partial \over \partial J_{\phi}}\right)
\left({\Psi^0_{l,j} \over l\Omega _R-2is} \right) \nonumber \\
 &+& l\tilde U_k{\partial \over \partial J_R}
\left({\bar{\tilde V}_{k-l}\Psi^0_{l,j} \over l\Omega _R-2is} \right)
+l\bar{\tilde U_k}{\partial \over \partial J_R}
\left({\tilde V_{k+l}\Psi^0_{l,j} \over l\Omega _R-2is} \right) 
\Biggr]_{J_{\phi}=0},
\end{eqnarray}
where
\begin{equation}
\tilde U_k={\tilde V_k \over k\Omega_R+m\Omega_{\phi}-\omega}, ~~
\bar{\tilde U_k}={\bar{\tilde V_k} \over k\Omega_R+m\Omega_{\phi}-\bar\omega}.
\end{equation}
Note that the solution for ${\cal W}_{2,2}=-{\cal K}_{2,2}$ intermingle 
different Fourier components, unlike ${\cal L}_2$, ${\cal W}_{2,1}$,
and ${\cal K}_{2,1}$ for which the Fourier components can be separated
as in equations (\ref{eq::W_2_1_ell}) and (\ref{eq::Ltwo}).

The partial derivatives of $\Psi^m_{l,j}(J_R,J_{\phi})$ needed for
equations (\ref{eq::hj-equation}) and (\ref{eq::hj-boundaryintegral})
can most easily be calculated simultaneously with the 
$\Psi^m_{l,j}(J_R,J_{\phi})$. For this we differentiate equation
(\ref{eq::fourier-coeffs}) partially with respect to an action to obtain
\begin{equation}
{\partial \Psi^m_{l,j}\over \partial J_{\nu}} = 
{1\over \pi} \int\limits_{0}^{\pi} \left \{ 
{\partial \psi^m_j \over \partial R}{\partial R \over \partial J_{\nu}}
               \cos [l\theta _R+m(\theta _\phi -\phi)] 
-m\psi^m_j(R){\partial \over \partial J_\nu}\left(\theta_{\phi} -\phi\right)
\sin [l\theta _R+m(\theta _\phi -\phi)] \right \} {\rm d}\theta _R.
\label{eq::deriv-fourier-coeff}
\end{equation}
The variables $R$, $(\theta _\phi -\phi)$, and $v_R$ are to be 
regarded as functions of $(J_R,J_{\phi},\theta_R)$; there is 
no $\theta_{\phi}$ dependence because of axisymmetry. We use the 
fact that $v_R=dR/dt=(\partial R/\partial \theta_R)\Omega_R$ to obtain
\begin{equation}
\label{eq::dRdJ}
\frac{{\rm d}}{{\rm d}t}\left[{\partial R \over \partial J_\nu}\right]=
{\partial^2 R \over \partial \theta_R \partial J_\nu}\frac{d\theta_R}{dt}
=\Omega_R\frac{\partial}{\partial J_\nu}
\left({\partial R \over \partial \theta_R}\right)
=\Omega_R\frac{\partial}{\partial J_\nu}\left(\frac{v_R}{\Omega_R}\right)
={\partial v_R \over \partial J_\nu}-\frac{v_R}{\Omega_R}
{\partial \Omega_R \over \partial J_\nu}.
\end{equation}
We obtain in a similar way the equations
\begin{eqnarray}
\frac{{\rm d}}{{\rm d}t}\left[{\partial \over \partial J_\nu}
 (\theta _\phi -\phi)\right]
&=& {\partial \Omega_{\phi} \over \partial J_\nu}
-\frac{\delta_{\nu\phi}}{R^2}
+{2J_{\phi} \over R^3}{\partial R \over \partial J_\nu}
-\frac{1}{\Omega_R}\left[\Omega_{\phi}-\frac{J_{\phi}}{R^2}\right]
{\partial \Omega_R \over \partial J_\nu}, \label{eq::dthetadJ} \\
\frac{{\rm d}}{{\rm d}t}\left[{\partial v_R \over \partial J_\nu}\right]
&=& {2J_{\phi} \over R^3}\delta_{\nu\phi}
-\left[{3J_{\phi}^2 \over R^4}+V_0^{\prime\prime}(R)\right]
{\partial R \over \partial J_\nu}
-\frac{1}{\Omega_R}\left[{J_{\phi}^2 \over R^3}-V_0^{\prime}(R)\right]
{\partial \Omega_R \over \partial J_\nu}. \label{eq::dvdJ}
\end{eqnarray}
Here $\delta_{\nu\phi}$ is 1 for $\nu=\phi$ and 0 for $\nu=R$.
The set of three equations (\ref{eq::dRdJ}) through (\ref{eq::dvdJ}) 
can be added to the others to be integrated
for an orbit, and they provide the additional values needed to evaluate
the partial derivatives (\ref{eq::deriv-fourier-coeff}). 
Initial values are $\partial v_R/\partial J_\nu=
\partial(\theta_{\phi}-\phi)/\partial J_\nu=0$ at $\theta_R=t=0$
where $R=R_{\rm min}$ because $v_R=\theta_{\phi}-\phi=0$ there for
all orbits. However the initial $R_{\rm min}$ values change with the
actions, and initial values for the derivatives of $R$ with respect
to the actions are
\begin{equation}
\left[ {\partial R \over \partial J_R}\right]_{R=R_{\rm min}}=
\frac{R^3_{\rm min}\Omega_R}
{R^3_{\rm min}V_0^{\prime}(R_{\rm min})-J^2_{\phi}}, ~~
\left[ {\partial R \over \partial J_{\phi}}\right]_{R=R_{\rm min}}=
\frac{R_{\rm min}(R^2_{\rm min}\Omega_{\phi}-J_{\phi})}
{R^3_{\rm min}V_0^{\prime}(R_{\rm min})-J^2_{\phi}}.
\end{equation}
They are obtained by differentiating the zeroth order energy equation.

\section{LAGRANGIAN DESCRIPTION, AND BOUNDARY INTEGRALS AS FLUXES}
\label{app::lagrangian}
The matrix analysis of \S\ref{sec::dynamical-theory} uses an Eulerian
description of phase space. Kalnajs (1977) gives an alternative
Lagrangian description. An advantage of using a Lagrangian description
of phase space is that it automatically includes any contributions from
the motion of boundaries in phase space. Kalnajs's Lagrangian analysis,
with our definition (\ref{eq::expansion-V}) of Fourier coefficients, 
yields the formula
\begin{equation}
\label{eq::matrix-elements-lagrange} 
M_{jk}(m,\omega) = - 4\pi^2 \sum_{l=-\infty}^{\infty} 
           \int {\rm d}{\bf J}f_0(J_R,J_{\phi})
\left(l{\partial \over \partial J_R}
            +m{\partial \over \partial J_{\phi}}\right)
\left[{\Psi^m_{l,j}\Psi^m_{l,k} \over l\Omega_R+m\Omega_{\phi}-\omega}\right],
\end{equation}
with the unperturbed DF $f_0$ undifferentiated. Our equation 
(\ref{eq::matrix-elements}) follows after integrating
(\ref{eq::matrix-elements-lagrange}) by parts with respect to $J_R$ and
$J_{\phi}$. Those integrations introduce boundary terms at any boundary
of the region of integration in action space unless one or other of $f_0$
and the term in square brackets vanish at that boundary. In the case of
the wholly prograde DF of equation (\ref{eq::sharp-cut-DF}), integration
by parts with respect to $J_{\phi}$ gives precisely the
boundary integral ${\bf M}^{\rm B}$ as defined in (\ref{eq::MsupB-elements}),
plus the area integral (\ref{eq::MsupA-elements}) 
after also integrating with respect to $J_R$.
This shows that the boundary integrals, which arise from the step function
in the Eulerian description of \S\ref{sec::boundary-integrals},
can be explained as due to the motion of that boundary. The boundary
integral ${\bf M}^{\rm B}$ arises from the perturbation of a non-zero
population of radial orbits.

Lynden-Bell \& Kalnajs (1972) had earlier used a Lagrangian analysis
to calculate the perturbed angular momentum ${\cal L}_2$ and derived
the result
\begin{equation}
{{\rm d}{\cal L}_2(t) \over {\rm d}t} = 
    2ms\pi^2e^{2st} \sum_{l=-\infty}^{\infty}\int {\rm d}{\bf J}
    f_0(J_R,J_{\phi})
\left(l{\partial \over \partial J_R}+m{\partial \over \partial J_{\phi}}\right)
\left[{|\tilde V_{l}|^2 \over |l\Omega _R+m\Omega _{\phi} -\omega|^2}\right].
\end{equation}
They then integrated by parts to get an expression for the contribution
$L(h_1,h_2)$ to ${\cal L}_2$ from stars with angular momenta in the range
$(h_1,h_2)$. Their result is
\begin{eqnarray}
{{\rm d} L(h_1,h_2) \over {\rm d}t} & = &
    -2ms\pi^2e^{2st} \sum_{l=-\infty}^{\infty}
           \int\limits_{h_1}^{h_2} {\rm d}J_{\phi}
           \int\limits_{0}^{\infty}  {\rm d}J_R
 \left(l{\partial f_0\over \partial J_R}
       +m{\partial f_0\over \partial J_{\phi}} \right)
\left[{|\tilde V_{l}|^2 \over |l\Omega _R+m\Omega _{\phi} -\omega|^2}\right]
     \nonumber \\
   & + & 2m^2s\pi^2e^{2st} \sum_{l=-\infty}^{\infty}
           \int\limits_{0}^{\infty}  {\rm d}J_R
           \left[{f_0(J_R,J_{\phi}) \vert \tilde V_{l} \vert ^2 
        \over |l\Omega _R+m\Omega _{\phi} -\omega|^2}\right]^{J_{\phi}=h_2}
         _{J_{\phi}=h_1}.
\end{eqnarray}
This formula of an area integral and two boundary integrals is the
same as one obtains from applying the Eulerian equation
(\ref{eq::dLtwodt}) to the DF
$f_0(J_R,J_{\phi})H(J_{\phi}-h_1)H(h_2-J_{\phi})$ which represents the
stars with angular momenta in the range $(h_1,h_2)$.  Lynden-Bell \&
Kalnajs interpret the boundary integrals as representing fluxes
through the two boundaries. That interpretation, together with
equations (\ref{eq::quadformL}) and (\ref{eq::quadformWtwoone}) which
give a physical significance to the real and imaginary parts of the
matrix ${\bf M}$, shows that neglecting the boundary integral terms
(\ref{eq::MsupB-elements}) for a prograde disk of stars means
neglecting the contributions to the total potential energy and angular
momentum which arise from the perturbation of radial orbits. The numerical
results reported in \S\ref{subsec::example-Kuz-disk} shows that this
neglect can cause large errors.

\end{document}